\pdfoutput=1
\documentclass[12pt,a4paper]{article}
\usepackage{ifthen} % for conditional statements
\usepackage{multirow}
\newboolean{pdflatex}
\setboolean{pdflatex}{true} % False for eps figures 

\newboolean{articletitles}
\setboolean{articletitles}{true} % False removes titles in references

\newboolean{uprightparticles}
\setboolean{uprightparticles}{false} %True for upright particle symbols

\newboolean{inbibliography}
\setboolean{inbibliography}{true} %True once you enter the bibliography

\def\paperauthors{LHCb collaboration} % Leave as is for PAPER and CONF
\def\paperasciititle{Measurement of psi(2S) production cross-sections in proton-proton collisions at 7 and 13 TeV} % Set ASCII title here
\def\papertitle{Measurement of $\psitwos$ production cross-sections in proton-proton collisions at $\sqs=7$ and $13\tev$} % Latex formatted title
\def\paperkeywords{{High Energy Physics}, {LHCb}} % Comma separated list
\def\papercopyright{\the\year\ CERN for the benefit of the LHCb collaboration} % new since 9/Apr/2018
\def\paperlicence{CC-BY-4.0 licence}
\def\paperlicenceurl{https://creativecommons.org/licenses/by/4.0/}

\RequirePackage[normalem]{ulem} %DIF PREAMBLE
\RequirePackage{color}\definecolor{RED}{rgb}{1,0,0}\definecolor{BLUE}{rgb}{0,0,1} %DIF PREAMBLE
\usepackage[top=1in, bottom=1.25in, left=1in, right=1in]{geometry}

\usepackage{microtype}
\usepackage{lineno}  % for line numbering during review
\usepackage{xspace} % To avoid problems with missing or double spaces after
\usepackage{caption} %these three command get the figure and table captions automatically small

\usepackage{graphicx}  % to include figures (can also use other packages)
\usepackage{color}
\usepackage{colortbl}
\graphicspath{{./figs/}} % Make Latex search fig subdir for figures
\DeclareGraphicsExtensions{.pdf,.PDF,png,.PNG}

\usepackage{amsmath} % Adds a large collection of math symbols
\usepackage{amssymb}
\usepackage{amsfonts}
\usepackage{upgreek} % Adds in support for greek letters in roman typeset

\newcommand*\patchAmsMathEnvironmentForLineno[1]{%
\expandafter\let\csname old#1\expandafter\endcsname\csname #1\endcsname
\expandafter\let\csname oldend#1\expandafter\endcsname\csname
end#1\endcsname
 \renewenvironment{#1}%
   {\linenomath\csname old#1\endcsname}%
   {\csname oldend#1\endcsname\endlinenomath}%
}
\newcommand*\patchBothAmsMathEnvironmentsForLineno[1]{%
  \patchAmsMathEnvironmentForLineno{#1}%
  \patchAmsMathEnvironmentForLineno{#1*}%
}
\AtBeginDocument{%
\patchBothAmsMathEnvironmentsForLineno{equation}%
\patchBothAmsMathEnvironmentsForLineno{align}%
\patchBothAmsMathEnvironmentsForLineno{flalign}%
\patchBothAmsMathEnvironmentsForLineno{alignat}%
\patchBothAmsMathEnvironmentsForLineno{gather}%
\patchBothAmsMathEnvironmentsForLineno{multline}%
\patchBothAmsMathEnvironmentsForLineno{eqnarray}%
}

\usepackage{hyperxmp}

\usepackage[pdftex,
            pdfauthor={\paperauthors},
            pdftitle={\paperasciititle},
            pdfkeywords={\paperkeywords},
            pdfcopyright={Copyright (C) \papercopyright},
            pdflicenseurl={\paperlicenceurl}]{hyperref}

\usepackage[colorinlistoftodos,textsize=scriptsize]{todonotes}

\usepackage[all]{hypcap} % Internal hyperlinks to floats.

\usepackage{xspace} 
\usepackage{upgreek}

\def\lhcb   {\mbox{LHCb}\xspace}

\def\cdf    {\mbox{CDF}\xspace}

\def\lhc    {\mbox{LHC}\xspace}

\def\MagUp {\mbox{\em Mag\kern -0.05em Up}\xspace}

\ifthenelse{\boolean{uprightparticles}}%
{

 \def\Pmu         {\ensuremath{\upmu}\xspace}

 \def\Pchi        {\ensuremath{\upchi}\xspace}                 
 \def\Ppsi        {\ensuremath{\uppsi}\xspace}

 \def\PDelta      {\ensuremath{\Delta}\xspace}                 
 \def\PXi         {\ensuremath{\Xi}\xspace}                 
 \def\PLambda     {\ensuremath{\Lambda}\xspace}                 
 \def\PSigma      {\ensuremath{\Sigma}\xspace}                 
 \def\POmega      {\ensuremath{\Omega}\xspace}                 
 \def\PUpsilon    {\ensuremath{\Upsilon}\xspace}

 \def\PB      {\ensuremath{\mathrm{B}}\xspace}                 
                  
 \def\PD      {\ensuremath{\mathrm{D}}\xspace}

 \def\PJ      {\ensuremath{\mathrm{J}}\xspace}                 
 \def\PK      {\ensuremath{\mathrm{K}}\xspace}

 \def\Pb      {\ensuremath{\mathrm{b}}\xspace}                 
 \def\Pc      {\ensuremath{\mathrm{c}}\xspace}                 
                  
 \def\Pe      {\ensuremath{\mathrm{e}}\xspace}

 \def\Pi      {\ensuremath{\mathrm{i}}\xspace}

 \def\thebaroffset{0.0em}
}
{

 \def\Pmu         {\ensuremath{\mu}\xspace}

 \def\Pchi        {\ensuremath{\chi}\xspace}                 
 \def\Ppsi        {\ensuremath{\psi}\xspace}                 
                  
 \mathchardef\PDelta="7101
 \mathchardef\PXi="7104
 \mathchardef\PLambda="7103
 \mathchardef\PSigma="7106
 \mathchardef\POmega="710A
 \mathchardef\PUpsilon="7107
                  
 \def\PB      {\ensuremath{B}\xspace}                 
                  
 \def\PD      {\ensuremath{D}\xspace}

 \def\PJ      {\ensuremath{J}\xspace}                 
 \def\PK      {\ensuremath{K}\xspace}

 \def\Pb      {\ensuremath{b}\xspace}                 
 \def\Pc      {\ensuremath{c}\xspace}                 
                  
 \def\Pe      {\ensuremath{e}\xspace}

 \def\Pi      {\ensuremath{i}\xspace}

 \def\thebaroffset{0.18em}
}
\newcommand{\offsetoverline}[2][\thebaroffset]{\kern #1\overline{\kern -#1 #2}}%

\makeatletter
\ifcase \@ptsize \relax% 10pt
  \newcommand{\miniscule}{\@setfontsize\miniscule{4}{5}}% \tiny: 5/6
\or% 11pt
  \newcommand{\miniscule}{\@setfontsize\miniscule{5}{6}}% \tiny: 6/7
\or% 12pt
  \newcommand{\miniscule}{\@setfontsize\miniscule{5}{6}}% \tiny: 6/7
\fi
\makeatother

\DeclareRobustCommand{\optbar}[1]{\shortstack{{\miniscule (\rule[.5ex]{1.25em}{.18mm})}
  \\ [-.7ex] $#1$}}

\def\epem       {{\ensuremath{\Pe^+\Pe^-}}\xspace}
\def\mumu       {{\ensuremath{\Pmu^+\Pmu^-}}\xspace}

\def\cquark    {{\ensuremath{\Pc}}\xspace}

\def\bquark    {{\ensuremath{\Pb}}\xspace}

\def\KorKbar {\kern \thebaroffset\optbar{\kern -\thebaroffset \PK}{}\xspace}

\def\DorDbar {\kern \thebaroffset\optbar{\kern -\thebaroffset \PD}\xspace}

\def\BorBbar {\kern \thebaroffset\optbar{\kern -\thebaroffset \PB}\xspace}

\def\jpsi     {{\ensuremath{{\PJ\mskip -3mu/\mskip -2mu\Ppsi\mskip 2mu}}}\xspace}
\def\psitwos  {{\ensuremath{\Ppsi{(2S)}}}\xspace}

\def\chicJ    {{\ensuremath{\Pchi_{\cquark J}}}\xspace}

\def\Y#1S{\ensuremath{\PUpsilon{(#1S)}}\xspace}

\def\LorLbar     {\kern \thebaroffset\optbar{\kern -\thebaroffset \PLambda}\xspace}

\def\BF         {{\ensuremath{\mathcal{B}}}\xspace}
\def\BR         {\BF}

\def\to                 {\ensuremath{\rightarrow}\xspace}

\newcommand{\etot}{{\ensuremath{\varepsilon_{\mathrm{ tot}}}}\xspace}

\def\AT#1     {\ensuremath{A_{\mathrm{T}}^{#1}}\xspace}           % 2

\def\C#1      {\ensuremath{\mathcal{C}_{#1}}\xspace}                       % 9
\def\Cp#1     {\ensuremath{\mathcal{C}_{#1}^{'}}\xspace}                    % 7
\def\Ceff#1   {\ensuremath{\mathcal{C}_{#1}^{\mathrm{(eff)}}}\xspace}        % 9  
\def\Cpeff#1  {\ensuremath{\mathcal{C}_{#1}^{'\mathrm{(eff)}}}\xspace}       % 7
\def\Ope#1    {\ensuremath{\mathcal{O}_{#1}}\xspace}                       % 2
\def\Opep#1   {\ensuremath{\mathcal{O}_{#1}^{'}}\xspace}                    % 7

\newcommand{\nospaceunit}[1]{\ensuremath{\text{#1}}}       
\newcommand{\aunit}[1]{\ensuremath{\text{\,#1}}}       
\newcommand{\tev}{\aunit{Te\kern -0.1em V}\xspace}
\newcommand{\gev}{\aunit{Ge\kern -0.1em V}\xspace}
\newcommand{\mev}{\aunit{Me\kern -0.1em V}\xspace}
\newcommand{\kev}{\aunit{ke\kern -0.1em V}\xspace}
\newcommand{\ev}{\aunit{e\kern -0.1em V}\xspace}
\newcommand{\mevc}{\ensuremath{\aunit{Me\kern -0.1em V\!/}c}\xspace}
\newcommand{\gevc}{\ensuremath{\aunit{Ge\kern -0.1em V\!/}c}\xspace}
\newcommand{\mevcc}{\ensuremath{\aunit{Me\kern -0.1em V\!/}c^2}\xspace}
\newcommand{\gevcc}{\ensuremath{\aunit{Ge\kern -0.1em V\!/}c^2}\xspace}
\def\mum  {\ensuremath{\,\upmu\nospaceunit{m}}\xspace}

\def\mub{\ensuremath{\,\upmu\nospaceunit{b}}\xspace}
\def\nb {\aunit{nb}\xspace}

\def\pb {\aunit{pb}\xspace}
\def\invpb {\ensuremath{\pb^{-1}}\xspace}

\def\ps   {\ensuremath{\aunit{ps}}\xspace}

\newcommand{\stat}{\aunit{(stat)}\xspace}
\newcommand{\syst}{\aunit{(syst)}\xspace}

\newcommand{\chisq}{\ensuremath{\chi^2}\xspace}

\def\deriv {\ensuremath{\mathrm{d}}}

\def\gsim{{~\raise.15em\hbox{$>$}\kern-.85em
          \lower.35em\hbox{$\sim$}~}\xspace}
\def\lsim{{~\raise.15em\hbox{$<$}\kern-.85em
          \lower.35em\hbox{$\sim$}~}\xspace}

\def\sPlot{\mbox{\em sPlot}\xspace}

\def\sqs   {\ensuremath{\protect\sqrt{s}}\xspace}

\def\pt         {\ensuremath{p_{\mathrm{T}}}\xspace}

\def\ptot       {\ensuremath{p}\xspace}

\newcommand{\lum} {\ensuremath{\mathcal{L}}\xspace}
\def\evtgen     {\mbox{\textsc{EvtGen}}\xspace}

\def\geant      {\mbox{\textsc{Geant4}}\xspace}

\def\photos     {\mbox{\textsc{Photos}}\xspace}

\def\pythia     {\mbox{\textsc{Pythia}}\xspace}

\xspace

\def\tell1  {TELL1\xspace}
\def\ukl1   {UKL1\xspace}

\def\pp    {{\ensuremath{pp}}\xspace}
\def\Qbar  {{\ensuremath{\overline Q}}\xspace}
\def\QQbar {{\ensuremath{Q\Qbar}}\xspace}
\def\ppbar    {{\ensuremath{p\overline{p}}}\xspace}
\def\psimumu {{\ensuremath{\psitwos\to\mumu}}\xspace}
\def\psiee {{\ensuremath{\psitwos\to\epem}}\xspace}
\def\psifromb   {{\ensuremath{\psitwos\text{-from-}b}}\xspace}
\def\bratiopsi {{\ensuremath{\mathcal{B}(b \rightarrow \psi(2S)X)}}} % Add in the predefined LHCb symbols

\usepackage{cite} % Allows for ranges in citations
\usepackage{mciteplus}

\def\promptresult  {\ensuremath{1.430\pm0.005\stat\pm0.099\syst\mub}\xspace}
\def\frombresult   {\ensuremath{0.426\pm0.002\stat\pm0.030\syst\mub}\xspace}
\def\promptresultseven  {\ensuremath{0.471\pm 0.001\stat\pm 0.025\syst\mub}\xspace}
\def\frombresultseven   {\ensuremath{0.126\pm 0.001\stat\pm 0.008\syst\mub}\xspace}
\def\gevcsquare    {\ensuremath{{\,(\mathrm{Ge\kern -0.1em V\!/}c)^2}}\xspace}
\def\mygevc        {\ensuremath{{\mathrm{Ge\kern -0.1em V\!/}c}}\xspace}
\def\Rpsijpsi      {\ensuremath{R_{\psitwos/\jpsi}}\xspace}
\def\xx    {\ensuremath{\kern 0.5em }}
\usepackage{longtable} % only for template; not usually to be used in PAPERs
\usepackage{rotating}
\usepackage{enumitem}

\begin{document}

%%%%%%%%%%%%%%%%%%%%%%%%%
%%%%% Title     %%%%%%%%%
%%%%%%%%%%%%%%%%%%%%%%%%%
\renewcommand{\thefootnote}{\fnsymbol{footnote}}
\setcounter{footnote}{1}

% %%%%%%% CHOOSE TITLE PAGE--------
%\onecolumn
%\input{title-LHCb-INT}
%\input{title-LHCb-ANA}
%\input{title-LHCb-CONF}
% $Id: title-LHCb-PAPER.tex 122889 2018-08-17 17:59:55Z pkoppenb $
% ===============================================================================
% Purpose: LHCb-PAPER journal paper title page template
% Author: 
% Created on: 2010-09-25
% ===============================================================================

%%%%%%%%%%%%%%%%%%%%%%%%%
%%%%%  TITLE PAGE  %%%%%%
%%%%%%%%%%%%%%%%%%%%%%%%%
\begin{titlepage}
\pagenumbering{roman}

% Header ---------------------------------------------------
\vspace*{-1.5cm}
\centerline{\large EUROPEAN ORGANIZATION FOR NUCLEAR RESEARCH (CERN)}
\vspace*{1.5cm}
\noindent
\begin{tabular*}{\linewidth}{lc@{\extracolsep{\fill}}r@{\extracolsep{0pt}}}
\ifthenelse{\boolean{pdflatex}}% Logo format choice
{\vspace*{-1.5cm}\mbox{\!\!\!\includegraphics[width=.14\textwidth]{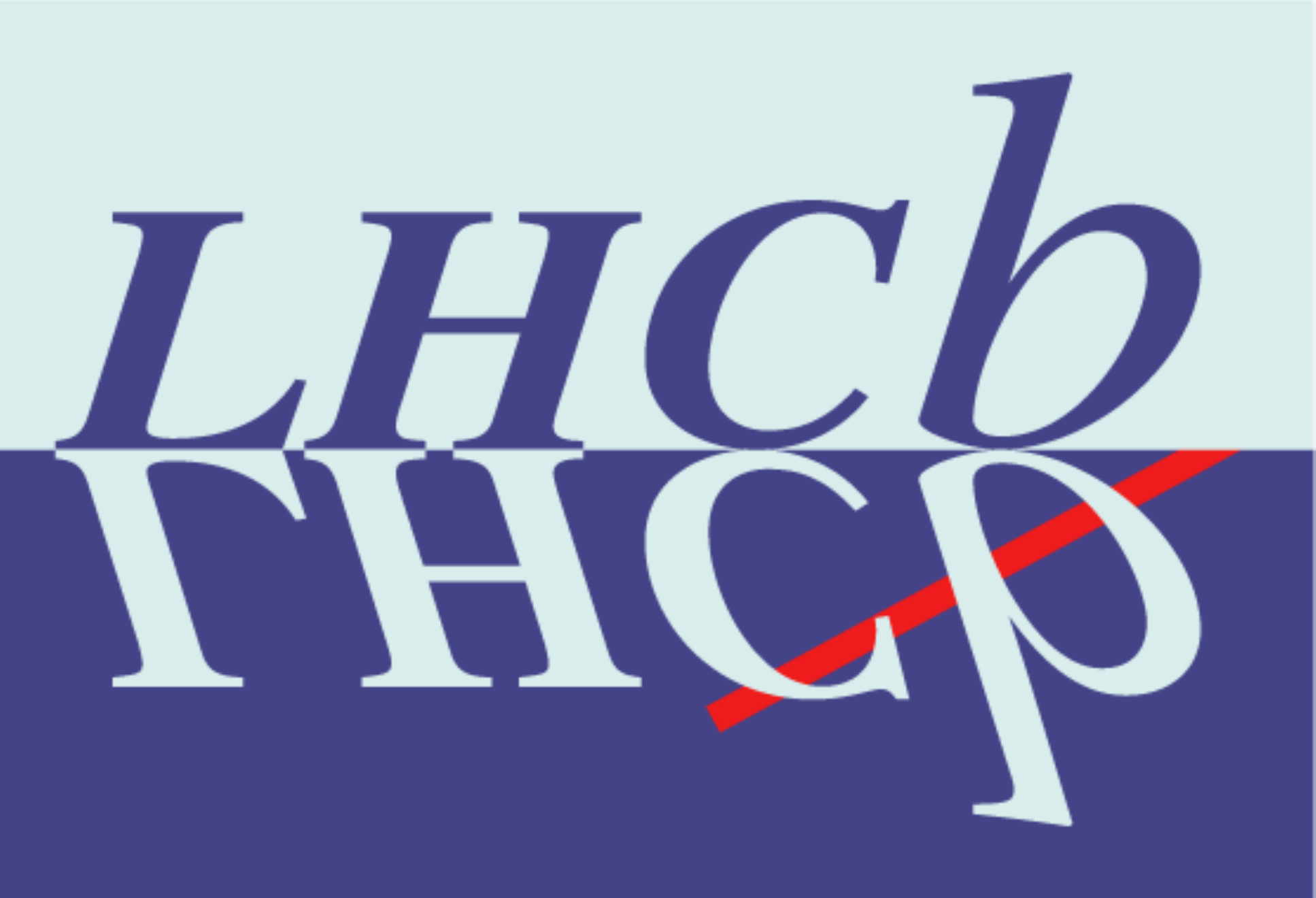}} & &}%
{\vspace*{-1.2cm}\mbox{\!\!\!\includegraphics[width=.12\textwidth]{lhcb-logo.eps}} & &}%
\\
 & & CERN-EP-2019-150 \\  % ID 
 & & LHCb-PAPER-2018-049 \\  % ID 
 & & 28 February 2020 \\ % Date - Can also hardwire e.g.: 23 March 2010 \today
\end{tabular*}

\vspace*{1.0cm}

% Title --------------------------------------------------
{\normalfont\bfseries\boldmath\huge
\begin{center}
% DO NOT EDIT HERE. Instead edit macro in main.tex to keep metadata correct
  \papertitle 
\end{center}
}

\vspace*{1.5cm}

% Authors -------------------------------------------------
\begin{center}
%In the footnote, replace 'paper' by 'Letter' in case of submission to PRL or PLB 
% Edit macro in main.tex to keep metadata correct
\paperauthors\footnote{Authors are listed at the end of this paper.}
\end{center}

\vspace{\fill}

% Abstract -----------------------------------------------
\begin{abstract}
  \noindent
The cross-sections of \psitwos meson production in proton-proton collisions at \mbox{$\sqs=13\tev$} are measured with a data sample collected by the \lhcb detector 
corresponding to an integrated luminosity of $275\invpb$.
The production cross-sections for prompt \psitwos mesons and those for \psitwos mesons from $b$-hadron decays ($\psitwos\mbox{-from-}\bquark$) are determined as functions of the transverse momentum, \pt, and the rapidity, $y$, of the \psitwos meson in the kinematic range $2<\pt<20\gevc$ and $2.0<y<4.5$.
The production cross-sections integrated over this kinematic region
are
    \begin{equation*}
        \begin{split}
            \sigma(\mbox{prompt }\psitwos,13\tev)&=\promptresult,\\ 
            \sigma(\psitwos\mbox{-from-}\bquark,13\tev)&=\frombresult.
        \end{split}
    \end{equation*}
A new measurement of \psitwos production cross-sections in $pp$ collisions at \mbox{$\sqrt{s}=7\tev$} is also performed using data collected in 2011, corresponding to an integrated luminosity of $614\invpb$. 
The integrated production cross-sections in the kinematic range \mbox{$3.5<\pt<14\gevc$} and \mbox{$2.0<y<4.5$} are
    \begin{equation*}
        \begin{split}
            \sigma(\mbox{prompt }\psitwos,7\tev)&=\promptresultseven,\\ 
            \sigma(\psitwos\mbox{-from-}\bquark,7\tev)&=\frombresultseven.
        \end{split}
    \end{equation*}
All results show reasonable agreement with theoretical calculations. 
  
\end{abstract}

\vspace*{2.0cm}

\begin{center}
  Published in Eur.~Phys.~J.~C~80 (2020) 185
\end{center}

\vspace{\fill}

{\footnotesize 
% Edit macro in main.tex to keep metadata correct
\centerline{\copyright~\papercopyright. \href{\paperlicenceurl}{\paperlicence}.}}
\vspace*{2mm}

\end{titlepage}

%%%%%%%%%%%%%%%%%%%%%%%%%%%%%%%%
%%%%%  EOD OF TITLE PAGE  %%%%%%
%%%%%%%%%%%%%%%%%%%%%%%%%%%%%%%%

%  empty page follows the title page ----
\newpage
\setcounter{page}{2}
\mbox{~}
%\newpage
%
%% Author List ----------------------------
%%  You need to get a new author list!
%\input{LHCb_authorlist.tex}
%
%The author list for journal publications is provided by the Membership Committee shortly after 'approval to go to paper' has been given.
%%It will be made available on the page
%%\verb!http://www.physik.uzh.ch/~strauman/forMemCo/LHCb-PAPER-XXXX-XXX/! .
%It will be sent to you by email shortly after a paper number has beens assigned.
%The author list should be included already at first circulation, 
%to allow new members of the collaboration to verify whether they have been included correctly.
%Occasionally a misspelled name is corrected or associated institutions become full members.
%In that case, a new author list will be sent to you.
%In case line numbering doesn't work well after including the authorlist, try moving the \verb!\bigskip! after the last author to a separate line.
%
%
%The authorship for Conference Reports should be ``The LHCb
%  collaboration'', with a footnote giving the name(s) of the contact
%  author(s), but without the full list of collaboration names.

\cleardoublepage

%\twocolumn
% %%%%%%%%%%%%% ---------

\renewcommand{\thefootnote}{\arabic{footnote}}
\setcounter{footnote}{0}

%%%%%%%%%%%%%%%%%%%%%%%%%%%%%%%%
%%%%%  Table of Content   %%%%%%
%%%%%%%%%%%%%%%%%%%%%%%%%%%%%%%%
%%%% Uncomment next 2 lines if desired
%\tableofcontents
%\cleardoublepage

%%%%%%%%%%%%%%%%%%%%%%%%%
%%%%% Main text %%%%%%%%%
%%%%%%%%%%%%%%%%%%%%%%%%%

\pagestyle{plain} % restore page numbers for the main text
\setcounter{page}{1}
\pagenumbering{arabic}

%% Uncomment during review phase. 
%% Comment before a final submission.
%%\linenumbers

% You can include short sections directly in the main tex file.
% However, for larger papers it is desirable to split the text into
% several semiautonomous files, which can be revised independently.
% This is especially useful when developing a document in
% collaboration with several people, since then different parts can be
% edited independently.  This type of file organization is shown here.
% 
\section{Introduction}
\label{sec:Introduction}
The study of hadronic production of heavy quarkonia can provide important information about quantum chromodynamics (QCD). 
The production of heavy quark pairs, $\QQbar$, can be calculated with perturbative QCD, while the hadronisation of $\QQbar$ pairs into heavy quarkonia is nonperturbative and must be determined using input from experimental results.
Heavy-quarkonium production therefore probes both perturbative and nonperturbative aspects of QCD by providing stringent tests of theoretical models.
Knowledge of hadronic production of heavy quarkonium has been significantly improved in the past forty years~\cite{Andronic:2015wma,Brambilla:2010cs}, but the mechanism behind it is still not fully understood.
Colour-singlet model calculations \cite{Carlson:1976cd,Donnachie:1976ue,Ellis:1976fj,Fritzsch:1977ay,Gluck:1977zm,Chang:1979nn,Baier:1981uk} require that the intermediate $\QQbar$ state is colourless and has the same $J^{PC}$ quantum numbers as those of the outgoing quarkonium state.
In the nonrelativistic QCD (NRQCD) approach~\cite{Bodwin:1994jh,Cho:1995vh,*Cho:1995ce}, intermediate $\QQbar$ states with all possible colour-spin-parity quantum numbers have nonzero probability to be transformed into the desired quarkonium. 
The transition probability of a $\QQbar$ pair into the quarkonium state is described by a long-distance matrix element (LDME), which is assumed to be universal and can be determined from experimental data.

In high-energy proton-proton ($pp$) collisions, charmonium states can be produced directly from hard collisions of partons inside the protons, through the feed-down from excited states, or via weak decays of $\bquark$ hadrons. The first two contributions, which cannot be distinguished experimentally, are referred to as prompt production; while the third component can be separated from prompt production by exploiting the lifetime of \bquark-hadrons. For prompt $\jpsi$ production the feed-down contribution is large, mostly from radiative decays of $\chicJ$ ($J=0,1,2$) mesons. This complicates the comparison between theoretical calculations and experimental results. On the contrary, the feed-down contribution to $\psitwos$ mesons is negligible~\cite{PDG2018}, thus theoretical calculations can be directly compared with measurements.

The studies of heavy quarkonium production are crucial to separate the contributions of single parton scattering~(SPS) \cite{LHCb-PAPER-2011-013} and double parton scattering~(DPS) \cite{LHCb-PAPER-2012-003} to multiple-quarkonium production. Multiple-quarkonium production through the SPS process shares the same LDMEs as the single quarkonium production, thus providing a new method to test the theoretical calculations. The DPS process can reveal the transverse profile of partons inside the proton.
Further theoretical and experimental works provide deeper insights on how to interpret the production mechanism of multiple quarkonia.
In particular, additional data help in improving the precision of LDME determination.

The differential cross-sections of inclusive \psitwos meson production in $\ppbar$ collisions at centre-of-mass energies of \mbox{$\sqs=1.8$} and \mbox{$1.96\tev$} were measured by the $\cdf$ experiment at the Fermilab Tevatron Collider~\cite{Abe:1997jz,Aaltonen:2009dm}, and in \pp collisions at  \mbox{$\sqs=7\tev$}~\cite{Chatrchyan:2011kc,LHCb-PAPER-2011-045,Abelev:2014qha,Aad:2014fpa,Khachatryan:2015rra,Aad:2015duc},
$8\tev$~\cite{Aad:2015duc},
and $13\tev$~\cite{Sirunyan:2017qdw} with \lhc data.
This paper presents measurements of \psitwos production cross-sections in $\pp$ collisions using a data sample collected by \lhcb in 2015 (2011) corresponding to an integrated luminosity of \mbox{$275\pm11\invpb$} at \mbox{$\sqs=13\tev$} (\mbox{$614\pm11\invpb$} at \mbox{$\sqs=7\tev$}).
The $\psitwos$ mesons from prompt production are abbreviated as ``prompt $\psitwos$", while those from $\bquark$-hadron decays are abbreviated as ``$\psitwos$-from-$\bquark$".
The \psitwos mesons are reconstructed through their decay mode $\psimumu$. 
The double-differential production cross-sections of prompt \psitwos and \psitwos-from-\bquark as functions of transverse momentum \pt and rapidity $y$
and their integrated production cross-sections are measured,
assuming zero polarisation of the \psitwos meson.
The kinematic region of the measurement at $13\tev$ ($7\tev$) is $2<\pt<20\gevc$ ($3.5<\pt<14\gevc$) and $2.0<y<4.5$.
Compared to the previous LHCb measurement at $7\tev$ using 2010 data~\cite{LHCb-PAPER-2011-045}, the new analysis at $7\tev$ has several advantages: 
the 2011 data sample is much larger than that in the previous measurement corresponding to an integrated luminosity of $36\invpb$,
the previous measurement did not provide the \psitwos production cross-section as a function of the rapidity $y$ because of the limited sample size, and
the same final state and offline selection criteria as those in the $13\tev$ measurement are used in the new $7\tev$ measurement.
This guarantees that the maximum number of systematic uncertainties cancel in the cross-section ratio between $13\tev$ and $7\tev$, which is measured in the present analysis. 
This represents a more stringent test of the theoretical models, since many of the experimental and theoretical uncertainties cancel.
Finally, the \psitwos meson differential production cross-sections are compared with those of the \jpsi meson at $\sqs = 13\tev$~\cite{LHCb-PAPER-2015-037}.

\section{Detector and simulation}
\label{sec:Detector}
The \lhcb detector~\cite{Alves:2008zz,LHCb-DP-2014-002} is a single-arm forward spectrometer covering the \mbox{pseudorapidity} range $2<\eta <5$, designed for the study of particles containing \bquark or \cquark quarks. 
The detector includes a high-precision tracking system consisting of a silicon-strip vertex detector surrounding the $pp$ interaction region~\cite{LHCb-DP-2014-001}, a large-area silicon-strip detector located upstream of a dipole magnet with a bending power of about $4{\mathrm{\,Tm}}$, and three stations of silicon-strip detectors and straw drift tubes~\cite{LHCb-DP-2013-003,LHCb-DP-2017-001} placed downstream of the magnet.
The tracking system provides a measurement of the momentum, \ptot, of charged particles with a relative uncertainty that varies from 0.5\% at low momentum to 1.0\% at 200\gevc. The minimum distance of a track to a primary vertex (PV), the impact parameter (IP), is measured with a resolution of $(15+29/\pt)\mum$,
where \pt is in\,\gevc.
Different types of charged hadrons are distinguished using information from two ring-imaging Cherenkov detectors~\cite{LHCb-DP-2012-003}. 
Photons, electrons and hadrons are identified by a calorimeter system consisting of scintillating-pad (SPD) and preshower detectors, an electromagnetic calorimeter and a hadronic calorimeter. 
Muons are identified by a system composed of alternating layers of iron and multiwire proportional chambers~\cite{LHCb-DP-2012-002}.
The online event selection is performed by a trigger~\cite{LHCb-DP-2012-004}, 
which consists of a hardware stage, based on information from the calorimeter and muon
systems, followed by a software stage, which applies a full event
reconstruction. 

Simulated samples are used to evaluate the \psitwos detection efficiency. 
In the simulation, $\pp$ collisions are generated using \pythia 8~\cite{Sjostrand:2007gs,Sjostrand:2006za} with a specific \lhcb configuration~\cite{LHCb-PROC-2010-056}. 
Decays of unstable particles are described by \evtgen~\cite{Lange:2001uf}, in which final-state radiation is generated using \photos~\cite{Golonka:2005pn}. 
Both the leading-order colour-singlet and colour-octet contributions are included in the generated prompt charmonium states~\cite{LHCb-PROC-2010-056,Bargiotti:2007zz}. These states are generated with zero polarisation.
The interaction of the generated particles with the detector, and its response, are implemented using the \geant toolkit~\cite{Allison:2006ve, *Agostinelli:2002hh} as described in Ref.~\cite{LHCb-PROC-2011-006}.  

\section{\boldmath Selection of \psitwos candidates}
\label{sec:selection}
The decay channel $\psitwos\to\mumu$ is used in the measurements of the \psitwos production cross-sections at both $13\tev$ and $7\tev$. The same strategy is used for both analyses, except for different trigger requirements.
The hardware trigger selects events that contain two tracks consistent with muon hypotheses, and the product of the transverse momenta of the two muons is required to be greater than $(1.3\gevc)^2$. 
At the software trigger stage the two muons are required to be oppositely charged, to have good track quality, to form a good-quality vertex, and to each have a momentum larger than $6\gevc$. The invariant mass of the \psitwos candidates is required to be within the range $3566<m_{\mumu}<3806\mevcc$. 
The transverse momentum of each muon is required to be larger than $0.3\gevc$ ($0.5\gevc$) and that of the \psitwos candidate is required to be larger than $2 \gevc$ ($3.5 \gevc$) for the $13\tev$ ($7\tev$) data trigger. 
Due to the different triggers, the \psitwos candidates are selected in different \pt ranges.
For the $13\tev$ data taking, an alignment and calibration of the detector is performed in near real-time~\cite{LHCb-PROC-2015-011} and updated constants are made available for the trigger. 

To suppress the background associated to random combination of tracks (combinatorial) more stringent criteria are applied offline on the $\psitwos$ vertex fit quality, the muon kinematics and particle identification requirements.
Each muon must have $\pt>1.2\gevc$ and $2.0<\eta<4.9$. 
At least one PV should be reconstructed in the event from at least four tracks in the vertex detector. 

For events with more than one PV, the $\psitwos$ candidate is associated to the PV
for which the difference in the $\chisq$ of the PV fit with and without the $\psitwos$ candidate is the smallest. This is equivalent to select the vertex with respect to which the signal candidate has the smallest impact parameter, compared to resolution. Simulation studies have shown that, using the above procedure, the fraction of candidates associated to the wrong PV is $0.3\%$, which is negligible.
To select \psitwos candidates, additional requirements on the pseudo decay time, $t_z$, $|t_z|<10\ps$, and its uncertainty, $\sigma_{t_z}$,  $\sigma_{t_z}<0.3\ps$, are applied. 
The pseudo decay time $t_z$ is defined as
\begin{equation}
t_z=\frac{\left(z_{\psitwos}-z_\mathrm{PV}\right)\times M_{\psitwos}}{p_z},
\end{equation}
where $z_\psitwos$ ($z_{\mathrm{PV}}$) is the $z$ coordinate of the reconstructed $\psitwos$ decay vertex (the PV),
    $p_z$ is the $z$-component of the measured $\psitwos$ momentum, 
    and $M_{\psitwos}$ is the world average $\psitwos$ mass~\cite{PDG2018}.
The $z$-axis is the direction of the proton beam pointing downstream into the LHCb acceptance~\cite{Alves:2008zz}.
The pseudo decay time defined above provides a good approximation of the $b$-hadron decay time \cite{Qian:1475409}
and is used to separate prompt \psitwos and \psitwos-from-\bquark candidates. 

\section{Cross-section determination}
\label{sec:fit}
The double-differential production cross-section for prompt \psitwos or \psitwos-from-$\bquark$ in a given $(\pt,y)$ bin is defined as 
\begin{equation}
  \frac{\deriv^2\sigma}{\deriv y\,\deriv \pt} 
  = \frac{N(\pt,y)}
         {\etot(\pt,y)\times\lum_{\text{int}}\times\BR\times\Delta y \times \Delta \pt}, 
\label{eqn:CrossSection}
\end{equation}
where $N(\pt,y)$ 
is the signal yield, $\etot(\pt,y)$ is the total detection efficiency of the $\psimumu$ decay evaluated independently for prompt \psitwos\ or \psitwos-from-$\bquark$ in the given $(\pt,y)$ bin, 
   $\lum_{\text{int}}$ is the integrated luminosity, 
   $\BR$ is the branching fraction of the decay $\psimumu$, 
and $\Delta\pt=1\gevc$ and $\Delta y=0.5$ are the bin widths.
The integrated luminosity is determined using the beam-gas imaging and, for the $7\tev$ data, also the van der Meer scan methods \cite{LHCb-PAPER-2014-047}.
Assuming lepton universality in electromagnetic decays,  \mbox{$\BR(\psiee)=(7.89\pm0.17)\times10^{-3}$}~\cite{PDG2016} is used in Eq.~\ref{eqn:CrossSection}, 
taking advantage of the much smaller uncertainty compared to the $\psimumu$ decay.
The difference of the two branching fractions introduced by the mass difference between electrons and muons is negligible. 

\begin{figure}[!tbp]
\centering
\begin{minipage}[t]{0.49\textwidth}
\centering
\includegraphics[width=1.0\textwidth]{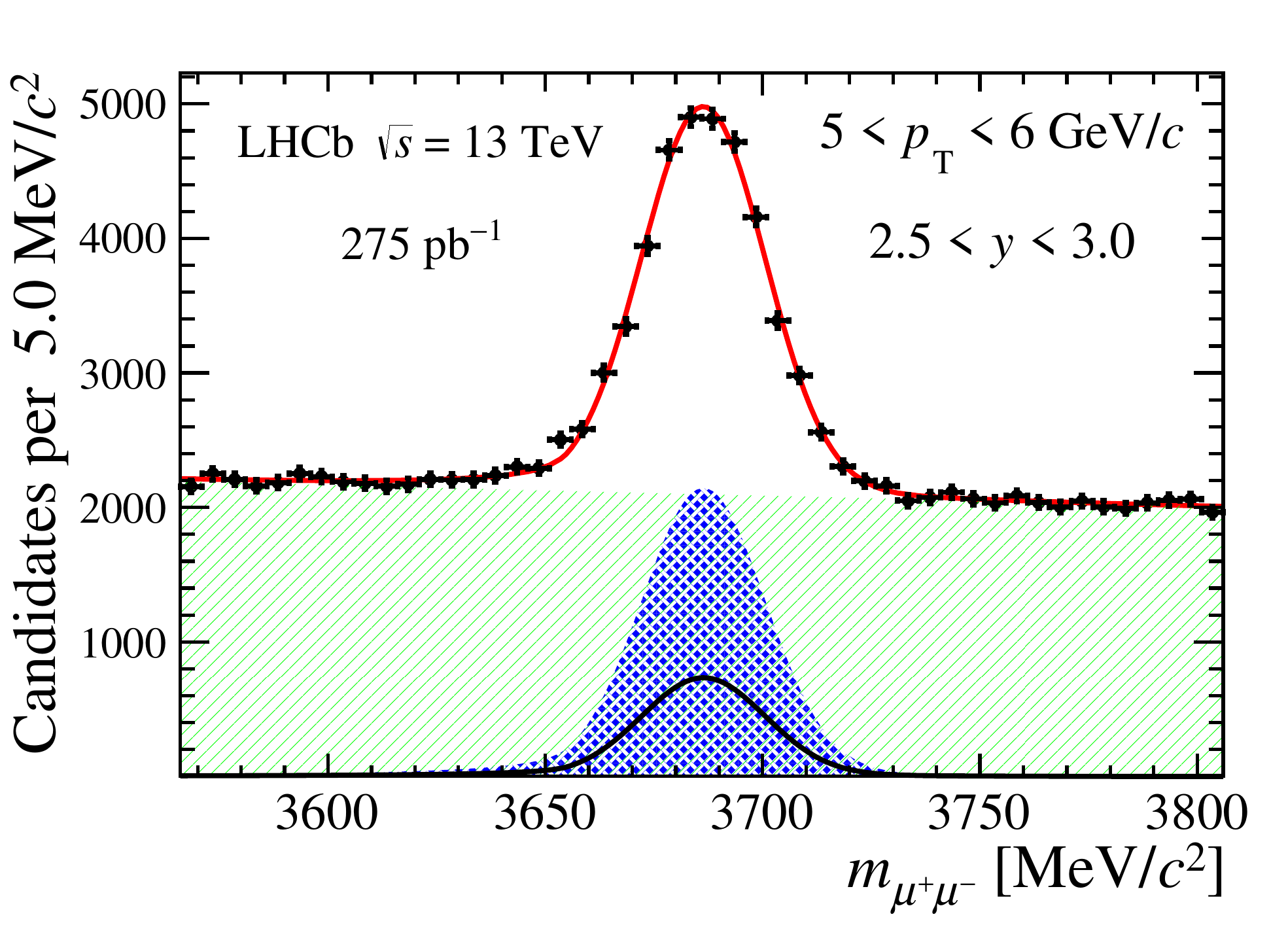}
\end{minipage}
\begin{minipage}[t]{0.49\textwidth}
\centering
\includegraphics[width=1.0\textwidth]{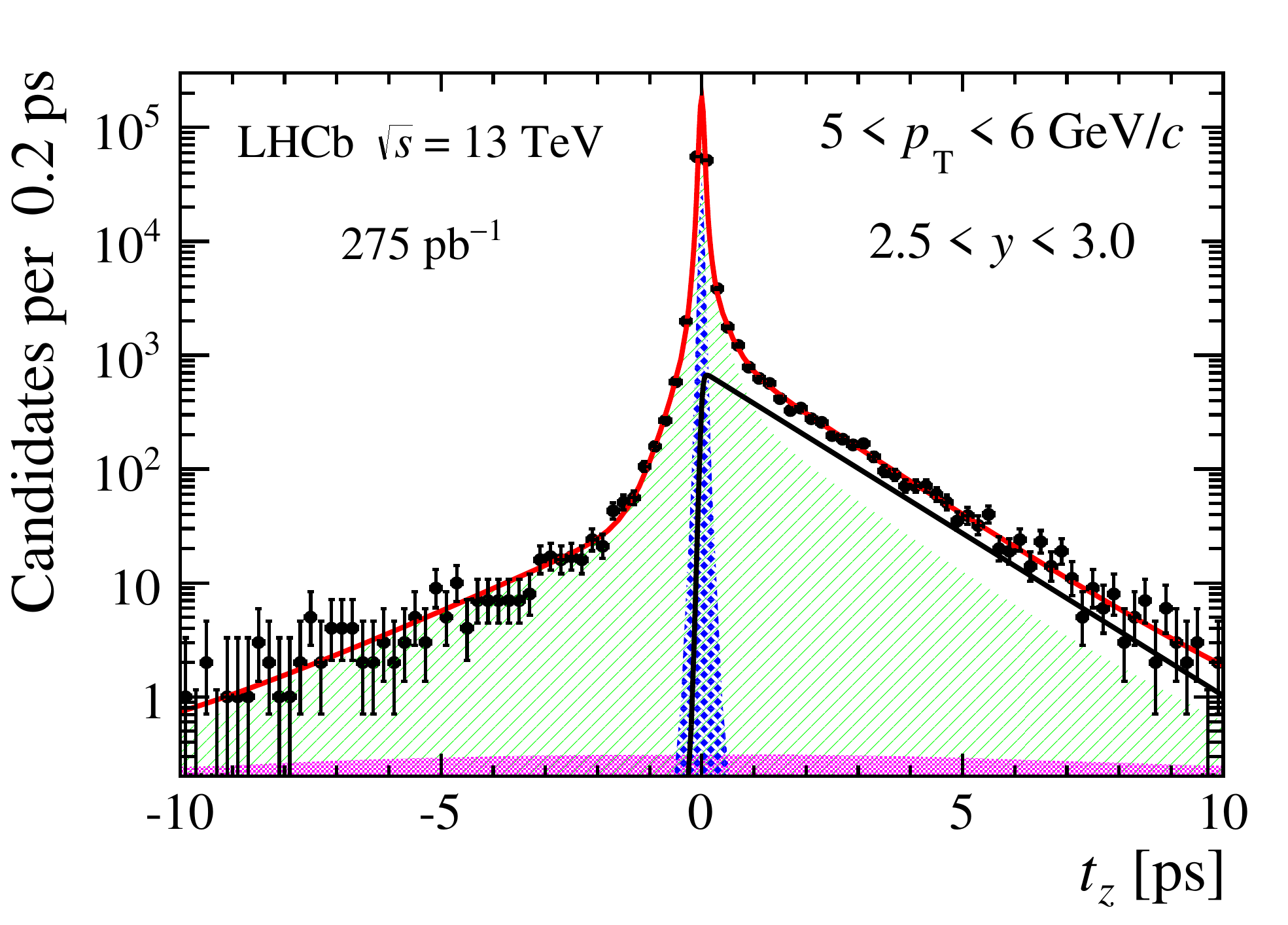}
\end{minipage}
\caption{Distributions of (left) the invariant mass $m_{\mumu}$ and (right) pseudo decay time $t_z$ of selected \psitwos candidates in the kinematic bin of $5<\pt<6\gevc$ and $2.5<y<3.0$ in the $13\tev$ data sample. Projections of the two-dimensional fit result are also shown. 
The solid (red) line is the total fit function, the shaded (green) area corresponds to the background component. 
The prompt \psitwos contribution is shown in cross-hatched (blue) area, \psitwos-from-$b$ in a solid (black) line and the tail contribution due to the association of \psitwos with the wrong PV is shown in filled (magenta) area. 
The tail contribution is invisible in the invariant-mass plot.}
\label{fig:FitResult}
\end{figure}

The yields of prompt \psitwos and \psitwos-from-\bquark candidates in each $(\pt,y)$ bin are determined from a two-dimensional extended unbinned maximum-likelihood fit to the distributions of the invariant mass, $m_{\mumu}$, and $t_z$ of the \psitwos candidates. The correlation between $m_{\mumu}$ and $t_z$ is found to be negligible. 
The invariant-mass distribution of the signal candidates in each bin is described by the sum of two Crystal Ball (CB) functions~\cite{Skwarnicki:1986xj} with a common mean value and different widths.
The parameters of the power-law tails, the relative fractions and the difference between the widths of the two CB functions are fixed to values obtained from simulation, leaving the mean value and the width of one of the CB functions as free parameters. 
The invariant-mass distribution of the combinatorial background is described by an exponential function with the slope parameter free to vary in the fit.
The $t_z$ distribution of prompt \psitwos mesons is described by a Dirac $\delta$ function at $t_z=0$, and that of \psitwos-from-$b$\ by an exponential function, both convolved with the sum of two Gaussian functions.
A \psitwos candidate can also be associated to a wrong PV, resulting in a long tail component in the $t_z$ distribution. 
This shape is modelled from data by calculating $t_z$ between the \psitwos candidate from a given event and the closest PV in the next event of the sample. 
The background $t_z$ distribution is parametrised with an empirical function based on the observed shape of the $t_z$ distribution in the \psitwos mass sidebands ($3566<m_{\mumu}<3620\mevcc$ and $3750<m_{\mumu}<3806\mevcc$). 
It is parametrised as the combination of a Dirac $\delta$ function and the sum of five exponential functions, three for positive $t_z$ and two for negative $t_z$.
This sum is convolved with the sum of two Gaussian functions.
All parameters of the background $t_z$ distribution are fixed to values determined from the \psitwos mass sidebands independently in each $(\pt,y)$ bin.
Figure~\ref{fig:FitResult} shows as an example the $m_{\mumu}$ and $t_z$ distributions in the kinematic bin corresponding to $5<\pt<6\gevc$ and $2.5<y<3.0$ for the $13\tev$ data sample. The one-dimensional projections of the fit result are also presented.
The total signal yields of prompt \psitwos and \psitwos-from-\bquark in the kinematic range for the $13\tev$ sample are $(440.7\pm 1.2) \times 10^3$ and $(140.0\pm 0.5) \times 10^3$, and for the $7\tev$ sample are $(433.9\pm 0.9) \times 10^3$ and $(115.1\pm 0.4) \times 10^3$, respectively.

%%%%%%%%%%%%%%%%%%%%%%%%%%%%%%%%%%%%%%%%%%%%%%%%%%%%%%%%%%%
%%%%%%%%%%%%%%%%%%%%%%%%%%%%%%%%%%%%%%%%%%%%%%%%%%%%%%%%%%

The total efficiency, $\etot$, in each kinematic bin is determined as the product of the geometrical acceptance of the detector and the efficiencies of particle reconstruction, event selection, muon identification and trigger requirements. 
The detector acceptance, selection and trigger efficiencies are calculated using simulated samples in each $(\pt,y)$ bin, independently for prompt \psitwos and \psitwos-from-$b$. 
The trigger efficiencies are also validated using data, as explained in Section~\ref{sec:systematics}. The track reconstruction and the muon-identification efficiencies are evaluated using simulated samples and calibrated with data. The efficiencies of prompt \psitwos and those of \psitwos-from-$b$ are very similar.
\section{Systematic uncertainties}
\label{sec:systematics}

A variety of sources of systematic uncertainty are studied as described below and are summarised in Table~\ref{tab:SystematicSummary}. For the uncertainties that vary in kinematic bins, the largest uncertainties always appear in the bins with small sample sizes.

\begin{table}[!bp]
\caption{Systematic uncertainties on the \psitwos cross-section measurements. The uncertainty from the $t_z$ fit only affects the $\psitwos$-from-$b$ result. Uncertainties labelled with ``$\ast$" are correlated between kinematic bins.}
\centering
\begin{tabular}{lcc}
\hline
    Source & $13\tev$ (\%) & $7\tev$ (\%) \\
\hline
    Signal mass shape$^{\ast}$ & $0.0$--$4.1$ & $0.0$--$8.5$  \\
    Radiative tail$^{\ast}$ & $1.0$ & $1.0$   \\
    Tracking$^{\ast}$ &  $(0.1$--$2.4)\oplus (2\times0.8)$ & $(0.7$--$3.0)\oplus(2\times0.4)$\\
    Muon ID$^{\ast}$ &  $(0.1$--$1.1)\oplus(0.1$--$4.6)$ & $(0.7$--$8.9)\oplus(0.4$--$5.4)$ \\
    Trigger$^{\ast}$   &  $0.1$--$9.3$ & $0.0$--$4.4$ \\
    Kinematic spectrum &  $0.0$--$2.0$ & $0.0$--$4.9$\\
    Luminosity$^{\ast}$ &  $3.9$ & $1.7$\\
    $\mathcal{B}(\psiee)^{\ast}$& $2.2$ & $2.2$\\
    Simulated sample size (prompt \psitwos) & $0.7$--$11.5$ & $1.3$--$13.1$ \\
    Simulated sample size (\psitwos-from-$\bquark$) & $0.8$--$5.7$ & $1.2$--$9.5$ \\
    $t_z$ fit$^{\ast}$ ($\psitwos$-from-$b$ only) &  $0.1$--$8.4$ & $0.1$--$9.2$\\
\hline
\end{tabular}\label{tab:SystematicSummary}
\end{table}

The uncertainty related to the modelling of the signal mass shape is studied by replacing the baseline model with a kernel-density estimated distribution \cite{Cranmer:2000du} obtained from the simulated sample in each kinematic bin. 
In order to account for the resolution difference between data and simulation, a Gaussian function is used to smear the shape of the distribution in simulation.
The relative difference of the signal yield in each kinematic bin, $0.0$--$4.1\%$ ($0.0$--$8.5\%$) for the $13\tev$ ($7\tev$) sample, is taken as the systematic uncertainty due to signal mass shape.

Due to the presence of final-state radiation in the \psimumu decay, a fraction of \psitwos candidates fall outside the mass window used to determine the signal yields.
The efficiency of the selection of the mass window is estimated using simulated samples, and the imperfect modelling of the radiation is studied by comparisons of the radiative tails between simulation and data, from which an uncertainty of $1.0\%$ is assigned to the cross-sections in all kinematic bins. 

The track detection efficiencies are determined from a simulated sample in each $(\pt,y)$ bin of the \psitwos meson, and are corrected by using $\jpsi \to \mumu$ decays reconstructed in a control data sample and in simulation.
These efficiencies are calculated as functions of $p$ and $\eta$ with a tag-and-probe approach~\cite{LHCb-DP-2013-002}.
The uncertainties due to the finite size of the control samples are propagated to the results using a large number of pseudoexperiments. In each pseudoexperiment, a new efficiency-correction ratio in each $(\pt,y)$ bin is generated according to a Gaussian distribution where the original ratio and its uncertainty are used as the Gaussian mean and standard deviation, respectively.
The contribution to the systematic uncertainty in each kinematic bin of \psitwos mesons varies from $0.1\%$ ($0.7\%$) to $2.4\%$ ($3.0\%$) for the $13\tev$ ($7\tev$) data sample.
The distribution of the number of SPD hits in simulation is weighted to match that in data to correct the effect of the detector occupancy. 
As a crosscheck the number of tracks is used as an alternative weighting variable. 
The tracking efficiencies are found to be different when different variables are used. Therefore, an additional systematic uncertainty of $0.8\%$ ($0.4\%$) per muon track is assigned for the $13\tev$ ($7\tev$) sample.

The muon identification efficiency is determined from simulation and calibrated with a data sample of $\jpsi\to\mumu$ decays.
The statistical uncertainty due to the finite size of the calibration sample is propagated to the final results using pseudoexperiments. 
The resulting uncertainties vary from $0.1\%$ ($0.7\%$) to $1.1\%$ ($8.9\%$) in different $(\pt,y)$ bins for the $13\tev$ ($7\tev$) sample.
The uncertainty related to the kinematic binning scheme of the calibration samples is studied by changing the size and the boundaries of the $p$ and $\eta$ bins, and number of SPD hits. 
This leads to systematic uncertainties of $0.1$--$4.6\%$ ($0.4$--$5.4\%$) for the $13\tev$ ($7\tev$) sample.

The trigger efficiency is determined from simulated samples.  To estimate the systematic uncertainty, a tag-and-probe method is used to estimate the trigger efficiencies in each ($\pt,y$) bin of a \psitwos data sample that is independent of the detection of \psitwos signals~\cite{LHCb-DP-2012-004}. The same procedure is applied to the simulated \psitwos samples, and the relative difference of efficiencies between data and simulation in each kinematic bin, $0.1$--$9.3\%$ ($0.0$--$4.4 \%$) for the $13\tev$ ($7\tev$) sample, is taken as a systematic uncertainty.

The \pt and $y$ distributions of \psitwos mesons in simulation and in data could be different within each kinematic bin due to the finite bin size, causing differences in efficiencies.
The possible discrepancy is studied by weighting the kinematic distribution in simulation to match that in data.
All efficiencies are recalculated, and the relative differences of the total efficiencies between the new and the nominal results,
which are found to be in the range $0.0$--$2.0\%$ ($0.0$--$4.9\%$) for the $13\tev$ ($7\tev$) sample, are taken as systematic uncertainties.

The integrated luminosity is determined using the beam-gas imaging method for the $13\tev$ data sample, and by a combination of the beam-gas imaging and van der Meer scan methods~\cite{LHCb-PAPER-2014-047} for the $7\tev$ data sample. 
The uncertainty associated with the luminosity determination is $3.9\%$ ($1.7\%$) for the $13\tev$ ($7\tev$) sample. 
The uncertainty of the branching fraction of the $\psiee$ decay, $2.2\%$, is taken as a systematic uncertainty~\cite{PDG2016}. 
The limited size of the simulated sample in each bin leads to uncertainties of \mbox{$0.7$--$11.5\%$} (\mbox{$1.3$--$13.1\%$}) for prompt \psitwos and \mbox{$0.8$--$5.7\%$} (\mbox{$1.2$--$9.5\%$}) for \psitwos-from-$\bquark$ for the $13\tev$ ($7\tev$) sample, and are smaller than or comparable with the data statistical uncertainty in each bin.

There are sources of systematic uncertainties that are related to the $t_z$ variable, the effects of which are notable for \psitwos-from-$b$ and are negligible for prompt \psitwos.
The modelling of the $t_z$ resolution is modified by adding a third Gaussian to the nominal resolution model.
The variation in the \psitwos-from-\bquark fraction $F_b$ is found to be negligible. An alternative method is adopted to estimate the systematic uncertainty due to the modelling of the background $t_z$ distribution. In this method, the background distribution is obtained with the $\sPlot$ technique~\cite{Pivk:2004ty} using the invariant mass as the discriminating variable. The $t_z$ distribution is then parametrised for the two-dimensional fits to obtain the fraction $F_b$. The relative difference of $F_b$ in each kinematic bin between the two methods is taken as a systematic uncertainty. The total systematic uncertainty on the \psifromb cross-section related to the $t_z$ fit model is $0.1$--$8.4\%$ ($0.1$--$9.2\%$) for the $13\tev$ ($7\tev$) sample.

\section{Results}
\label{sec:results}
\subsection{Production cross-sections}
\begin{figure}[!tbp]
\centering
\begin{minipage}[t]{0.49\textwidth}
\centering
\includegraphics[width=1.0\textwidth]{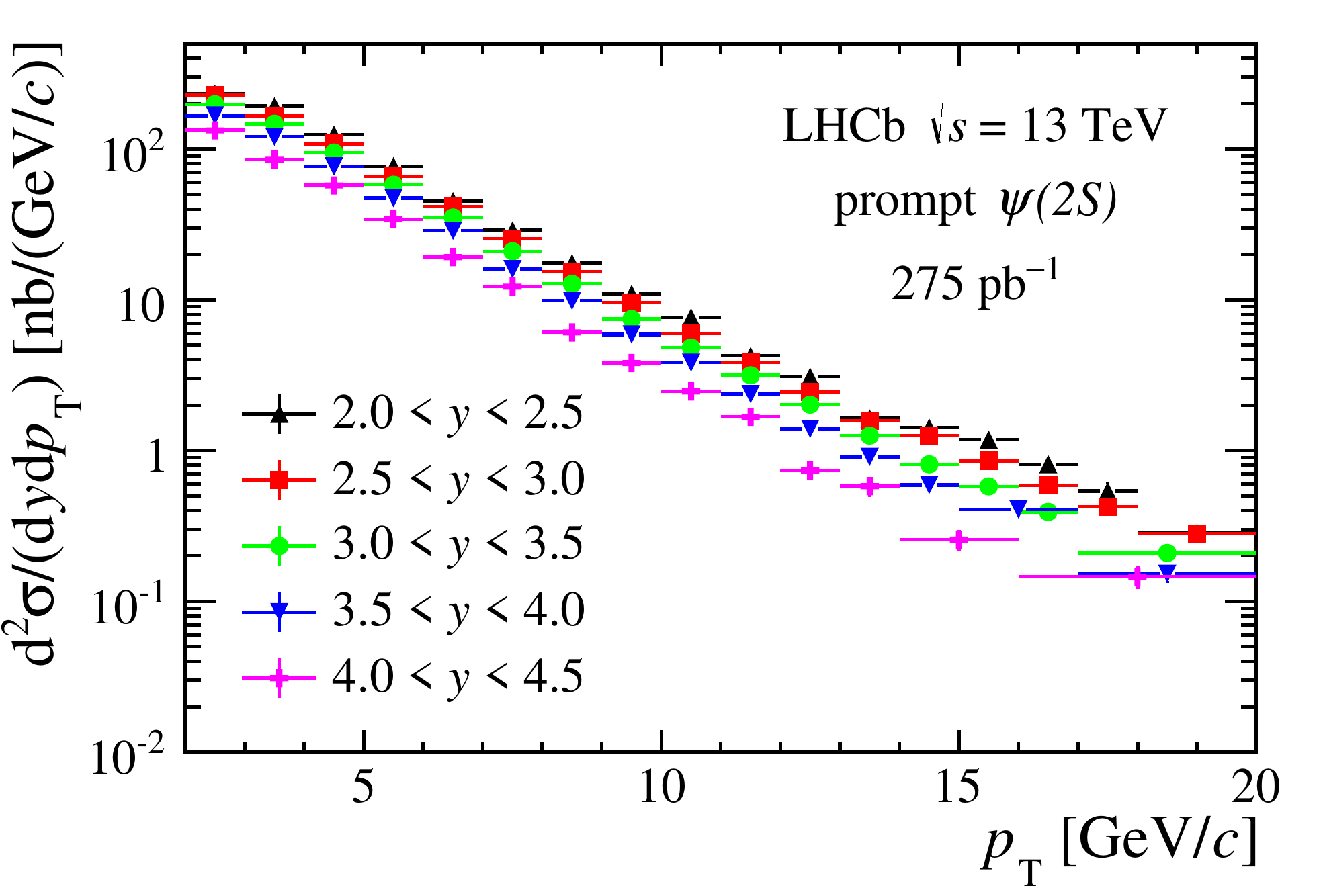}
\end{minipage}
\begin{minipage}[t]{0.49\textwidth}
\centering
\includegraphics[width=1.0\textwidth]{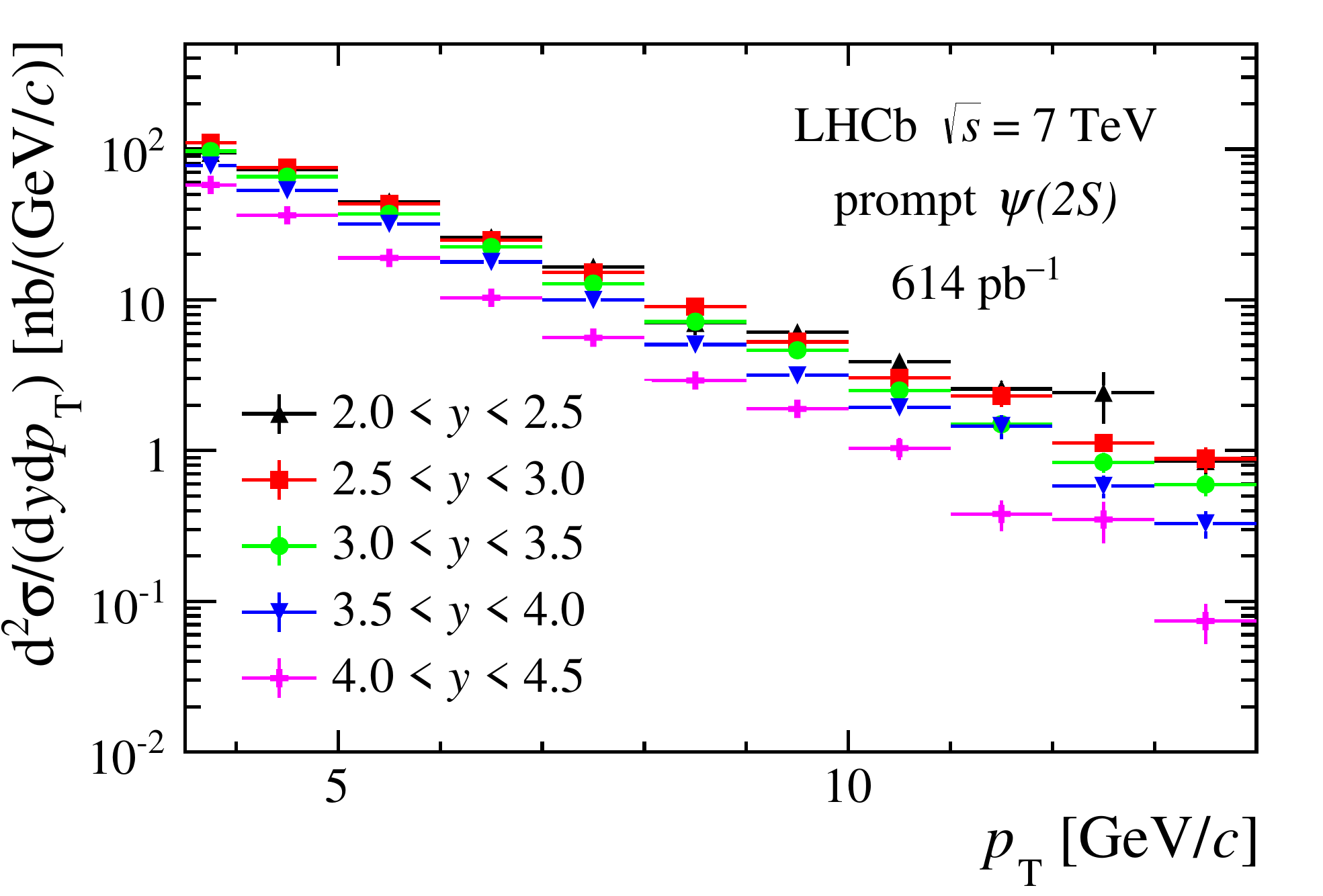}
\end{minipage}
    \caption{Double-differential production cross-sections of prompt \psitwos as functions of \pt in bins of $y$ at (left) $13\tev$ and (right) $7\tev$. The statistical and systematic uncertainties are added in quadrature. The \psitwos meson is assumed to be produced unpolarised.} 
\label{fig:results_prompt}
\end{figure}

\begin{figure}[!tbp]
\centering
\begin{minipage}[t]{0.49\textwidth}
\centering
\includegraphics[width=1.0\textwidth]{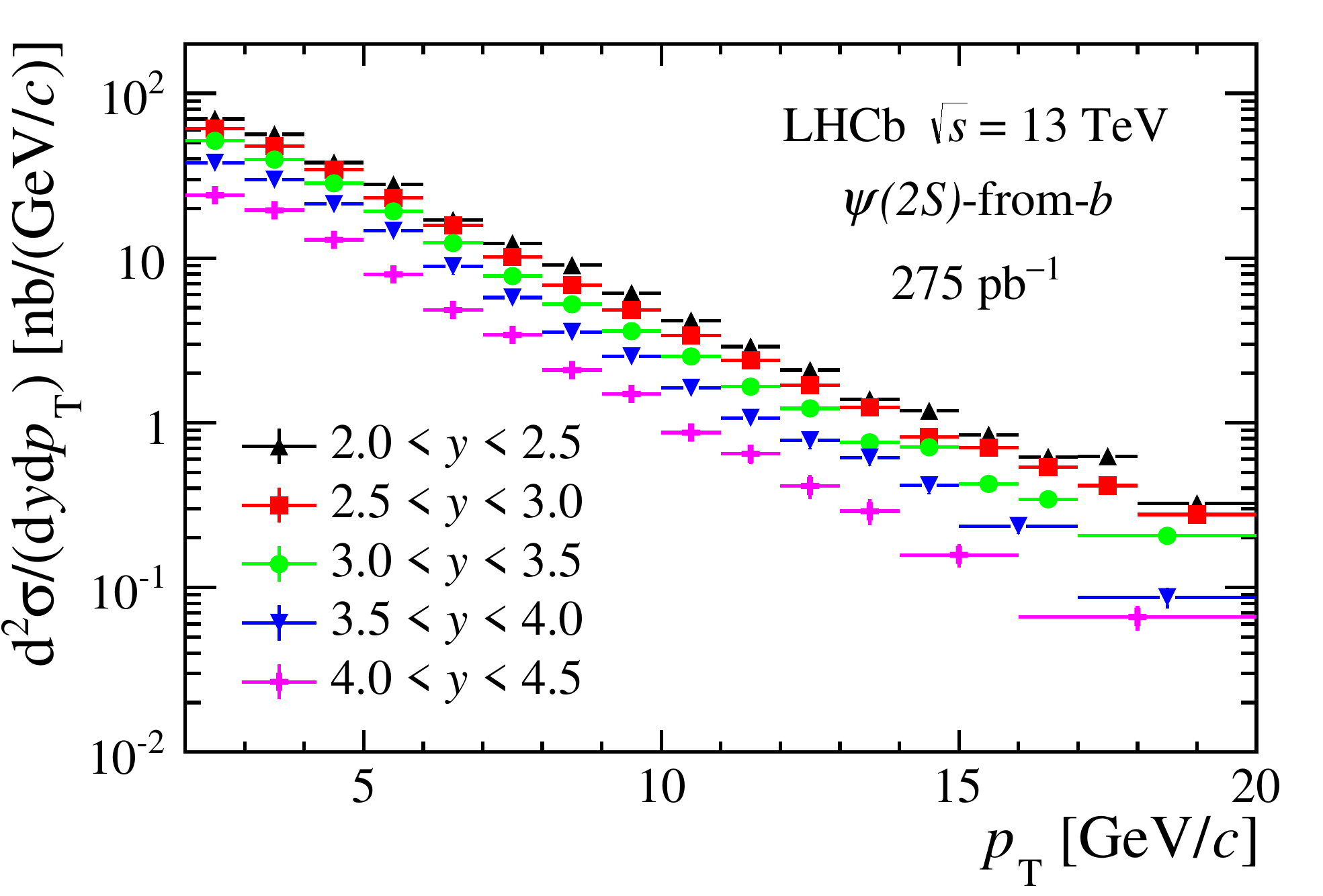}
\end{minipage}
\begin{minipage}[t]{0.49\textwidth}
\centering
\includegraphics[width=1.0\textwidth]{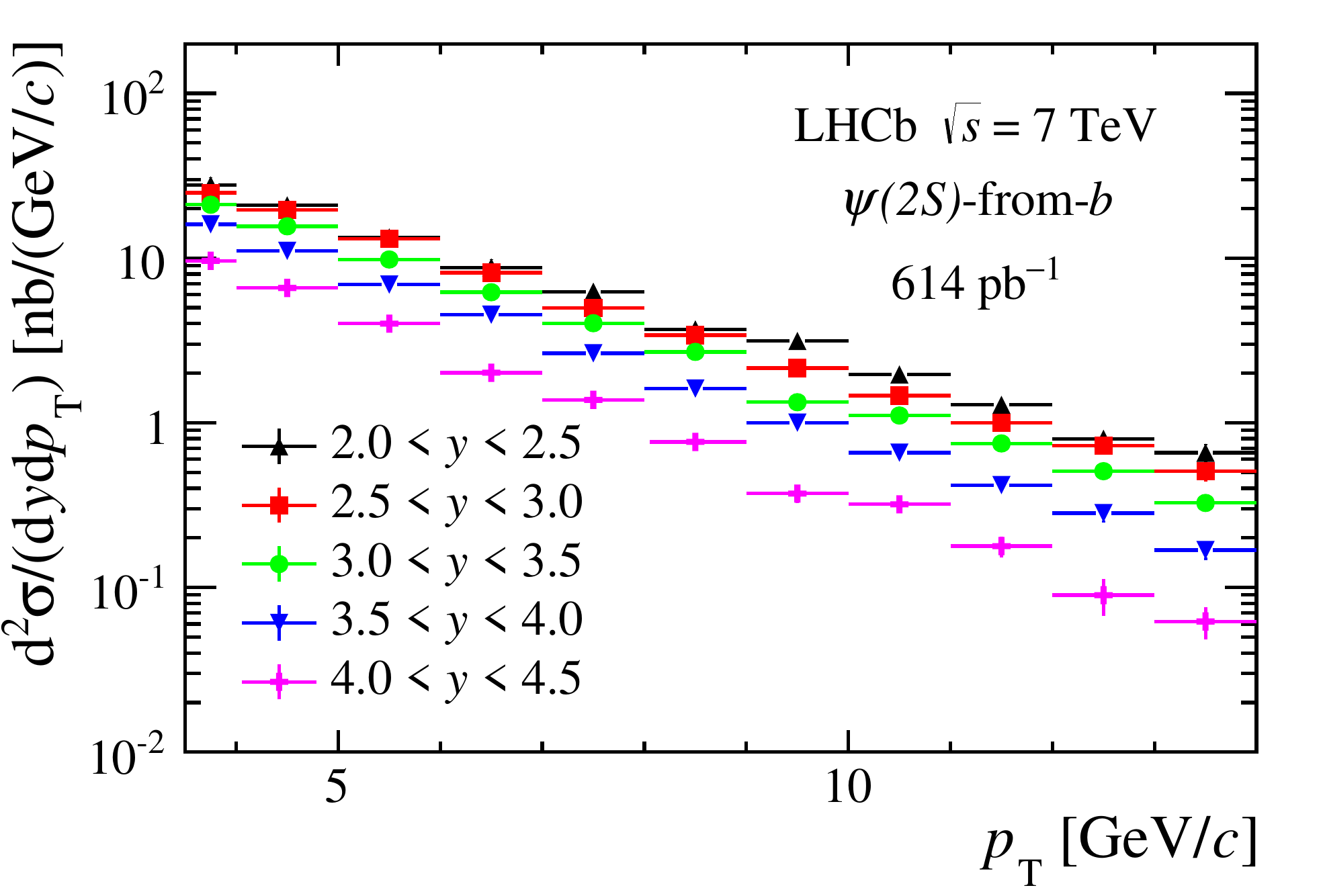}
\end{minipage}
    \caption{Double-differential production cross-sections of \psitwos-from-\bquark as functions of $\pt$ in bins of $y$ at (left) $13\tev$ and (right) $7\tev$. The statistical and systematic uncertainties are added in quadrature. The \psitwos meson is assumed to be produced unpolarised.} 
\label{fig:results_fromb}
\end{figure}

The double-differential production cross-sections for prompt \psitwos and \psitwos-from-$b$
are measured as functions of \pt and $y$ assuming no polarisation of \psitwos mesons. The results are shown in Figs.~\ref{fig:results_prompt} and~\ref{fig:results_fromb}, respectively.
The corresponding values are listed in Tables~\ref{tab:results_prompt},~\ref{tab:results_fromb},~\ref{tab7TeV:results_prompt}, and~\ref{tab7TeV:results_fromb}
in Appendix~\ref{sec:tables}.
%%%%%%%%%%%%%%%%%%%%%%%%%%%%%%%%%%%%%%%%%%%%%%%%%%%%%%%%%%%
%%%%%%%%%%%%%%%%%%%%%%%%%%%%%%%%%%%%%%%%%%%%%%%%%%%%%%%%%%%
%%%%%%%%%%%%%%%%%%%%%%%%%%%%%%%%%%%%%%%%%%%%%%%%%%%%%%%%%%%
%%%%%%%%%%%%%%%%%%%%%%%%%%%%%%%%%%%%%%%%%%%%%%%%%%%%%%%%%%%

\begin{figure}[!tbp]
\centering
\begin{minipage}[t]{0.49\textwidth}
\centering
\includegraphics[width=1.0\textwidth]{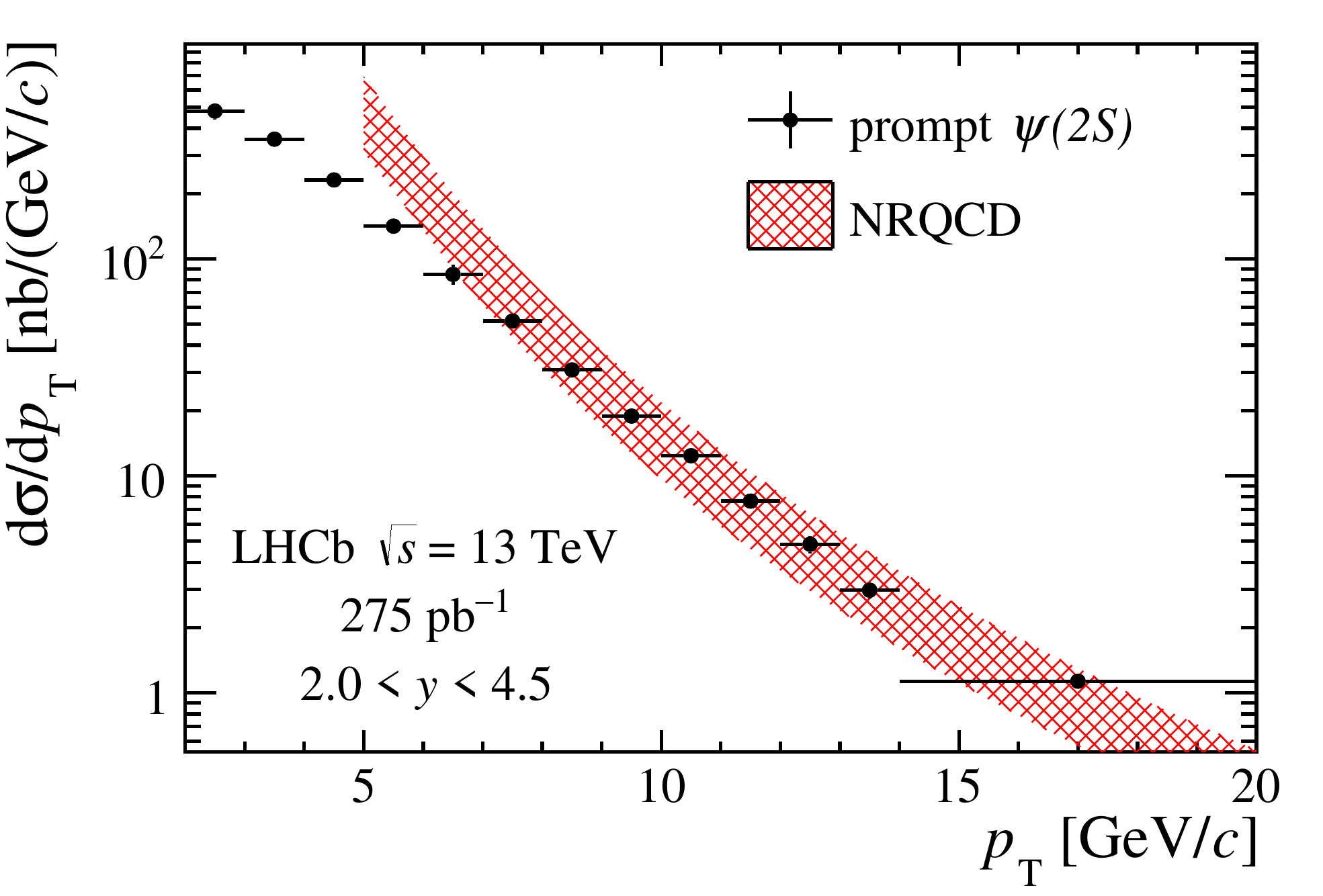}
\end{minipage}
\begin{minipage}[t]{0.49\textwidth}
\centering
\includegraphics[width=1.0\textwidth]{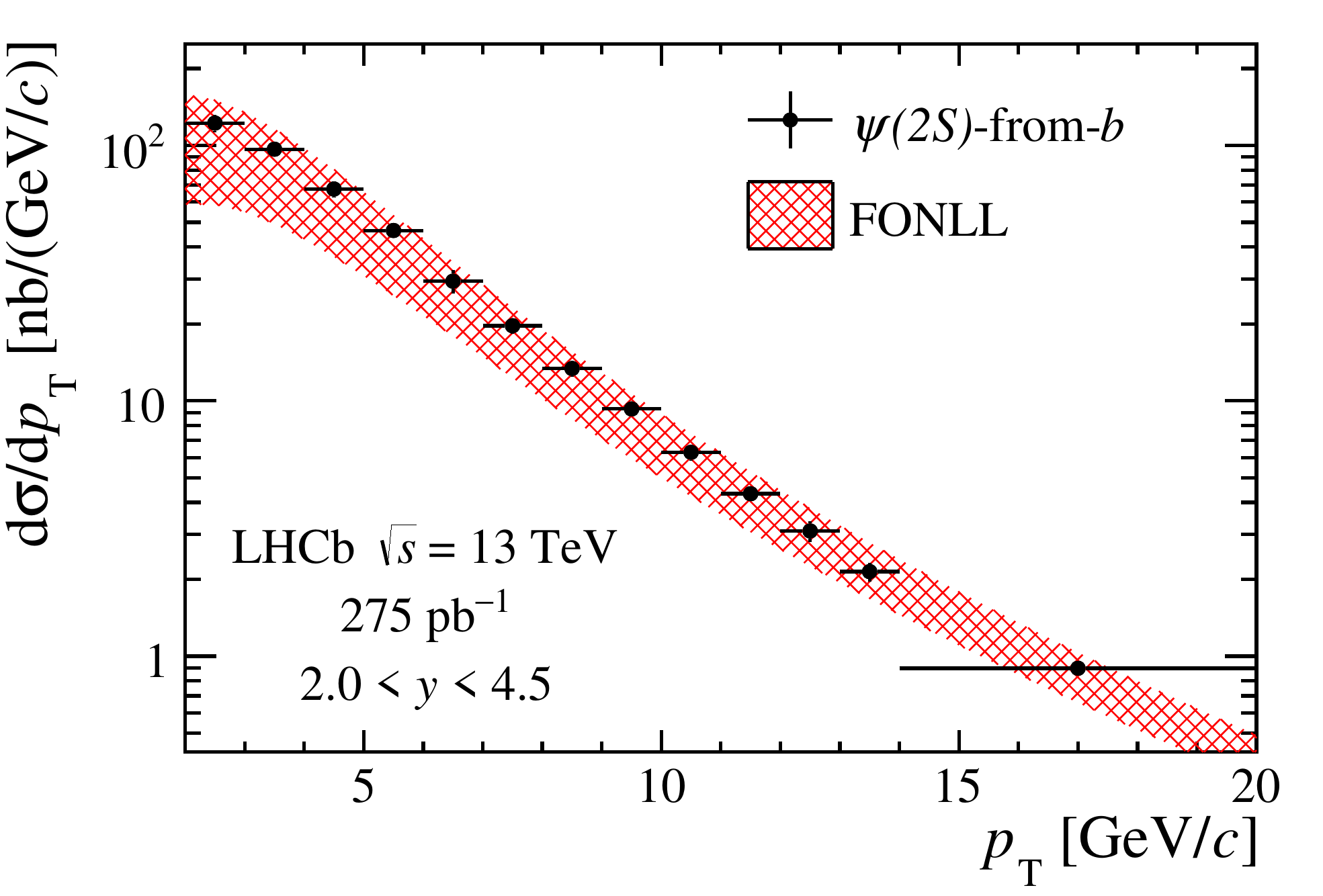}
\end{minipage}
\begin{minipage}[t]{0.49\textwidth}
\centering
\includegraphics[width=1.0\textwidth]{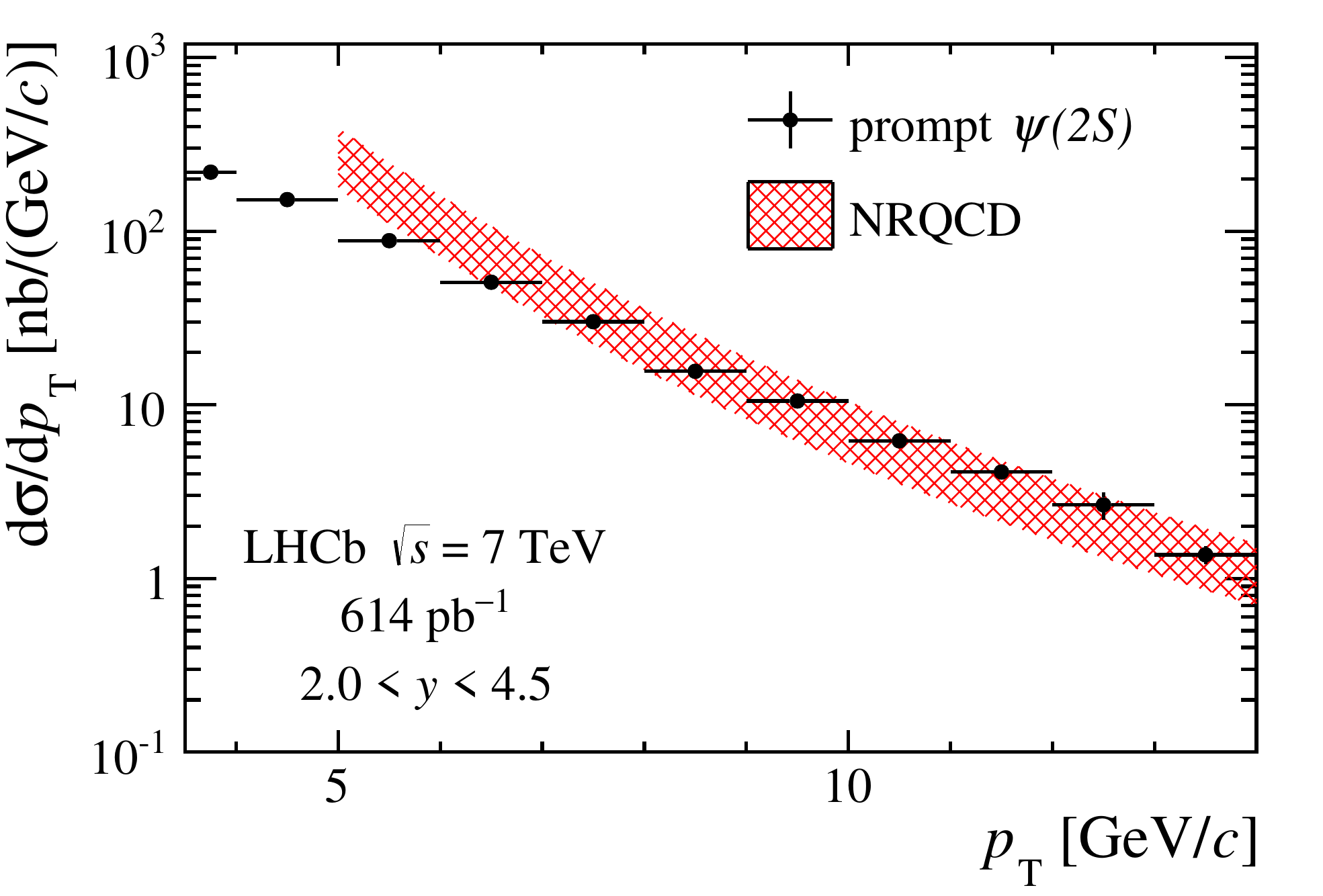}
\end{minipage}
\begin{minipage}[t]{0.49\textwidth}
\centering
\includegraphics[width=1.0\textwidth]{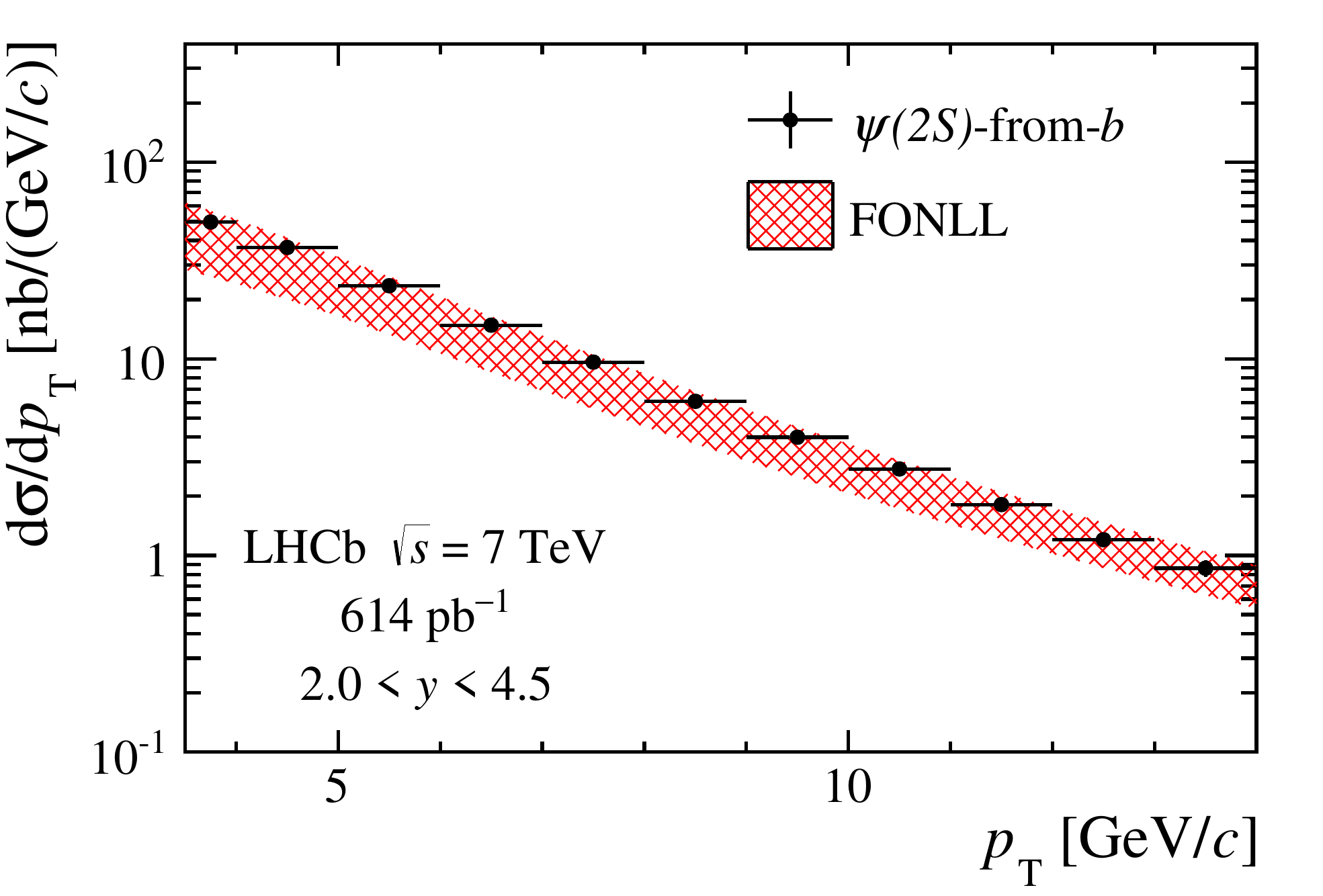}
\end{minipage}
\caption{
    Differential production cross-sections as functions of \pt in the range $2.0<y<4.5$ for the (top) $13\tev$ and (bottom) $7\tev$ samples.
    The left-hand figures are for prompt \psitwos and the results are compared with the NRQCD calculations~\cite{Shao:2014yta}; the right-hand figures are for \psitwos-from-\bquark and the results are compared with the FONLL calculations~\cite{Cacciari:1998it}.} 
\label{fig:results_PT}
\end{figure}

\begin{figure}[!tbp]
\centering
\begin{minipage}[t]{0.49\textwidth}
\centering
\includegraphics[width=1.0\textwidth]{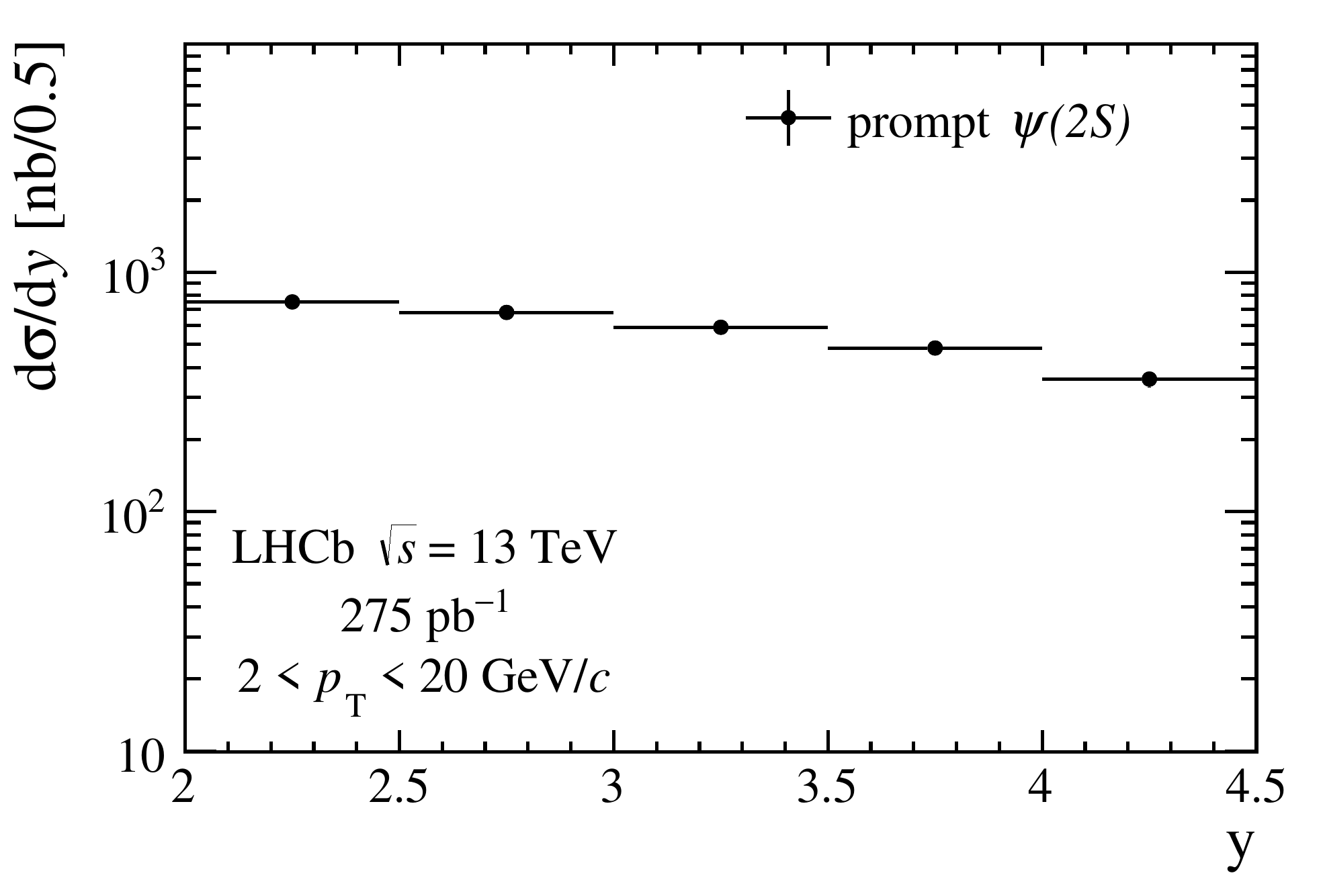}
\end{minipage}
\begin{minipage}[t]{0.49\textwidth}
\centering
\includegraphics[width=1.0\textwidth]{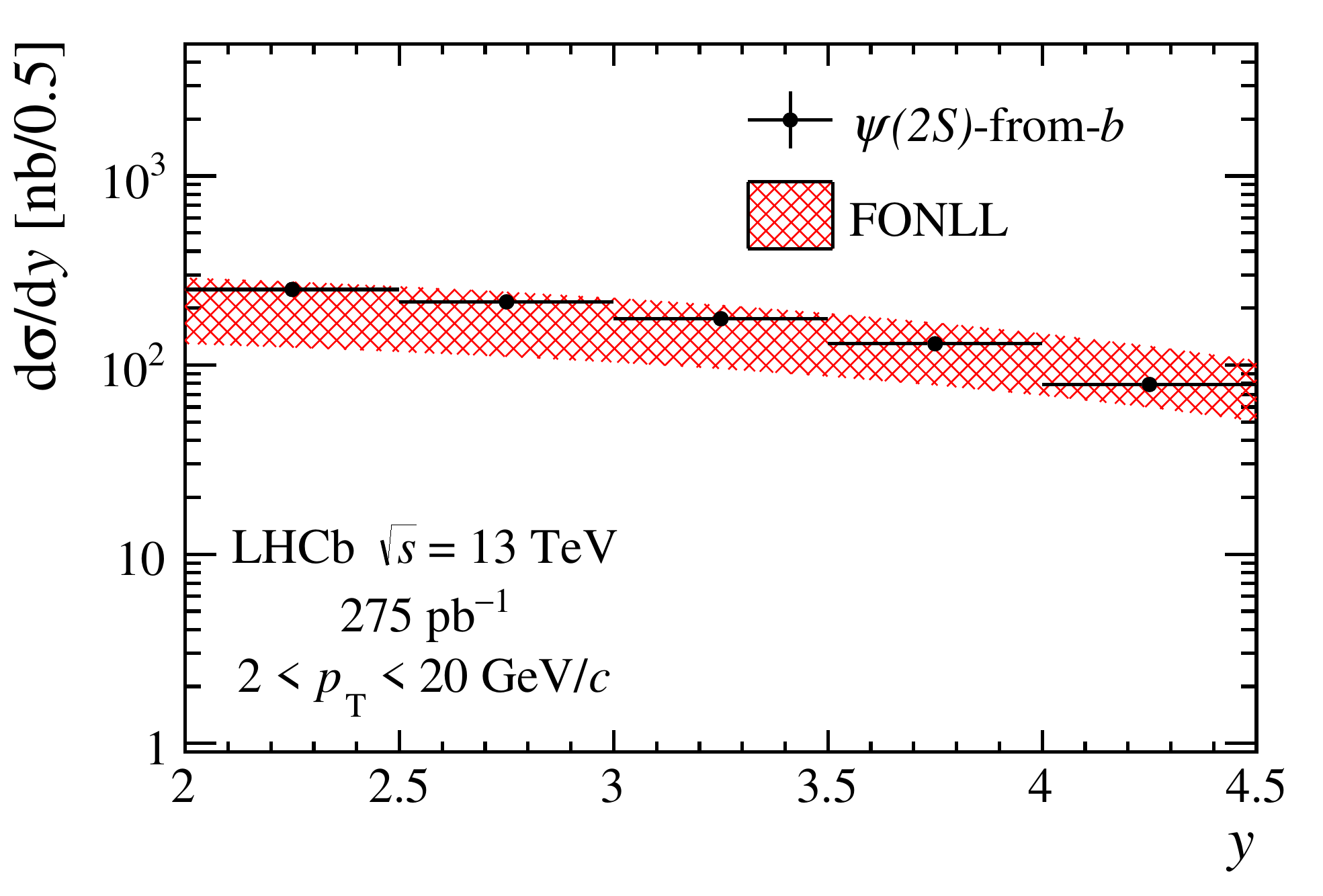}
\end{minipage}
\begin{minipage}[t]{0.49\textwidth}
\centering
\includegraphics[width=1.0\textwidth]{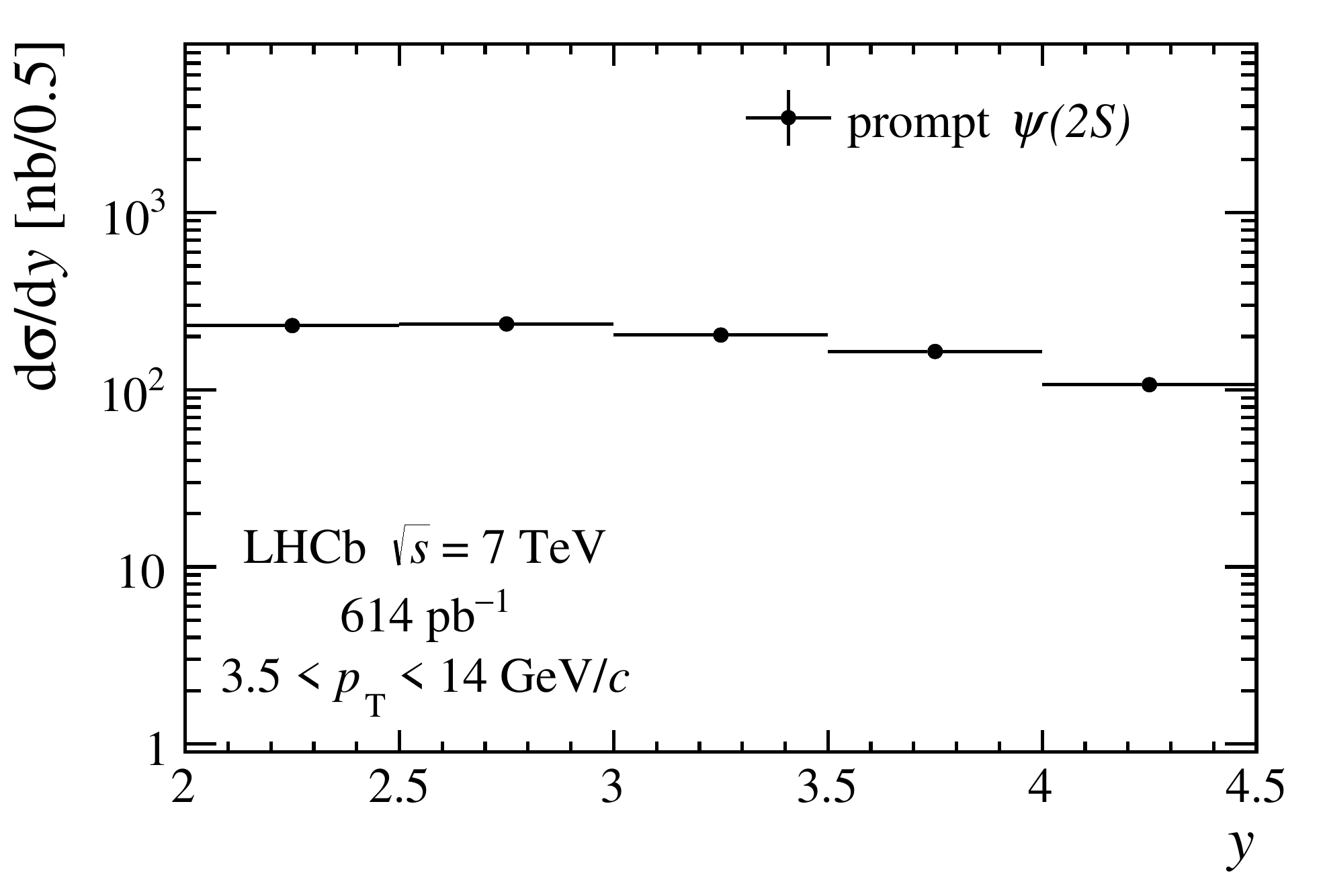}
\end{minipage}
\begin{minipage}[t]{0.49\textwidth}
\centering
\includegraphics[width=1.0\textwidth]{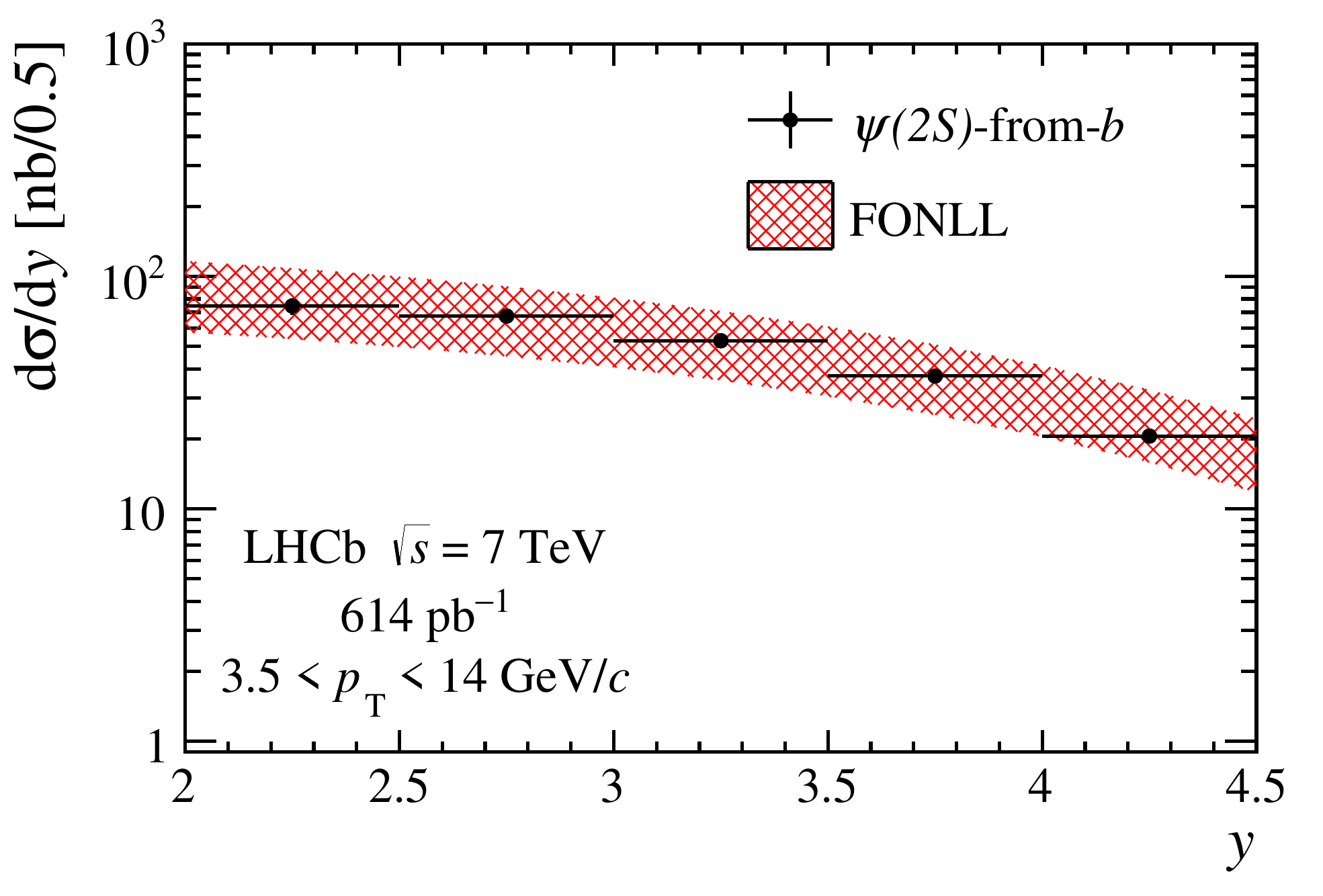}
\end{minipage}
\caption{
    Differential production cross-sections as functions of $y$ in the range \mbox{$2<\pt<20\gevc$} for the $13\tev$ sample (top) and in the range $3.5<\pt<14\gevc$ for the $7\tev$ sample (bottom).
    The left figures are for prompt \psitwos,
    the right figures are for \psitwos-from-\bquark compared with the FONLL calculations~\cite{Cacciari:1998it}.} 
\label{fig:results_Y}
\end{figure}

By integrating the double-differential results over $y$ in the range $2.0<y<4.5$, the differential production cross-sections of prompt \psitwos and \psitwos-from-\bquark as functions of $\pt$ are shown in Fig.~\ref{fig:results_PT}. 
The results of prompt \psitwos production are compared with the theoretical calculations based on NRQCD~\cite{Shao:2014yta}, and those of \mbox{\psitwos-from-\bquark} are compared with the fixed-order-plus-next-leading-logarithm (FONLL) calculations~\cite{Cacciari:1998it}.
The differential cross-section as function of $y$ at $13\tev$ ($7\tev$) is obtained by integrating the double-differential results over $\pt$ in the range \mbox{$2<\pt<20\gevc$} (\mbox{$3.5<\pt<14\gevc$}). The results are presented in Fig.~\ref{fig:results_Y}. The theoretical calculations based on FONLL are shown for \mbox{\psitwos-from-$b$}. The NRQCD calculations are omitted since they are not reliable in the low \pt region \cite{FACCIOLI201498}.
The values of the differential cross-sections are shown in Tables~\ref{tab:results_PT}, \ref{tab:results_Y}, \ref{tab7TeV:results_PT}, and \ref{tab7TeV:results_Y}
in Appendix~\ref{sec:tables}.
In the NRQCD calculations, only the dominant uncertainties associated with the LDMEs are considered \cite{Shao:2014yta}.
The FONLL calculations include the uncertainty due to $b$-quark mass and the scales of renormalisation and factorisation.
The NRQCD calculations show reasonable agreement with experimental data for $\pt>7\gevc$. The FONLL calculations agree well with the measurements.
The production cross-sections of prompt \psitwos and \psitwos-from-$b$ integrated in the kinematic range $2.0<y<4.5$ and \mbox{$2<\pt<20\gevc$} at $13\tev$, are measured to be:
\begin{equation*}
\begin{split}
\sigma(\mbox{prompt }\psitwos,13\tev) &= \promptresult,\\
\sigma(\psitwos\mbox{-from-}\bquark,13\tev) &= \frombresult.
\end{split}
\end{equation*}
The production cross-sections of prompt \psitwos and \psitwos-from-$b$ integrated in the kinematic range $2.0<y<4.5$ and \mbox{$3.5<\pt<14\gevc$} at $7\tev$, are measured to be:
\begin{equation*}
\begin{split}
\sigma(\mbox{prompt }\psitwos,7\tev) &= \promptresultseven,\\
\sigma(\psitwos\mbox{-from-}\bquark,7\tev) &= \frombresultseven.
\end{split}
\end{equation*}

As mentioned above, these results are obtained under the assumption of zero polarisation of \psitwos mesons. Possible polarisation of \psitwos meson would affect the detection efficiency. This effect is studied for extreme cases of fully transverse and fully longitudinal polarisation corresponding to the parameter $\alpha$ be equal to $+1$ or $-1$, respectively, within the helicity frame~\cite{Jacob:1959at} approach. Also the polarisation case of $\alpha = -0.2$, corresponding to a conservative limit of the \psitwos polarisation measured at $7\tev$~\cite{LHCb-PAPER-2013-067}, is considered. Resulting scaling factors for prompt \psitwos production cross-sections are listed in Appendix \ref{sec:tables_pol} in Tables \ref{tab:corr_factors_13_+1}, \ref{tab:corr_factors_13_-1} and \ref{tab:corr_factors_13_-0.2} for $13\tev$ results, and in Tables \ref{tab:corr_factors_7_+1}, \ref{tab:corr_factors_7_-1} and \ref{tab:corr_factors_7_-0.2} for $7\tev$ results, respectively.
%%%%%%%%%%%%%%%%%%%%%%%%%%%%%%%%%%%%%%%%%%%%%%%%%%%%%%%%%%%
%%%%%%%%%%%%%%%%%%%%%%%%%%%%%%%%%%%%%%%%%%%%%%%%%%%%%%%%%%

%%%%%%%%%%%%%%%%%%%%%%%%%%%%%%%%%%%%%%%%%%%%%%%%%%%%%%%%%%%
%%%%%%%%%%%%%%%%%%%%%%%%%%%%%%%%%%%%%%%%%%%%%%%%%%%%%%%%%%

\subsection{\boldmath Fraction of \psitwos-from-$\bquark$ mesons}
\label{sec:Fb}
%%%%%%%%%%%%%%%%%%%%%%%%%%%%%%%%%%%%%%%%%%%%%%%%%%%%%%%%%%%
The fraction of \psitwos-from-$\bquark$ is \mbox{$F_b\equiv N_b/(N_b+N_p)$}, where $N_p$ is the efficiency-corrected signal yield of prompt \psitwos and $N_b$ is that of \psitwos-from-$\bquark$.
The fractions $F_b$ as functions of \pt and $y$ are shown in Fig.~\ref{fig:Fb}. 
The corresponding values are presented in Table~\ref{tab:Fb} in Appendix~\ref{sec:tables}.
Only statistical uncertainties are shown 
owing to the cancellation of most systematic contributions, except for that due to the $t_z$ fit, which is negligible.
For each $y$ bin, the fraction increases with increasing \pt of the \psitwos mesons. 
For each \pt bin, the fraction decreases with increasing $y$ of the \psitwos mesons.

%%%%%%%%%%%%%%%%%%%%%%%%%%%%%%%%%%%%%%%%%%%%%%%%%%%%%%%%%%%
\begin{figure}[!tbp]
\centering
\begin{minipage}[t]{0.49\textwidth}
\centering
\includegraphics[width=1.0\textwidth]{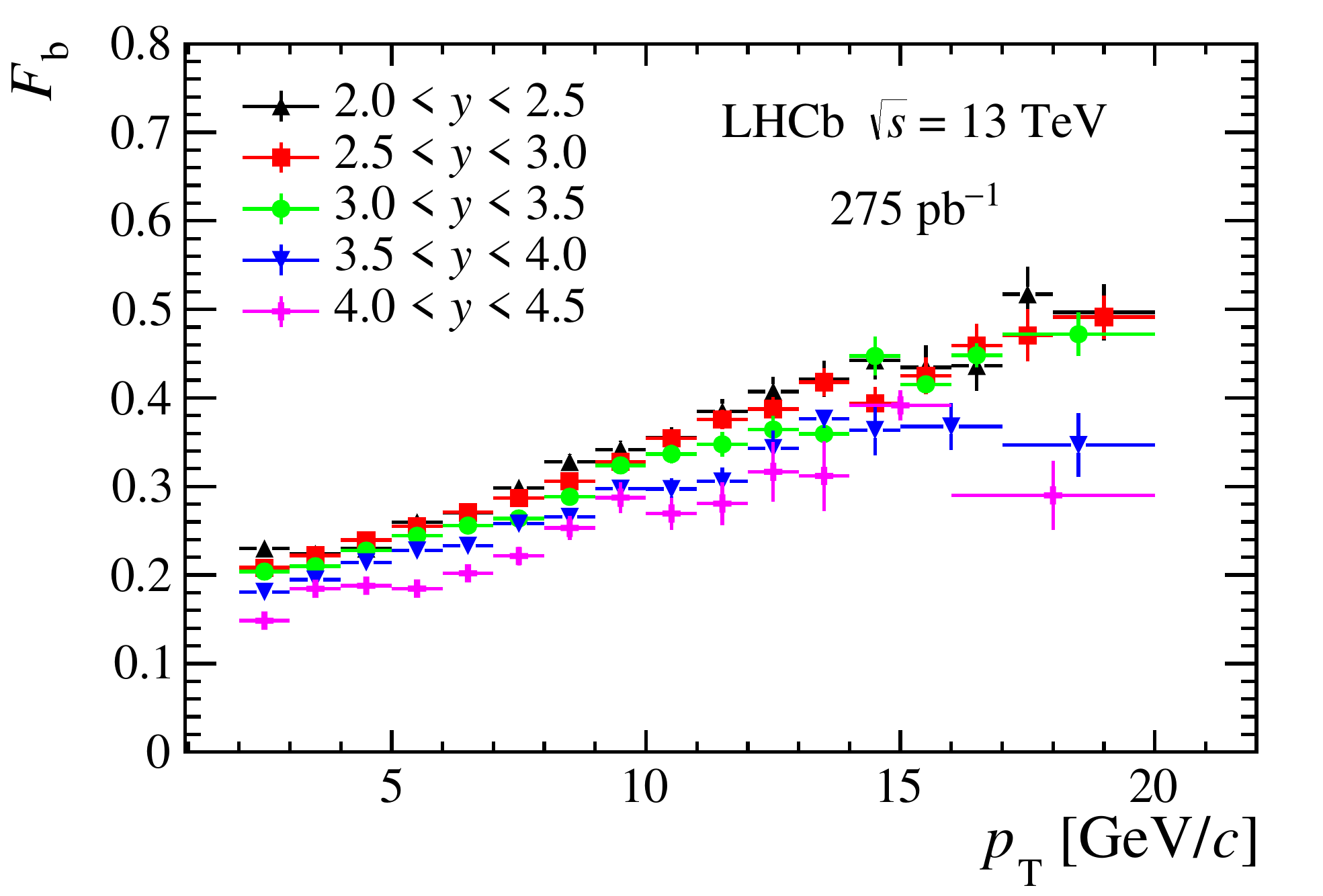}
\end{minipage}
\begin{minipage}[t]{0.49\textwidth}
\centering
\includegraphics[width=1.0\textwidth]{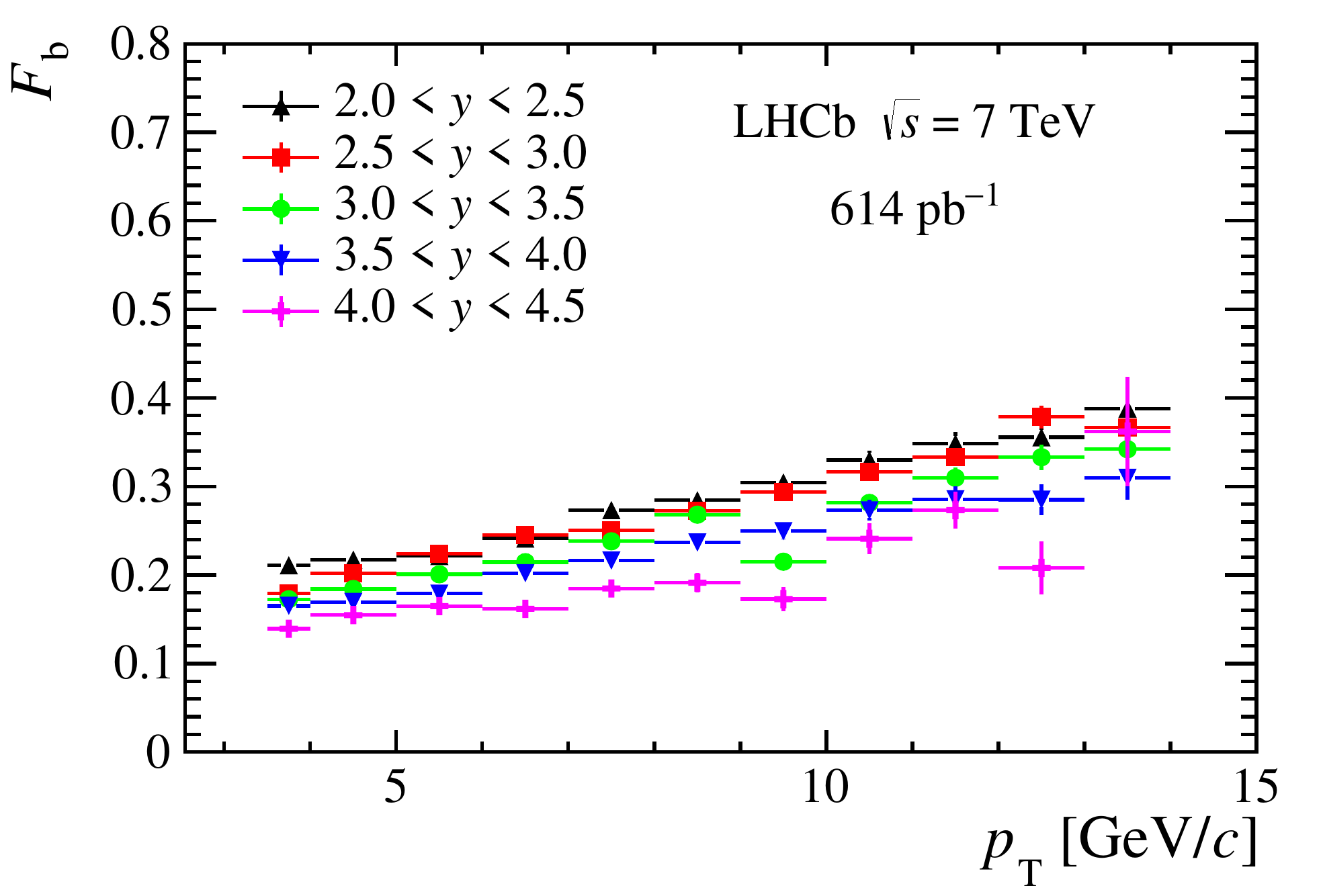}
\end{minipage}
\caption{
    Fractions of $\psitwos$-from-$\bquark$ in bins of \pt and $y$ for the (left) $13\tev$ and (right) $7\tev$ samples. The error bars represent the statistical uncertainties.}
\label{fig:Fb}
\end{figure}
%%%%%%%%%%%%%%%%%%%%%%%%%%%%%%%%%%%%%%%%%%%%%%%%%%%%%%%%%%

\subsection{\boldmath Comparison with $\jpsi$ results at $13\tev$}
\label{sec:comparisonwithJpsi}
\begin{figure}[!tbp]
\centering
\begin{minipage}[t]{0.49\textwidth}
\centering
\includegraphics[width=1.0\textwidth]{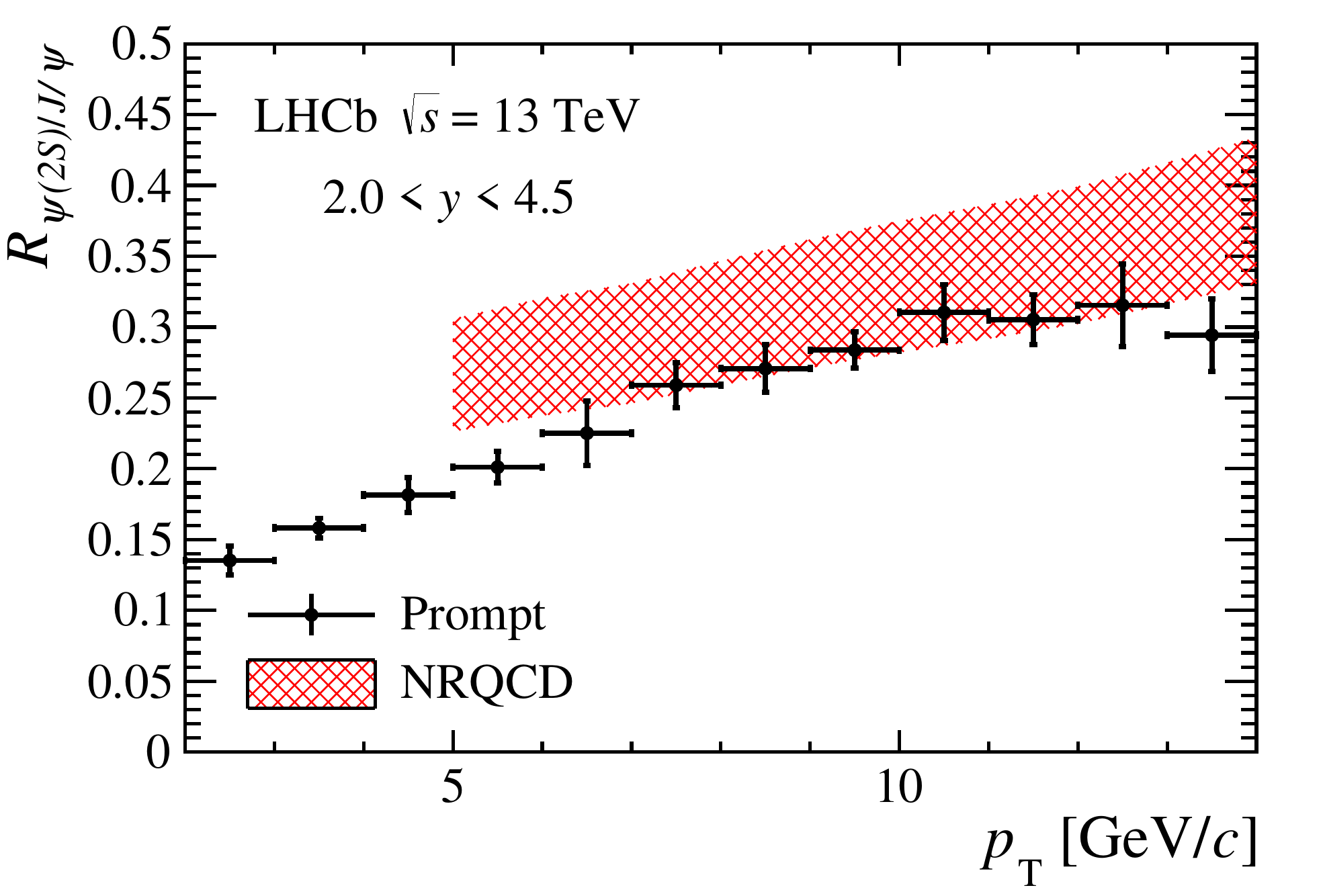}
\end{minipage}
\begin{minipage}[t]{0.49\textwidth}
\centering
\includegraphics[width=1.0\textwidth]{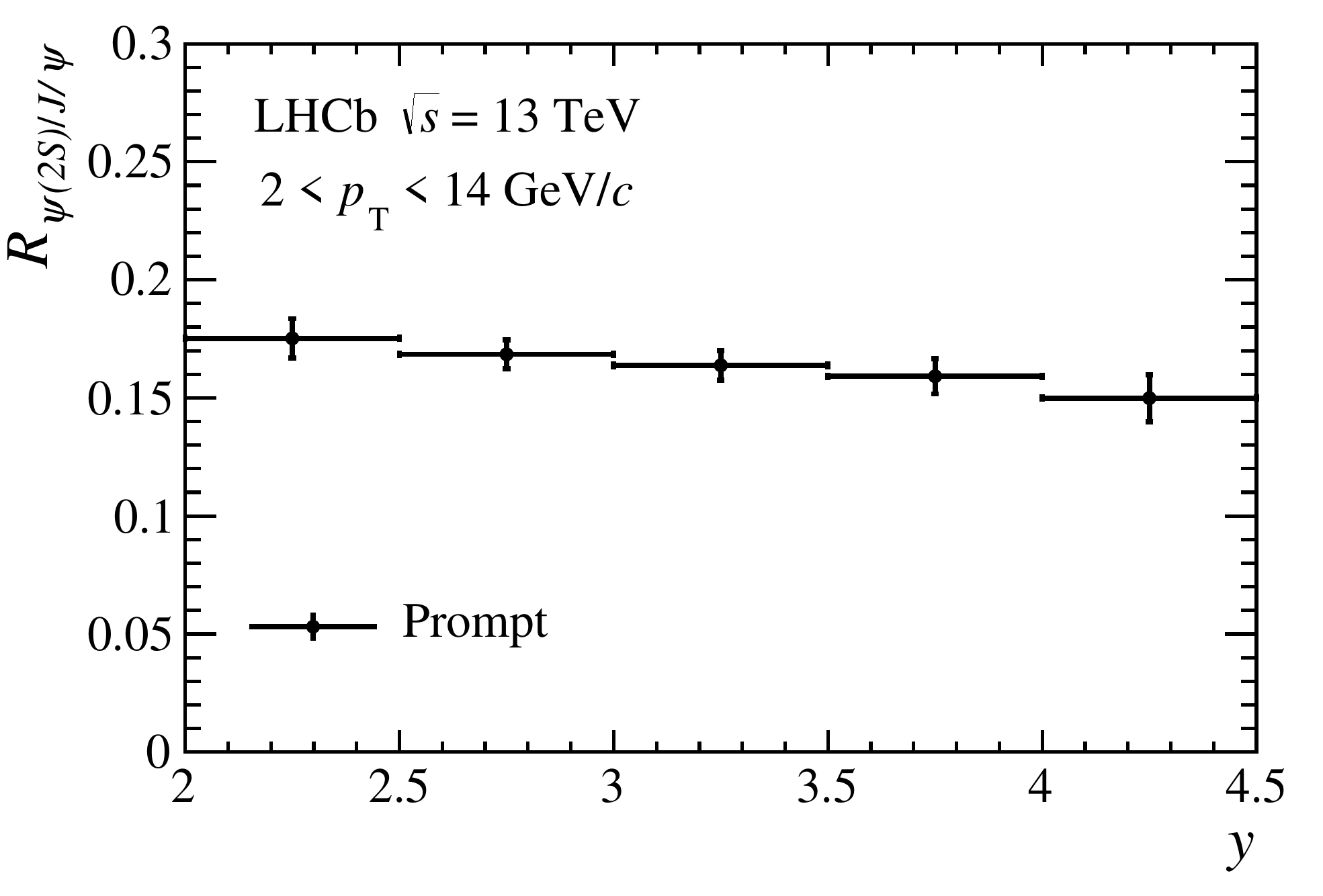}
\end{minipage}
\caption{Ratios of differential cross-sections between prompt \psitwos and prompt \jpsi mesons at $13\tev$ as functions of (left) \pt and (right) $y$. The NRQCD predicted ratio~\cite{Shao:2014yta} is shown in the left panel for comparison.}
\label{fig:Ratio_jpsi_PT_prompt}
\end{figure}

\begin{figure}[!tbp]
\centering
\begin{minipage}[t]{0.49\textwidth}
\centering
\includegraphics[width=1.0\textwidth]{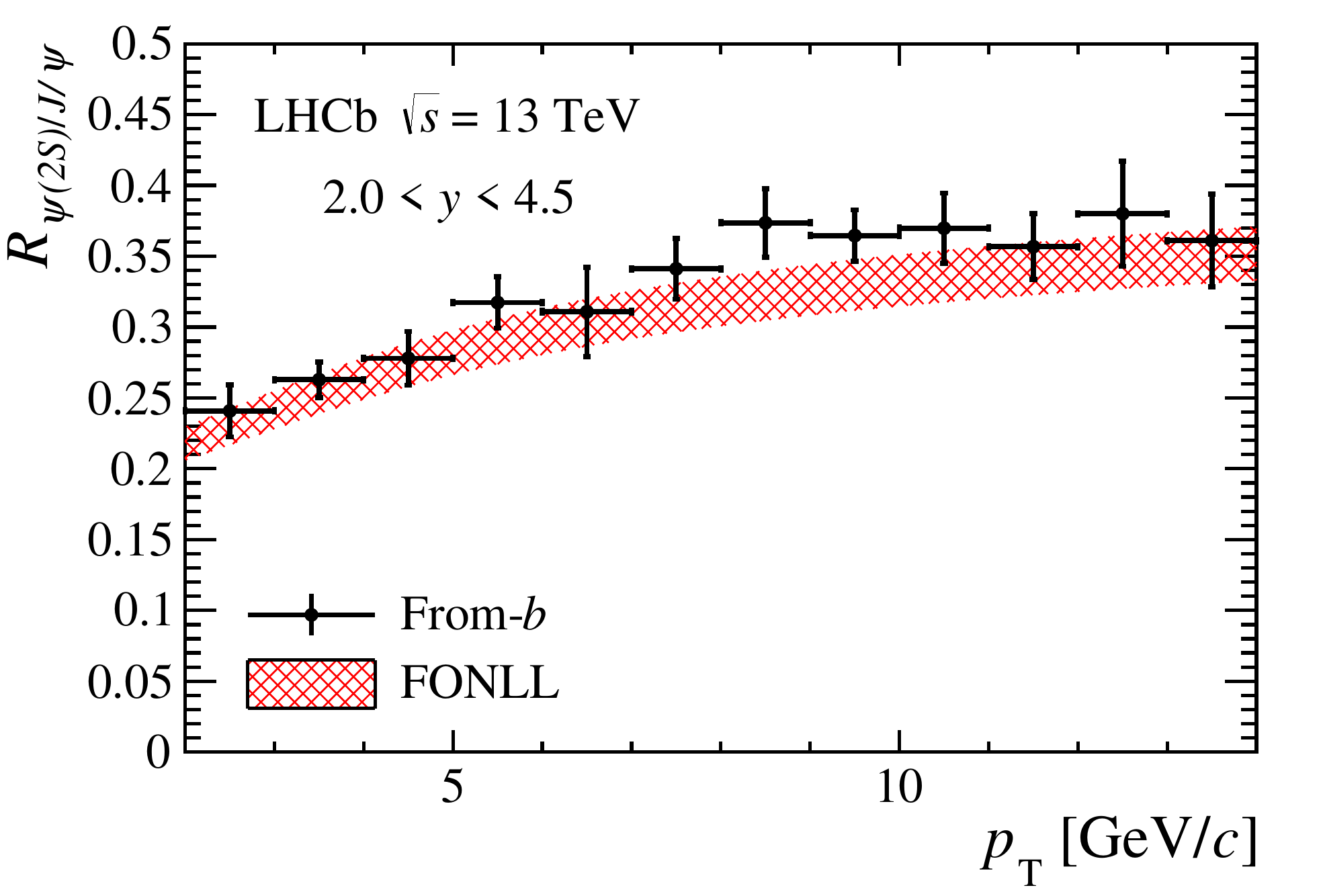}
\end{minipage}
\begin{minipage}[t]{0.49\textwidth}
\centering
\includegraphics[width=1.0\textwidth]{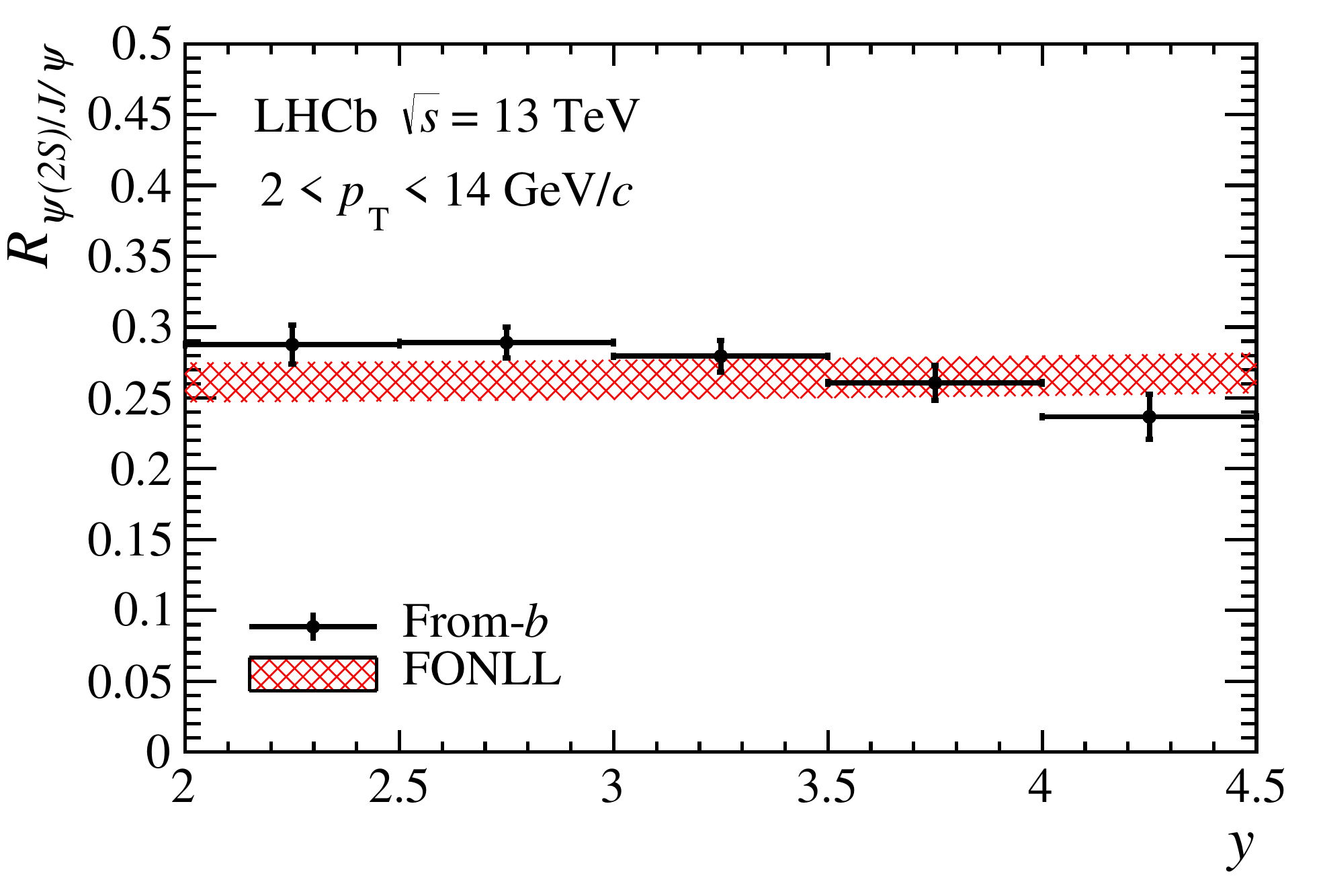}
\end{minipage}
\caption{Ratios of differential cross-sections between \psitwos-from-\bquark and \jpsi mesons from \mbox{\bquark-hadron} decays at $13\tev$ as functions of (left) \pt and (right) $y$. The FONLL calculations~\cite{Cacciari:2015fta} are shown for comparison.}
\label{fig:Ratio_jpsi_PT_bdecay}
\end{figure}

The production cross-sections of \psitwos mesons at $13\tev$ are compared with those of \jpsi mesons measured by \lhcb at $\sqs=13\tev$ in the range $0<\pt<14\gevc$ and $2.0<y<4.5$~\cite{LHCb-PAPER-2015-037}, where the $\jpsi$ meson is also assumed to be produced with zero polarisation.
The ratio, \Rpsijpsi, of the differential production cross-sections in the common range between prompt \psitwos and prompt \jpsi mesons is shown in Fig.~\ref{fig:Ratio_jpsi_PT_prompt} as a function of $\pt$~($y$) integrated over $2.0<y<4.5$~($2<\pt<14\gevc$).
The NRQCD calculation of \Rpsijpsi for prompt productions~\cite{Shao:2014yta} is also shown.
The ratio of production cross-sections between \psitwos-from-\bquark and \jpsi-from-\bquark is shown in Fig.~\ref{fig:Ratio_jpsi_PT_bdecay} as a function of $\pt$~($y$) integrated over $2.0<y<4.5$~($2<\pt<14\gevc$).
The FONLL calculations~\cite{Cacciari:2015fta} are compared to the measured values.
To calculate these ratios from the measured cross-sections of \psitwos and \jpsi mesons, 
the systematic uncertainties due to the luminosity, the tracking correction, and the fit model are considered to be fully correlated. All other uncertainties are assumed to be uncorrelated.
The numerical results of the measured ratios are listed in Tables~\ref{tab:Ratio_jpsi_PT} and~\ref{tab:Ratio_jpsi_Y} in Appendix~\ref{sec:tables}.
The FONLL prediction agrees well with the experimental data for the production cross-section ratio between \psitwos-from-\bquark and  \jpsi mesons from \bquark-hadron decays, while the NRQCD predictions show reasonable agreement with the measurements for prompt \psitwos and prompt \jpsi.
%%%%%%%%%%%%%%%%%%%%%%%%%%%%%%%%%%%%%%%%%%%%%%%%%%%%%%%%%%%
%%%%%%%%%%%%%%%%%%%%%%%%%%%%%%%%%%%%%%%%%%%%%%%%%%%%%%%%%%%
%%%%%%%%%%%%%%%%%%%%%%%%%%%%%%%%%%%%%%%%%%%%%%%%%%%%%%%%%%%

\subsection{\boldmath Comparison between $13\tev$ and $7\tev$}
\begin{figure}[!tbp]
\centering
\begin{minipage}[t]{0.49\textwidth}
\centering
\includegraphics[width=1.0\textwidth]{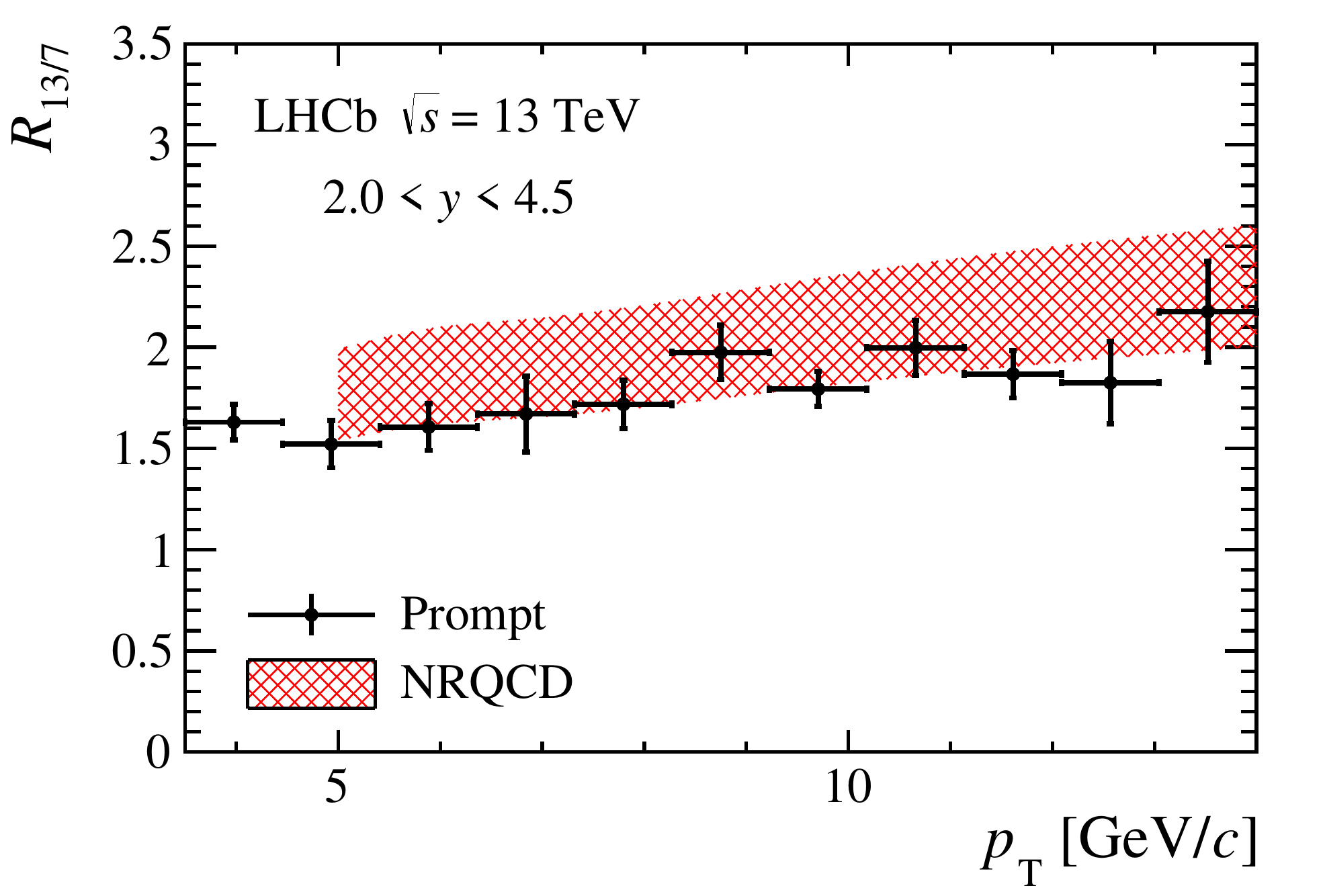}
\end{minipage}
\begin{minipage}[t]{0.49\textwidth}
\centering
\includegraphics[width=1.0\textwidth]{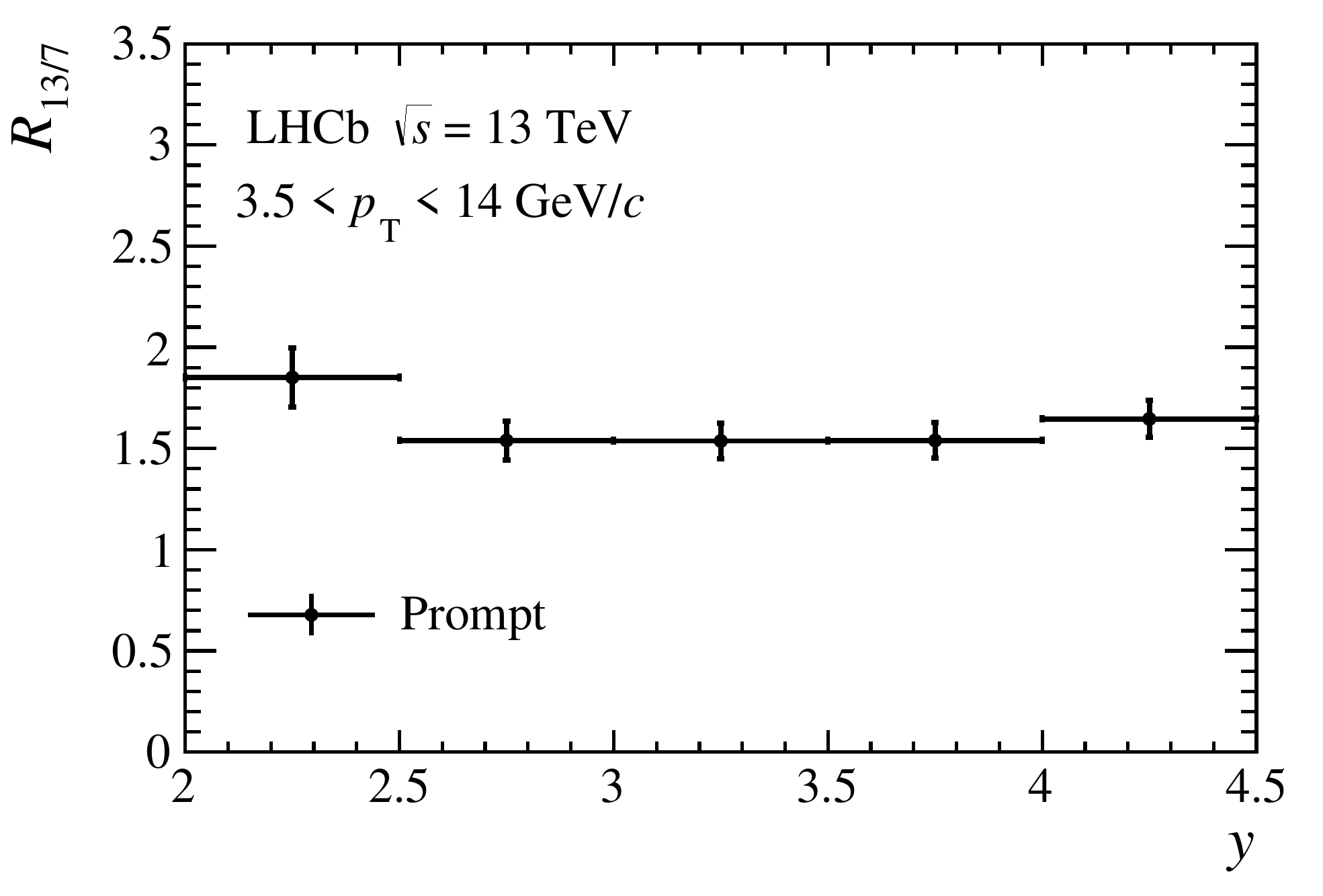}
\end{minipage}
    \caption{Ratio of differential production cross-sections between the $13\tev$ and $7\tev$ measurements as a function of (left) \pt integrated over $y$ and (right) $y$ integrated over $\pt$ for prompt \psitwos\ production. Theoretical calculations of NRQCD~\cite{Shao:2014yta} are compared to the data on the left side. } 
\label{fig:Ratio_7TeV_prompt}
\end{figure}

\begin{figure}[!tbp]
\centering
\begin{minipage}[t]{0.49\textwidth}
\centering
\includegraphics[width=1.0\textwidth]{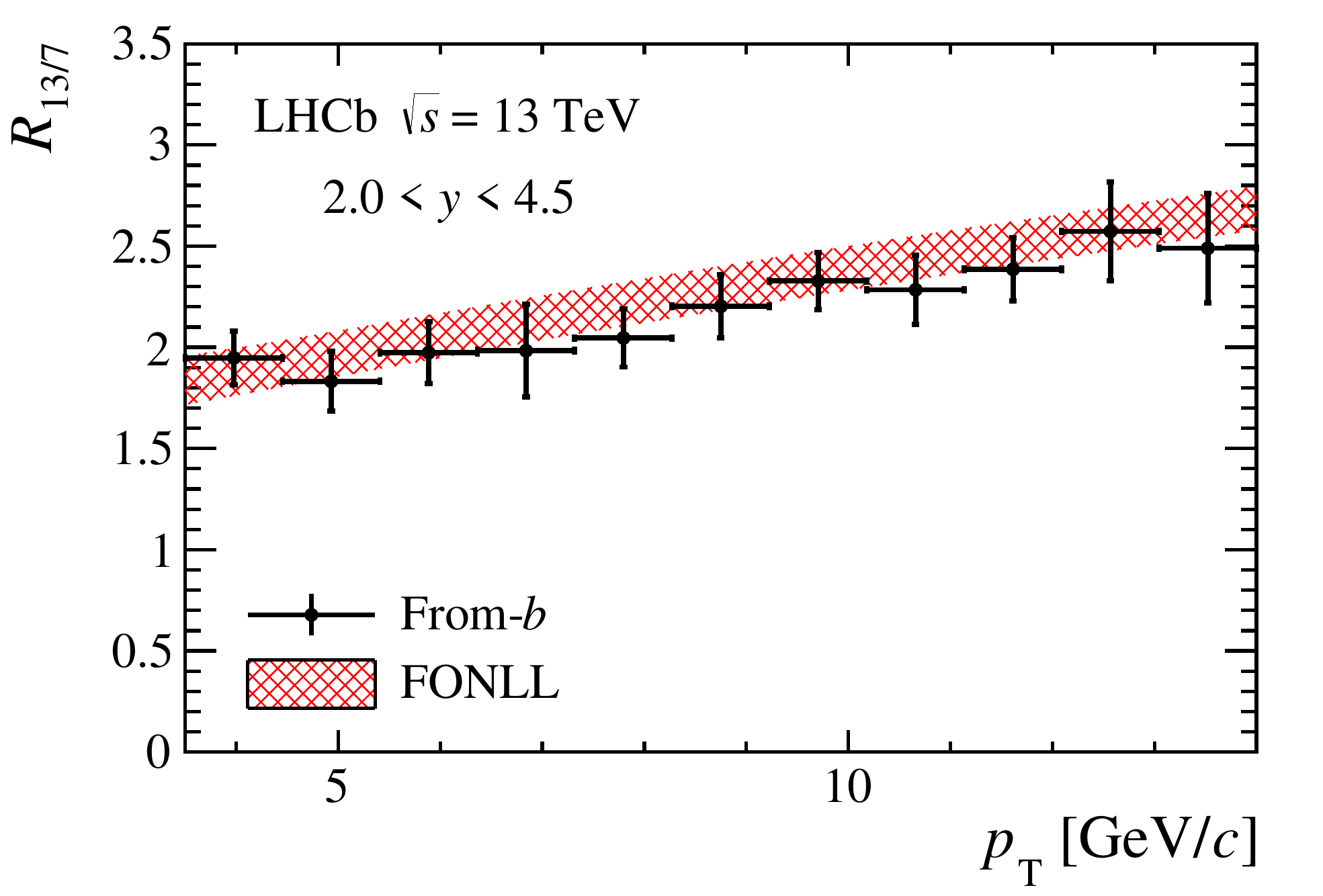}
\end{minipage}
\begin{minipage}[t]{0.49\textwidth}
\centering
\includegraphics[width=1.0\textwidth]{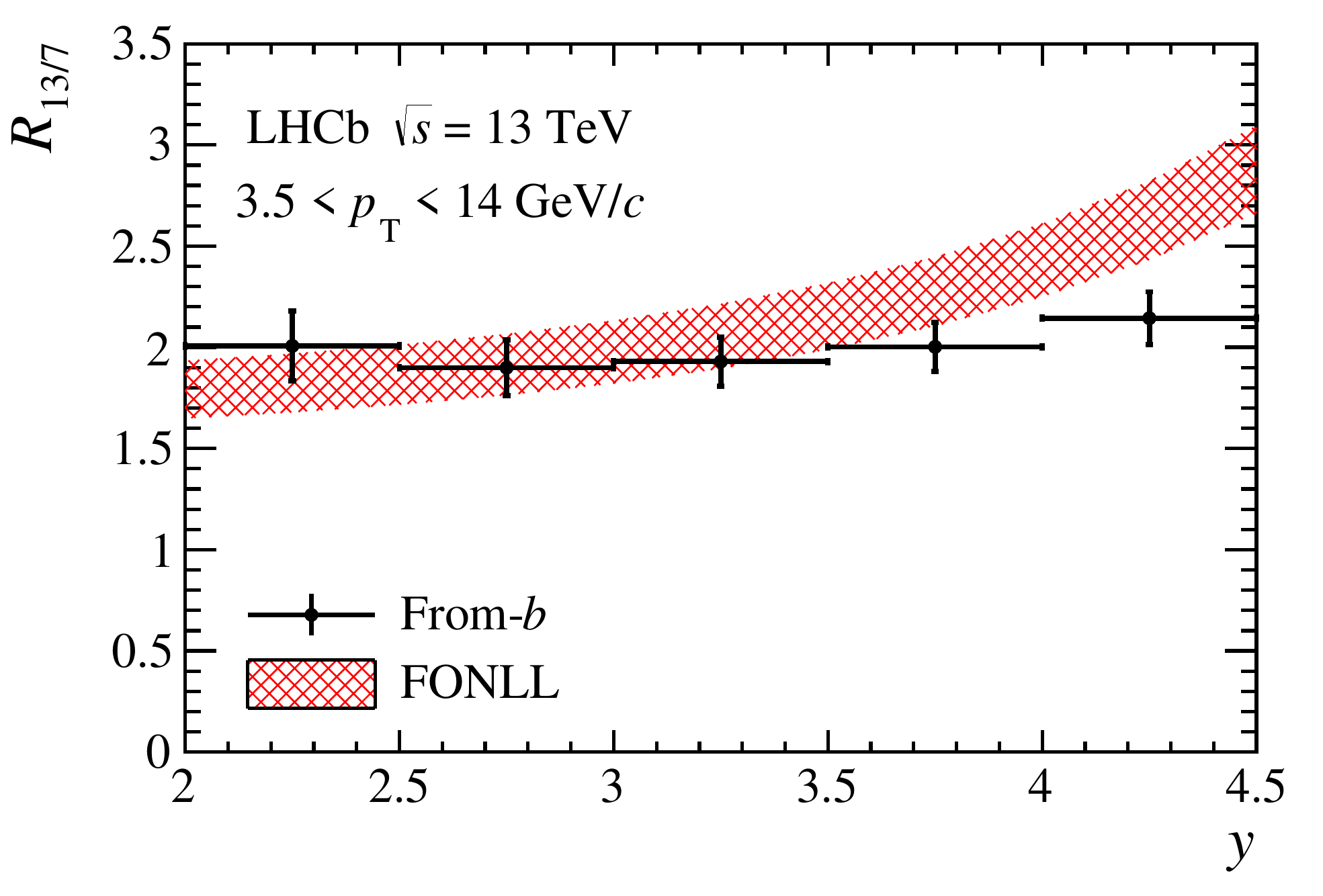}
\end{minipage}
    \caption{Ratio of differential production cross-sections between the $13\tev$ and the $7\tev$ measurements as a function of (left) \pt integrated over $y$ and (right) $y$ integrated over $\pt$ for \psitwos-from-\bquark. Theoretical FONLL calculations~\cite{Cacciari:2015fta} are compared to the data. } 
\label{fig:Ratio_7TeV_bdecay}
\end{figure}

The production cross-sections of $\psitwos$ mesons in $pp$ collisions at $13\tev$ and $7\tev$ are compared by means of their ratio, $R_{13/7}$.
Figures~\ref{fig:Ratio_7TeV_prompt} and~\ref{fig:Ratio_7TeV_bdecay} show the ratios as functions of $\pt$ integrated over $2.0<y<4.5$ and as functions of $y$ integrated over \mbox{$3.5<\pt<14\gevc$} for prompt \psitwos and \mbox{\psitwos-from-\bquark}.
The NRQCD (FONLL) calculations of $R_{13/7}$ for prompt \psitwos (\mbox{\psitwos-from-$b$}) are also shown in the left (right) panel for comparison. 
Both FONLL and NRQCD predictions on $R_{13/7}$ agree well with the corresponding experimental data.
The measured ratios are also presented in Tables~\ref{tab:Ratio_7TeV_PT} and~\ref{tab:Ratio_7TeV_Y} in Appendix~\ref{sec:tables}.

For both the theoretical calculations and the experimental measurements, some of the uncertainties in the ratio cancel, which allows for a more precise comparison to theory.
In the calculation of these ratios from the measured \psitwos production cross-sections at $13\tev$ and $7\tev$ the systematic uncertainty related to the branching fraction is cancelled.
The uncertainties due to the luminosity, the fit model and the tracking correction are partially correlated. 
Other uncertainties are assumed to be uncorrelated.

%%%%%%%%%%%%%%%%%%%%%%%%%%%%%%%%%%%%%%%%%%%%%%%%%%%%%%%%%%%
%%%% figure 10
%%%%%%%%%%%%%%%%%%%%%%%%%%%%%%%%%%%%%%%%%%%%%%%%%%%%%%%%%%%
%%%%%%%%%%%%%%%%%%%%%%%%%%%%%%%%%%%%%%%%%%%%%%%%%%%%%%%%%%%

%%%%%%%%%%%%%%%%%%%%%%%%%%%%%%%%%%%%%%%%%%%%%%%%%%%%%%%%%%
\subsection{\boldmath Measurement of the inclusive $\bquark\to\psitwos X$ branching fraction}\label{branching_fraction}
The reported results of the cross-section $\sigma(\psitwos\mbox{-from-}\bquark,13\tev)$, in combination with the previous results about \jpsi production \cite{LHCb-PAPER-2015-037}, can be used to determine the inclusive branching fraction \bratiopsi. To achieve this, both results must be extrapolated to the full phase space, as they are measured only for a limited range of phase space. The extrapolation factors $\alpha_{4\pi}(\psitwos)$ and $\alpha_{4\pi}(\jpsi)$ are determined with LHCb-tuned versions of Pythia 8~\cite{Sjostrand:2007gs} for the \psitwos and of Pythia 6~\cite{Sjostrand:2006za} for the \jpsi. The factors $\alpha_{4 \pi}(\psitwos)$ and $\alpha_{4\pi}(\jpsi)$ are found to be 7.29 and 5.20, respectively. 
In the ratio of the two factors,
\begin{equation*}
    \xi \equiv \frac{\alpha_{4\pi}(\psitwos)}{\alpha_{4\pi}(\jpsi)} = 1.402,
\end{equation*}
most of the theoretical uncertainties are expected to cancel. Alternatively, the correction factor $\xi$ can be obtained using FONLL calculations which uses different parton distribution functions. The values of $\xi$ obtained from the two methods differ by 2.89 \%. 

%The differences between the $\xi$ values obtained from Pythia and FONLL, using different parton distribution functions, are small. The maximum difference is less than 3\% between the available models.

With the definition of the ratio $\xi$, the $\mathcal{B}(b \rightarrow \psitwos X)$ branching fraction can be obtained from the ratio

\begin{equation}
\frac{\mathcal{B}(b \rightarrow \psitwos X)}{\mathcal{B}(b \rightarrow \jpsi X)} = 
\xi \ \frac{\sigma(\psitwos\mbox{-from-}\bquark, \mathrm{13~TeV})}{\sigma(\jpsi\mbox{-from-}\bquark, \mathrm{13~TeV})}.
 \end{equation}
By inserting the value $\sigma(\jpsi\mbox{-from-}\bquark,\,\mathrm{13~TeV}) = 2.25 \pm 0.01(\mathrm{stat}) \pm 0.14(\mathrm{syst})\mub$ \cite{LHCb-PAPER-2015-037} and the value of $\xi$, the ratio of the branching fractions is 
\begin{equation*}
\frac{\mathcal{B}(b \rightarrow \psitwos X)}{\mathcal{B}(b \rightarrow \jpsi X)} = 0.265 \pm 0.002(\mathrm{stat}) \pm 0.015(\mathrm{syst}) \pm 0.006(\mathrm{\BF})
,
 \end{equation*}
where possible correlations between uncertainties originating from \psifromb and \mbox{\jpsi-from-\bquark}, respectively, are taken into account. 
The last uncertainty is from the uncertainty of the branching fractions $\mathcal{B}(\psitwos \rightarrow e^+e^-)$ and $\mathcal{B}(\jpsi \rightarrow \mu^+ \mu^-)$. 
Using the known value $\mathcal{B}(b \rightarrow \jpsi X) = (1.16 \pm 0.10) \times 10^{-2}$~\cite{PDG2018}, one obtains
\begin{equation*}
\mathcal{B}(b \rightarrow \psitwos X) = (3.08 \pm 0.02(\mathrm{stat}) \pm 0.18(\mathrm{syst}) \pm 0.27(\mathrm{\BF})) \times 10^{-3}
.
 \end{equation*}
This result is in agreement with the world-average value~\cite{PDG2018}. The $\mathcal{B}(b \rightarrow \jpsi X)$ uncertainty dominates the total uncertainty from the branching fractions.  
%\cleardoublepage
\section{Conclusions}
\label{sec:conclusion}
The production cross-sections of \psitwos mesons in proton-proton collisions at a \mbox{centre-of-mass energy} of $13\tev$ are reported with a data sample corresponding to an integrated luminosity of $275\pm 11\invpb$, collected by the \lhcb detector in 2015. 
The double-differential cross-sections, as functions of \pt and $y$ of the \psitwos meson in the range of \mbox{$2<\pt<20\gevc$} and \mbox{$2.0<y<4.5$}, are determined for prompt \psitwos mesons and \psitwos mesons from $\bquark$-hadron decays. A new measurement of the branching fraction \bratiopsi~is presented, which is in agreement with the world average~\cite{PDG2018}.
The measured prompt \psitwos production cross-section as a function of transverse momentum is in good agreement in the high \pt region with theoretical calculations in the NRQCD framework. 
Theoretical predictions based on the FONLL calculations describe well the measured cross-sections for \psitwos mesons from $\bquark$-hadron decays.

A new measurement of \psitwos production cross-sections at $7\tev$ is performed using the 2011 data sample corresponding to an integrated luminosity of $614\pm11\invpb$.
The new result provides a significantly reduced uncertainty compared to the previous independent LHCb result~\cite{LHCb-PAPER-2011-045}.

The cross section ratios between $13\tev$ and $7\tev$ show reasonable agreement with theoretical calculations.

%
% Comment this in for paper drafts; do not include this in analysis note and conference reports
\section*{Acknowledgements}
\noindent We thank Kuang-Ta Chao and Yan-Qing Ma for frequent and interesting discussions on the production of $\psitwos$ mesons.
We express our gratitude to our colleagues in the CERN
accelerator departments for the excellent performance of the LHC. We
thank the technical and administrative staff at the LHCb
institutes.
We acknowledge support from CERN and from the national agencies:
CAPES, CNPq, FAPERJ and FINEP (Brazil); 
MOST and NSFC (China); 
CNRS/IN2P3 (France); 
BMBF, DFG and MPG (Germany); 
INFN (Italy); 
NWO (Netherlands); 
MNiSW and NCN (Poland); 
MEN/IFA (Romania); 
MSHE (Russia); 
MinECo (Spain); 
SNSF and SER (Switzerland); 
NASU (Ukraine); 
STFC (United Kingdom); 
DOE NP and NSF (USA).
We acknowledge the computing resources that are provided by CERN, IN2P3
(France), KIT and DESY (Germany), INFN (Italy), SURF (Netherlands),
PIC (Spain), GridPP (United Kingdom), RRCKI and Yandex
LLC (Russia), CSCS (Switzerland), IFIN-HH (Romania), CBPF (Brazil),
PL-GRID (Poland) and OSC (USA).
We are indebted to the communities behind the multiple open-source
software packages on which we depend.
Individual groups or members have received support from
AvH Foundation (Germany);
EPLANET, Marie Sk\l{}odowska-Curie Actions and ERC (European Union);
ANR, Labex P2IO and OCEVU, and R\'{e}gion Auvergne-Rh\^{o}ne-Alpes (France);
Key Research Program of Frontier Sciences of CAS, CAS PIFI, and the Thousand Talents Program (China);
RFBR, RSF and Yandex LLC (Russia);
GVA, XuntaGal and GENCAT (Spain);
the Royal Society
and the Leverhulme Trust (United Kingdom).

% This should be taken out in the final paper
%\input{supplementary-app}

\cleardoublepage
%\appendix
\clearpage
{\noindent\normalfont\bfseries\Large Appendices}
\appendix

\section{Result tables}
\label{sec:tables}

\begin{sidewaystable}[h]
\caption{Double-differential production cross-sections (in \nb/(\mygevc)) of prompt \psitwos mesons at $13\tev$ in bins of (\pt, $y$). The first uncertainties are statistical, the second are the uncorrelated systematic uncertainties between bins, and the last are the correlated systematic uncertainties between bins. Adjacent bins with large statistical uncertainties have been merged.}
\centering
\resizebox{\linewidth}{!}{
\begin{tabular}{l|ccccc}
\hline
\pt(\mygevc)& $2.0<y<2.5$& $2.5<y<3.0$& $3.0<y<3.5$& $3.5<y<4.0$& $4.0<y<4.5$\\
\hline
\kern 0.5em2--3 & $232.86\pm5.80\pm5.39\pm18.24$ & $228.46\pm3.53\pm1.65\pm17.70$ & $198.36\pm2.86\pm1.34\pm15.33$ & $167.07\pm2.19\pm1.30\pm12.93$ & $133.13\pm1.99\pm1.62\pm10.42$\\
\kern 0.5em3--4 & $192.93\pm3.96\pm4.02\pm\xx9.74$ & $166.62\pm2.36\pm1.27\pm\xx8.22$ & $147.39\pm1.90\pm1.07\pm\xx7.25$ & $120.42\pm1.49\pm0.96\pm\xx6.00$ & $\xx85.03\pm1.42\pm1.09\pm\xx4.37$\\
\kern 0.5em4--5 & $124.73\pm2.38\pm2.70\pm\xx9.15$ & $108.56\pm1.41\pm0.88\pm\xx7.92$ & $\xx94.79\pm1.09\pm0.74\pm\xx6.90$ & $\xx76.73\pm0.93\pm0.68\pm\xx5.59$ & $\xx57.41\pm1.00\pm0.82\pm\xx4.28$\\
\kern 0.5em5--6 & $\xx77.27\pm1.46\pm1.78\pm\xx4.89$ & $\xx66.05\pm0.82\pm0.59\pm\xx4.16$ & $\xx58.12\pm0.65\pm0.52\pm\xx3.65$ & $\xx47.25\pm0.44\pm0.48\pm\xx2.98$ & $\xx34.35\pm0.66\pm0.58\pm\xx2.29$\\
\kern 0.5em6--7 & $\xx45.13\pm0.91\pm1.11\pm\xx4.73$ & $\xx41.43\pm0.53\pm0.43\pm\xx4.34$ & $\xx35.20\pm0.43\pm0.37\pm\xx3.69$ & $\xx28.70\pm0.39\pm0.36\pm\xx3.01$ & $\xx19.22\pm0.47\pm0.38\pm\xx2.08$\\
\kern 0.5em7--8 & $\xx28.87\pm0.62\pm0.77\pm\xx1.95$ & $\xx25.38\pm0.35\pm0.32\pm\xx1.70$ & $\xx21.00\pm0.30\pm0.27\pm\xx1.41$ & $\xx15.97\pm0.27\pm0.25\pm\xx1.07$ & $\xx12.23\pm0.32\pm0.31\pm\xx0.91$\\
\kern 0.5em8--9 & $\xx17.52\pm0.42\pm0.49\pm\xx1.19$ & $\xx15.39\pm0.25\pm0.24\pm\xx1.04$ & $\xx12.75\pm0.22\pm0.20\pm\xx0.86$ & $\xx\xx9.86\pm0.20\pm0.19\pm\xx0.67$ & $\xx\xx6.08\pm0.21\pm0.18\pm\xx0.47$\\
\kern 0.5em9--10 & $\xx10.94\pm0.29\pm0.34\pm\xx0.56$ & $\xx\xx9.63\pm0.18\pm0.18\pm\xx0.48$ & $\xx\xx7.46\pm0.16\pm0.14\pm\xx0.38$ & $\xx\xx5.90\pm0.15\pm0.15\pm\xx0.30$ & $\xx\xx3.81\pm0.16\pm0.14\pm\xx0.26$\\
10--11 & $\xx\xx7.66\pm0.23\pm0.28\pm\xx0.50$ & $\xx\xx5.98\pm0.14\pm0.13\pm\xx0.39$ & $\xx\xx4.84\pm0.13\pm0.11\pm\xx0.31$ & $\xx\xx3.83\pm0.12\pm0.11\pm\xx0.25$ & $\xx\xx2.47\pm0.11\pm0.11\pm\xx0.19$\\
11--12 & $\xx\xx4.25\pm0.16\pm0.17\pm\xx0.23$ & $\xx\xx3.86\pm0.11\pm0.10\pm\xx0.21$ & $\xx\xx3.16\pm0.10\pm0.09\pm\xx0.17$ & $\xx\xx2.37\pm0.09\pm0.08\pm\xx0.13$ & $\xx\xx1.67\pm0.10\pm0.09\pm\xx0.13$\\
12--13 & $\xx\xx3.09\pm0.13\pm0.16\pm\xx0.27$ & $\xx\xx2.44\pm0.13\pm0.07\pm\xx0.21$ & $\xx\xx2.02\pm0.08\pm0.07\pm\xx0.18$ & $\xx\xx1.39\pm0.07\pm0.06\pm\xx0.12$ & $\xx\xx0.74\pm0.06\pm0.05\pm\xx0.07$\\
13--14 & $\xx\xx1.63\pm0.09\pm0.08\pm\xx0.12$ & $\xx\xx1.57\pm0.07\pm0.06\pm\xx0.12$ & $\xx\xx1.26\pm0.06\pm0.05\pm\xx0.09$ & $\xx\xx0.91\pm0.05\pm0.04\pm\xx0.07$ & $\xx\xx0.58\pm0.06\pm0.04\pm\xx0.05$\\  \cline{6-6}
14--15 &   $\xx\xx1.42\pm0.08\pm0.08\pm\xx0.07$ &   $\xx\xx1.26\pm0.06\pm0.05\pm\xx0.06$ &   $\xx\xx0.81\pm0.05\pm0.04\pm\xx0.04$ &   $\xx\xx0.59\pm0.04\pm0.04\pm\xx0.03$ &  \multirow{2}{*}{ $0.26\pm0.03\pm0.02\pm0.02$}   \\  \cline{5-5}
15--16 &   $\xx\xx1.18\pm0.08\pm0.09\pm\xx0.06$ &   $\xx\xx0.86\pm0.05\pm0.04\pm\xx0.04$ &   $\xx\xx0.58\pm0.04\pm0.03\pm\xx0.03$ & \multirow{2}{*}{ $0.41\pm0.03\pm0.02\pm0.02$} &   \\  \cline{6-6}
16--17 &    $\xx\xx0.81\pm0.06\pm0.07\pm\xx0.04$ &    $\xx\xx0.59\pm0.04\pm0.03\pm\xx0.03$ &    $\xx\xx0.39\pm0.03\pm0.03\pm\xx0.02$ &     & \multirow{4}{*}{ $0.15\pm0.01\pm0.02\pm0.01$} \\ \cline{4-5}
17--18 &    $\xx\xx0.54\pm0.05\pm0.06\pm\xx0.03$ &    $\xx\xx0.42\pm0.03\pm0.03\pm\xx0.02$ & \multirow{3}{*}{ $0.21\pm0.01\pm0.01\pm0.01$} & \multirow{3}{*}{ $0.15\pm0.01\pm0.01\pm0.01$} &   \\  \cline{2-3}
18--19 & \multirow{2}{*}{ $0.29\pm0.02\pm0.02\pm0.01$} & \multirow{2}{*}{ $0.28\pm0.02\pm0.02\pm0.02$} &    &    &   \\
19--20 &    &    &    &    &   \\ \cline{2-4}
 \hline
\end{tabular}
}
\label{tab:results_prompt}
\end{sidewaystable}

\begin{sidewaystable}[!htbp]
\caption{Double-differential production cross-sections (in $\nb/(\mygevc)$) of \psitwos-from-$b$ mesons at $13\tev$ in bins of (\pt,$y$). The first uncertainties are statistical, the second are the uncorrelated systematic uncertainties between bins, and the last are the correlated systematic uncertainties between bins. 
Adjacent bins with large statistical uncertainties have been merged.}
\centering
\resizebox{\linewidth}{!}{
\begin{tabular}{l|ccccc}
\hline
\pt(\mygevc)& $2.0<y<2.5$& $2.5<y<3.0$& $3.0<y<3.5$& $3.5<y<4.0$& $4.0<y<4.5$\\
\hline
\kern 0.5em 2--3 & $69.99\pm1.94\pm1.65\pm5.51$ & $61.15\pm1.14\pm0.48\pm4.78$ & $51.47\pm0.97\pm0.40\pm3.99$ & $37.86\pm0.87\pm0.35\pm2.95$ & $24.04\pm0.94\pm0.36\pm1.89$\\
\kern 0.5em 3--4 & $56.37\pm1.45\pm1.17\pm3.05$ & $47.89\pm0.83\pm0.36\pm2.45$ & $39.58\pm0.70\pm0.31\pm2.45$ & $29.92\pm0.64\pm0.27\pm1.57$ & $19.49\pm0.70\pm0.29\pm1.01$\\
\kern 0.5em 4--5 & $37.99\pm1.00\pm0.78\pm2.80$ & $34.38\pm0.58\pm0.25\pm2.51$ & $28.42\pm0.48\pm0.21\pm2.08$ & $21.25\pm0.45\pm0.19\pm1.55$ & $12.89\pm0.50\pm0.20\pm0.97$\\
\kern 0.5em 5--6 & $27.93\pm0.72\pm0.60\pm1.77$ & $23.18\pm0.40\pm0.18\pm1.46$ & $19.18\pm0.34\pm0.15\pm1.24$ & $14.62\pm0.24\pm0.14\pm0.94$ & $\xx7.94\pm0.35\pm0.14\pm0.56$\\
\kern 0.5em 6--7 & $16.99\pm0.49\pm0.37\pm1.78$ & $15.74\pm0.29\pm0.13\pm1.65$ & $12.33\pm0.24\pm0.11\pm1.29$ & $\xx8.91\pm0.22\pm0.10\pm0.94$ & $\xx4.83\pm0.25\pm0.09\pm0.52$\\
\kern 0.5em 7--8 & $12.29\pm0.37\pm0.28\pm0.83$ & $10.14\pm0.21\pm0.09\pm0.68$ & $\xx7.77\pm0.18\pm0.08\pm0.53$ & $\xx5.75\pm0.17\pm0.07\pm0.39$ & $\xx3.41\pm0.18\pm0.08\pm0.26$\\
\kern 0.5em 8--9 & $\xx9.05\pm0.29\pm0.20\pm0.62$ & $\xx6.84\pm0.16\pm0.07\pm0.46$ & $\xx5.24\pm0.14\pm0.06\pm0.36$ & $\xx3.56\pm0.13\pm0.05\pm0.24$ & $\xx2.09\pm0.13\pm0.05\pm0.16$\\
\kern 0.5em 9--10 & $\xx6.12\pm0.22\pm0.13\pm0.32$ & $\xx4.85\pm0.13\pm0.06\pm0.24$ & $\xx3.60\pm0.11\pm0.05\pm0.19$ & $\xx2.53\pm0.10\pm0.04\pm0.13$ & $\xx1.50\pm0.10\pm0.05\pm0.10$\\
10--11 & $\xx4.17\pm0.16\pm0.11\pm0.27$ & $\xx3.38\pm0.11\pm0.05\pm0.22$ & $\xx2.52\pm0.09\pm0.04\pm0.16$ & $\xx1.63\pm0.08\pm0.03\pm0.11$ & $\xx0.87\pm0.06\pm0.03\pm0.07$\\
11--12 & $\xx2.90\pm0.13\pm0.07\pm0.16$ & $\xx2.38\pm0.09\pm0.04\pm0.13$ & $\xx1.66\pm0.08\pm0.03\pm0.11$ & $\xx1.07\pm0.06\pm0.03\pm0.06$ & $\xx0.65\pm0.06\pm0.03\pm0.06$\\
12--13 & $\xx2.09\pm0.11\pm0.06\pm0.19$ & $\xx1.68\pm0.07\pm0.03\pm0.15$ & $\xx1.22\pm0.06\pm0.03\pm0.11$ & $\xx0.78\pm0.05\pm0.02\pm0.07$ & $\xx0.41\pm0.05\pm0.02\pm0.04$\\
13--14 & $\xx1.39\pm0.08\pm0.04\pm0.11$ & $\xx1.24\pm0.06\pm0.03\pm0.09$ & $\xx0.76\pm0.05\pm0.02\pm0.06$ & $\xx0.61\pm0.04\pm0.02\pm0.05$ & $\xx0.29\pm0.04\pm0.02\pm0.03$\\  \cline{6-6}
14--15 & $\xx1.18\pm0.07\pm0.04\pm0.06$ & $\xx0.82\pm0.05\pm0.02\pm0.04$ &   $\xx0.71\pm0.05\pm0.02\pm0.04$ &   $\xx0.42\pm0.04\pm0.02\pm0.02$ &  \multirow{2}{*}{ $0.16\pm0.02\pm0.01\pm0.02$}   \\  \cline{5-5}
15--16 & $\xx0.84\pm0.06\pm0.03\pm0.04$ & $\xx0.70\pm0.05\pm0.02\pm0.04$ &   $\xx0.42\pm0.03\pm0.01\pm0.03$ & \multirow{2}{*}{ $0.24\pm0.02\pm0.01\pm0.02$} &   \\  \cline{6-6}
16--17 & $\xx0.62\pm0.05\pm0.02\pm0.03$ & $\xx0.54\pm0.04\pm0.02\pm0.03$ &    $\xx0.34\pm0.03\pm0.01\pm0.02$ &     & \multirow{4}{*}{ $0.07\pm0.01\pm0.00\pm0.01$} \\ \cline{4-5}
17--18 & $\xx0.63\pm0.05\pm0.03\pm0.04$ & $\xx0.41\pm0.03\pm0.02\pm0.02$ & \multirow{3}{*}{ $0.21\pm0.01\pm0.01\pm0.01$} & \multirow{3}{*}{ $0.09\pm0.01\pm0.00\pm0.01$} &   \\  \cline{2-3}
18--19 & \multirow{2}{*}{ $0.32\pm0.03\pm0.01\pm0.02$} & \multirow{2}{*}{ $0.28\pm0.02\pm0.01\pm0.02$} &    &    &   \\
19--20 &    &    &    &    &   \\ \cline{2-4}
 \hline
\end{tabular}
}
\label{tab:results_fromb}
\end{sidewaystable}

%%%%%%%%%%%%%%%%%%%%%%%%%%%%%%%%%%%%%%%%%%%%%%%%%%%%%%%%%%
\begin{sidewaystable}[h]
\caption{
    Double-differential production cross-section in \nb/(\mygevc) of prompt \psitwos mesons at $7\tev$ in bins of (\pt,$y$). The first uncertainty is statistical, the second is the uncorrelated systematic uncertainties shared between bins and the last is the correlated systematic uncertainties.}
\centering
\resizebox{\linewidth}{!}{
\begin{tabular}{l|ccccc}
\hline
\pt(\gevc)& $2<y<2.5$& $2.5<y<3$& $3<y<3.5$& $3.5<y<4$& $4<y<4.5$\\
\hline
3.5--4 & $94.34\pm2.49\pm5.59\pm8.65$ & $109.85\pm1.38\pm2.50\pm5.04$ & $97.07\pm0.99\pm1.90\pm4.33$ & $77.87\pm0.78\pm1.51\pm3.11$ & $57.90\pm0.80\pm1.57\pm2.09$\\
\kern 0.75em 4--5 & $73.08\pm1.21\pm4.35\pm4.40$ & $\xx75.42\pm0.66\pm1.32\pm3.53$ & $65.66\pm0.47\pm0.85\pm2.64$ & $53.33\pm0.40\pm0.88\pm2.23$ & $36.34\pm0.40\pm0.80\pm1.30$\\
\kern 0.75em 5--6 & $44.86\pm0.71\pm2.47\pm2.24$ & $\xx43.26\pm0.38\pm1.04\pm4.14$ & $37.21\pm0.30\pm0.58\pm1.54$ & $31.96\pm0.27\pm0.86\pm1.31$ & $18.98\pm0.26\pm0.93\pm0.65$\\
\kern 0.75em 6--7 & $25.99\pm0.45\pm1.64\pm2.58$ & $\xx24.94\pm0.25\pm0.63\pm1.20$ & $22.42\pm0.21\pm0.58\pm0.86$ & $17.86\pm0.19\pm0.61\pm0.69$ & $10.33\pm0.17\pm0.39\pm0.50$\\
\kern 0.75em 7--8 & $16.58\pm0.30\pm1.26\pm1.04$ & $\xx15.21\pm0.17\pm0.41\pm0.66$ & $12.82\pm0.14\pm0.40\pm0.50$ & $\xx9.98\pm0.13\pm0.33\pm0.44$ & $\xx5.61\pm0.12\pm0.30\pm0.26$\\
\kern 0.75em 8--9 & $\xx7.03\pm0.14\pm1.16\pm0.35$ & $\xx\xx9.02\pm0.12\pm0.39\pm0.41$ & $\xx7.16\pm0.10\pm0.31\pm0.33$ & $\xx5.05\pm0.09\pm0.28\pm0.23$ & $\xx2.92\pm0.08\pm0.18\pm0.15$\\
\kern 0.75em 9--10 & $\xx6.10\pm0.14\pm0.63\pm0.28$ & $\xx\xx5.26\pm0.09\pm0.26\pm0.23$ & $\xx4.63\pm0.08\pm0.20\pm0.18$ & $\xx3.16\pm0.07\pm0.20\pm0.13$ & $\xx1.89\pm0.07\pm0.20\pm0.12$\\
\kern 0.25em10--11 & $\xx3.89\pm0.11\pm0.36\pm0.23$ & $\xx\xx3.03\pm0.06\pm0.17\pm0.14$ & $\xx2.50\pm0.05\pm0.17\pm0.11$ & $\xx1.93\pm0.06\pm0.18\pm0.07$ & $\xx1.04\pm0.05\pm0.15\pm0.06$\\
\kern 0.25em11--12 & $\xx2.56\pm0.08\pm0.29\pm0.13$ & $\xx\xx2.30\pm0.06\pm0.35\pm0.10$ & $\xx1.50\pm0.04\pm0.16\pm0.13$ & $\xx1.46\pm0.05\pm0.25\pm0.10$ & $\xx0.38\pm0.02\pm0.08\pm0.02$\\
\kern 0.25em12--13 & $\xx2.42\pm0.10\pm0.90\pm0.11$ & $\xx\xx1.12\pm0.04\pm0.09\pm0.05$ & $\xx0.84\pm0.03\pm0.12\pm0.04$ & $\xx0.58\pm0.03\pm0.08\pm0.05$ & $\xx0.35\pm0.03\pm0.10\pm0.04$\\
\kern 0.25em13--14 & $\xx0.85\pm0.04\pm0.16\pm0.04$ & $\xx\xx0.88\pm0.03\pm0.12\pm0.11$ & $\xx0.60\pm0.03\pm0.09\pm0.04$ & $\xx0.33\pm0.02\pm0.06\pm0.02$ & $\xx0.07\pm0.01\pm0.02\pm0.01$\\ 
\hline
\end{tabular}
}
\label{tab7TeV:results_prompt}
\end{sidewaystable}
%%%%%%%%%%%%%%%%%%%%%%%%%%%%%%%%%%%%%%%%%%%%%%%%%%%%%%%%%%

%%%%%%%%%%%%%%%%%%%%%%%%%%%%%%%%%%%%%%%%%%%%%%%%%%%%%%%%%%
\begin{sidewaystable}[!htbp]
\caption{Double-differential production cross-section in \nb/(\mygevc) of \psitwos-from-$b$ mesons at $7\tev$ in bins of (\pt,$y$). The first uncertainties are statistical, the second are the uncorrelated systematic uncertainties shared between bins and the last are the correlated systematic uncertainties.}
\centering
\resizebox{\linewidth}{!}{
\begin{tabular}{l|ccccc}
\hline
\pt(\gevc)& $2<y<2.5$& $2.5<y<3$& $3<y<3.5$& $3.5<y<4$& $4<y<4.5$\\
\hline
3.5--4 & $27.65\pm1.10\pm1.64\pm2.86$ & $24.88\pm0.55\pm0.44\pm1.66$ & $21.06\pm0.42\pm0.35\pm0.98$ & $16.01\pm0.36\pm0.43\pm0.65$ & $9.62\pm0.36\pm0.32\pm0.53$\\
\kern 0.75em4--5 & $20.85\pm0.54\pm1.12\pm1.36$ & $19.60\pm0.29\pm0.26\pm1.14$ & $15.58\pm0.22\pm0.19\pm0.74$ & $11.05\pm0.19\pm0.18\pm0.46$ & $6.58\pm0.18\pm0.18\pm0.23$\\
\kern 0.75em5--6 & $13.31\pm0.36\pm0.96\pm0.93$ & $13.04\pm0.19\pm0.19\pm1.31$ & $\xx9.80\pm0.15\pm0.13\pm0.43$ & $\xx6.90\pm0.13\pm0.12\pm0.28$ & $4.00\pm0.13\pm0.14\pm0.14$\\
\kern 0.75em6--7 & $\xx8.77\pm0.24\pm0.45\pm0.93$ & $\xx8.15\pm0.14\pm0.12\pm0.40$ & $\xx6.20\pm0.11\pm0.09\pm0.26$ & $\xx4.53\pm0.10\pm0.08\pm0.18$ & $2.01\pm0.09\pm0.08\pm0.14$\\
\kern 0.75em7--8 & $\xx6.24\pm0.18\pm0.33\pm0.41$ & $\xx4.96\pm0.10\pm0.08\pm0.25$ & $\xx4.02\pm0.08\pm0.07\pm0.16$ & $\xx2.64\pm0.07\pm0.05\pm0.15$ & $1.38\pm0.07\pm0.06\pm0.07$\\
\kern 0.75em8--9 & $\xx3.68\pm0.12\pm0.19\pm0.23$ & $\xx3.40\pm0.08\pm0.07\pm0.16$ & $\xx2.69\pm0.07\pm0.05\pm0.14$ & $\xx1.61\pm0.05\pm0.04\pm0.08$ & $0.76\pm0.05\pm0.04\pm0.04$\\
\kern 0.75em9--10 & $\xx3.14\pm0.11\pm0.21\pm0.19$ & $\xx2.15\pm0.06\pm0.04\pm0.09$ & $\xx1.33\pm0.05\pm0.04\pm0.12$ & $\xx1.00\pm0.04\pm0.03\pm0.04$ & $0.37\pm0.03\pm0.02\pm0.04$\\
\kern 0.25em10--11 & $\xx1.96\pm0.08\pm0.12\pm0.12$ & $\xx1.46\pm0.05\pm0.04\pm0.07$ & $\xx1.11\pm0.04\pm0.04\pm0.07$ & $\xx0.66\pm0.03\pm0.02\pm0.03$ & $0.32\pm0.03\pm0.02\pm0.02$\\
\kern 0.25em11--12 & $\xx1.28\pm0.06\pm0.07\pm0.07$ & $\xx1.00\pm0.04\pm0.03\pm0.05$ & $\xx0.75\pm0.03\pm0.03\pm0.07$ & $\xx0.42\pm0.02\pm0.02\pm0.03$ & $0.18\pm0.02\pm0.02\pm0.01$\\
\kern 0.25em12--13 & $\xx0.80\pm0.04\pm0.05\pm0.05$ & $\xx0.73\pm0.03\pm0.02\pm0.04$ & $\xx0.51\pm0.03\pm0.02\pm0.02$ & $\xx0.28\pm0.02\pm0.01\pm0.03$ & $0.09\pm0.02\pm0.01\pm0.01$\\
\kern 0.25em13--14 & $\xx0.66\pm0.04\pm0.07\pm0.04$ & $\xx0.51\pm0.03\pm0.02\pm0.06$ & $\xx0.33\pm0.02\pm0.01\pm0.02$ & $\xx0.17\pm0.02\pm0.01\pm0.01$ & $0.06\pm0.01\pm0.01\pm0.01$\\
 \hline
\end{tabular}
}
\label{tab7TeV:results_fromb}
\end{sidewaystable}
%%%%%%%%%%%%%%%%%%%%%%%%%%%%%%%%%%%%%%%%%%%%%%%%%%%%%%%%%%%%%%%%%

%%%%%%%%%%%%%%%%%%%%%%%%%%%%%%%%%%%%%%%%%%%%%%%%%%%%%%%%%%
\begin{table}[htp]
\caption{
    Differential production cross-sections $\deriv\sigma/\deriv\pt$ (in \nb/(\mygevc)) of prompt \psitwos and \psitwos-from-$\bquark$ mesons at $13\tev$. The first uncertainties are statistical and the second (third) are uncorrelated (correlated) systematic uncertainties amongst bins.}
\centering
%\small
\begin{tabular}{c|cc}
\hline
\pt(\mygevc)& Prompt \psitwos & \psitwos-from-$\bquark$ \\
\hline
\kern 0.5em2--3 & $479.94\pm3.97\pm3.08\pm37.31$ & $122.26\pm1.38\pm0.91\pm9.56$\\
\kern 0.5em3--4 & $356.19\pm2.69\pm2.29\pm17.79$ & $\xx96.62\pm1.02\pm0.66\pm5.27$\\
\kern 0.5em4--5 & $231.12\pm1.64\pm1.56\pm16.92$ & $\xx67.47\pm0.71\pm0.45\pm4.95$\\
\kern 0.5em5--6 & $141.52\pm0.98\pm1.04\pm\xx8.99$ & $\xx46.43\pm0.49\pm0.34\pm2.98$\\
\kern 0.5em6--7 & $\xx84.84\pm0.65\pm0.68\pm\xx8.92$ & $\xx29.40\pm0.35\pm0.21\pm3.10$\\
\kern 0.5em7--8 & $\xx51.72\pm0.44\pm0.48\pm\xx3.52$ & $\xx19.68\pm0.26\pm0.16\pm1.34$\\
\kern 0.5em8--9 & $\xx30.80\pm0.30\pm0.32\pm\xx2.11$ & $\xx13.38\pm0.20\pm0.12\pm0.92$\\
\kern 0.5em9--10 & $\xx18.87\pm0.22\pm0.23\pm\xx0.99$ & $\xx\xx9.29\pm0.16\pm0.08\pm0.49$\\
10--11 & $\xx12.39\pm0.17\pm0.18\pm\xx0.82$ & $\xx\xx6.29\pm0.12\pm0.07\pm0.42$\\
11--12 & $\xx\xx7.65\pm0.13\pm0.13\pm\xx0.44$ & $\xx\xx4.33\pm0.10\pm0.05\pm0.26$\\
12--13 & $\xx\xx4.84\pm0.11\pm0.10\pm\xx0.43$ & $\xx\xx3.09\pm0.08\pm0.04\pm0.28$\\
13--14 & $\xx\xx2.98\pm0.07\pm0.06\pm\xx0.23$ & $\xx\xx2.14\pm0.06\pm0.03\pm0.17$\\
14--20 & $\xx\xx1.13\pm0.02\pm0.02\pm\xx0.06$ & $\xx\xx0.90\pm0.02\pm0.01\pm0.05$\\
\hline
\end{tabular}
\label{tab:results_PT}
\end{table}
%%%%%%%%%%%%%%%%%%%%%%%%%%%%%%%%%%%%%%%%%%%%%%%%%%%%%%%%%%%%%%%%%%

%%%%%%%%%%%%%%%%%%%%%%%%%%%%%%%%%%%%%%%%%%%%%%%%%%%%%%%%%%%%%%%%%%
\begin{table}[htp]
\caption{
    Differential production cross-sections $\deriv\sigma/\deriv y$ (in \nb) of prompt \psitwos and \psitwos-from-$\bquark$ mesons at $13\tev$ per rapidity unit. The first uncertainties are statistical and the second (third) are uncorrelated (correlated) systematic uncertainties amongst bins.}
\centering
%\small
\begin{tabular}{c|cc}
\hline
$y$& Prompt \psitwos & \psitwos-from-$\bquark$ \\
\hline
2.0--2.5 & $751.4\pm7.7\pm7.6\pm51.8$ & $251.2\pm2.8\pm2.3\pm17.6$\\
2.5--3.0 & $679.1\pm4.6\pm2.4\pm46.7$ & $215.9\pm1.6\pm0.7\pm15.0$\\
3.0--3.5 & $588.7\pm3.7\pm2.0\pm40.4$ & $175.8\pm1.4\pm0.6\pm12.7$\\
3.5--4.0 & $482.3\pm2.9\pm1.9\pm33.2$ & $129.6\pm1.2\pm0.5\pm\xx9.1$\\
4.0--4.5 & $357.8\pm2.8\pm2.3\pm25.6$ & $\xx79.0\pm1.4\pm0.5\pm\xx5.7$\\
 \hline
\end{tabular}
\label{tab:results_Y}
\end{table}
%%%%%%%%%%%%%%%%%%%%%%%%%%%%%%%%%%%%%%%%%%%%%%%%%%%%%%%%%%%

%%%%%%%%%%%%%%%%%%%%%%%%%%%%%%%%%%%%%%%%%%%%%%%%%%%%%%%%%%%%%%%%%
\begin{table}[htp]
\caption{
    Differential cross-sections $\deriv\sigma/\deriv\pt$ (in \nb/(\mygevc)) of prompt \psitwos and \psitwos-from-$b$ mesons at $7\tev$, integrated over $y$ between 2.0 and 4.5. The first uncertainties are statistical and the second (third) are uncorrelated (correlated) systematic uncertainties amongst bins.}
\centering
%\small
\begin{tabular}{c|cc}
\hline
\pt(\gevc)& Prompt \psitwos & \psitwos-from-$b$ \\
\hline
3.5-4 & $218.52\pm1.61\pm3.38\pm11.61$ & $49.61\pm0.70\pm0.91\pm3.34$\\
\kern 0.75em4--5 & $151.91\pm0.78\pm2.39\pm\xx7.05$ & $36.83\pm0.35\pm0.60\pm1.97$\\
\kern 0.75em5--6 & $\xx88.13\pm0.47\pm1.51\pm\xx4.94$ & $23.52\pm0.24\pm0.50\pm1.55$\\
\kern 0.75em6--7 & $\xx50.77\pm0.30\pm0.99\pm\xx2.92$ & $14.83\pm0.16\pm0.24\pm0.95$\\
\kern 0.75em7--8 & $\xx30.10\pm0.21\pm0.73\pm\xx1.45$ & $\xx9.62\pm0.12\pm0.18\pm0.52$\\
\kern 0.75em8--9 & $\xx15.59\pm0.12\pm0.65\pm\xx0.74$ & $\xx6.08\pm0.09\pm0.11\pm0.32$\\
\kern 0.75em9--10 & $\xx10.52\pm0.10\pm0.38\pm\xx0.47$ & $\xx3.99\pm0.07\pm0.11\pm0.24$\\
\kern 0.25em10--11 & $\xx\xx6.20\pm0.08\pm0.25\pm\xx0.30$ & $\xx2.75\pm0.05\pm0.07\pm0.16$\\
\kern 0.25em11--12 & $\xx\xx4.10\pm0.06\pm0.27\pm\xx0.24$ & $\xx1.81\pm0.04\pm0.04\pm0.12$\\
\kern 0.25em12--13 & $\xx\xx2.65\pm0.06\pm0.46\pm\xx0.15$ & $\xx1.20\pm0.03\pm0.03\pm0.08$\\
\kern 0.25em13--14 & $\xx\xx1.37\pm0.03\pm0.12\pm\xx0.11$ & $\xx0.86\pm0.03\pm0.04\pm0.07$\\
 \hline
\end{tabular}
\label{tab7TeV:results_PT}
\end{table}

\begin{table}[htp]
    \caption{
        Differential cross-sections $\deriv\sigma/\deriv y$ (in \nb) of prompt \psitwos and \psitwos-from-$b$ mesons at $7\tev$, integrated over \pt between 3.5 and 14\gevc. The first uncertainties are statistical and the second (third) are uncorrelated (correlated) systematic uncertainties amongst bins.}
\centering
%\small
\begin{tabular}{ccc}
\hline
$y$& Prompt \psitwos & \psitwos-from-$b$ \\
\hline
2.0--2.5 & $230.5\pm2.0\pm6.3\pm15.7$ & $74.5\pm0.9\pm1.8\pm5.8$\\
2.5--3.0 & $235.4\pm1.1\pm2.3\pm13.1$ & $67.4\pm0.5\pm0.4\pm4.4$\\
3.0--3.5 & $203.9\pm0.8\pm1.6\pm\xx8.5$ & $52.8\pm0.4\pm0.3\pm2.5$\\
3.5--4.0 & $164.6\pm0.7\pm1.7\pm\xx6.8$ & $37.3\pm0.3\pm0.3\pm1.6$\\
4.0--4.5 & $106.9\pm0.7\pm1.6\pm\xx4.2$ & $20.6\pm0.3\pm0.3\pm1.0$\\
 \hline
\end{tabular}
\label{tab7TeV:results_Y}
\end{table}

%%%%%%%%%%%%%%%%%%%%%%%%%%%%%%%%%%%%%%%%%%%%%%%%%%%%%%%%%%
\begin{table}[htp]
\caption{Fractions of \psitwos-from-$\bquark$ (in \%) at $13\tev$ in bins of $(\pt,y)$ of \psitwos mesons. 
    The uncertainties are statistical only. The systematic uncertainties are negligible.
Adjacent bins with large statistical uncertainty have been merged.}
\centering
%\small
\begin{tabular}{l|ccccc}
\hline
\pt(\mygevc)& $2.0<y<2.5$& $2.5<y<3.0$& $3.0<y<3.5$& $3.5<y<4.0$& $4.0<y<4.5$\\
\hline
\kern 0.5em2--3 & $23.0\pm0.6$ & $20.8\pm0.4$ & $20.4\pm0.4$ & $18.1\pm0.4$ & $14.8\pm0.5$\\
\kern 0.5em3--4 & $22.4\pm0.6$ & $22.2\pm0.4$ & $21.0\pm0.4$ & $19.5\pm0.4$ & $18.4\pm0.6$\\
\kern 0.5em4--5 & $23.0\pm0.6$ & $23.9\pm0.4$ & $22.8\pm0.4$ & $21.4\pm0.4$ & $18.8\pm0.7$\\
\kern 0.5em5--6 & $25.9\pm0.6$ & $25.5\pm0.4$ & $24.4\pm0.4$ & $22.8\pm0.5$ & $18.5\pm0.7$\\
\kern 0.5em6--7 & $27.1\pm0.7$ & $27.1\pm0.5$ & $25.6\pm0.5$ & $23.3\pm0.5$ & $20.2\pm1.0$\\
\kern 0.5em7--8 & $29.8\pm0.8$ & $28.7\pm0.5$ & $26.4\pm0.6$ & $25.8\pm0.7$ & $22.1\pm1.0$\\
\kern 0.5em8--9 & $32.8\pm0.9$ & $30.6\pm0.6$ & $28.8\pm0.7$ & $26.6\pm0.8$ & $25.3\pm1.4$\\
\kern 0.5em9--10 & $34.2\pm1.0$ & $32.8\pm0.8$ & $32.4\pm0.9$ & $29.7\pm1.0$ & $28.7\pm1.7$\\
10--11 & $35.5\pm1.2$ & $35.4\pm0.9$ & $33.7\pm1.1$ & $29.7\pm1.2$ & $26.9\pm1.8$\\
11--12 & $38.5\pm1.5$ & $37.6\pm1.1$ & $34.8\pm1.4$ & $30.6\pm1.6$ & $28.1\pm2.4$\\
12--13 & $40.7\pm1.7$ & $38.7\pm1.4$ & $36.4\pm1.5$ & $34.3\pm2.0$ & $31.7\pm3.4$\\
13--14 & $42.1\pm2.1$ & $41.8\pm1.6$ & $35.9\pm1.9$ & $37.6\pm1.0$ & $31.2\pm3.9$\\  \cline{6-6}
14--15 & $44.2\pm2.1$ & $39.4\pm1.8$ & $44.7\pm2.2$ & $36.4\pm2.8$ &  \multirow{2}{*}{ $39.2\pm1.7$}   \\  \cline{5-5}
15--16 & $43.4\pm2.5$ & $42.5\pm2.1$ & $41.5\pm1.0$ & \multirow{2}{*}{ $36.8\pm2.6$} &   \\  \cline{6-6}
16--17 & $43.6\pm2.8$ & $45.9\pm2.5$ & $44.8\pm1.3$ &     & \multirow{4}{*}{ $29.0\pm3.9$} \\ \cline{4-5}
17--18 & $51.7\pm3.1$ & $47.1\pm3.0$ & \multirow{3}{*}{ $47.2\pm2.5$} & \multirow{3}{*}{ $34.7\pm3.6$} &   \\  \cline{2-3}
18--19 & \multirow{2}{*}{ $49.7\pm3.2$} & \multirow{2}{*}{ $49.1\pm2.5$} &    &    &   \\
19--20 &    &    &    &    &   \\ \cline{2-4}
 \hline
\end{tabular}
\label{tab:Fb}
\end{table}
%%%%%%%%%%%%%%%%%%%%%%%%%%%%%%%%%%%%%%%%%%%%%%%%%%%%%%%%%%

%%%%%%%%%%%%%%%%%%%%%%%%%%%%%%%%%%%%%%%%%%%%%%%%%%%%%%%%%%
\begin{table}[htp]
\caption{Fractions of \psitwos-from-$\bquark$ (in \%) at $7\tev$ in bins of $(\pt,y)$ of \psitwos mesons. 
    The uncertainties are statistical only. The systematic uncertainties are negligible.}
\centering
%\small
\begin{tabular}{l|ccccc}
\hline
\pt(\mygevc)& $2.0<y<2.5$& $2.5<y<3.0$& $3.0<y<3.5$& $3.5<y<4.0$& $4.0<y<4.5$\\
\hline
3.5--4 & $21.11\pm0.79$ & $17.91\pm0.37$ & $17.26\pm0.33$ & $16.52\pm0.34$ & $13.93\pm0.50$\\
\kern 0.75em4--5 & $21.73\pm0.53$ & $20.21\pm0.28$ & $18.42\pm0.25$ & $16.95\pm0.27$ & $15.47\pm0.40$\\
\kern 0.75em5--6 & $22.15\pm0.55$ & $22.40\pm0.31$ & $20.08\pm0.29$ & $17.89\pm0.32$ & $16.49\pm0.50$\\
\kern 0.75em6--7 & $24.12\pm0.61$ & $24.49\pm0.37$ & $21.45\pm0.35$ & $20.21\pm0.41$ & $16.18\pm0.65$\\
\kern 0.75em7--8 & $27.34\pm0.68$ & $25.04\pm0.44$ & $23.81\pm0.44$ & $21.64\pm0.52$ & $18.48\pm0.83$\\
\kern 0.75em8--9 & $28.44\pm0.80$ & $27.26\pm0.54$ & $26.81\pm0.57$ & $23.68\pm0.70$ & $19.12\pm1.10$\\
\kern 0.75em9--10 & $30.46\pm0.08$ & $29.40\pm0.68$ & $21.50\pm0.69$ & $24.95\pm0.88$ & $17.28\pm1.38$\\
\kern 0.25em10--11 & $32.97\pm1.08$ & $31.67\pm0.83$ & $28.14\pm0.87$ & $27.30\pm1.17$ & $24.10\pm1.77$\\
\kern 0.25em11--12 & $34.86\pm1.33$ & $33.31\pm1.02$ & $30.98\pm1.16$ & $28.55\pm1.44$ & $27.33\pm2.10$\\
\kern 0.25em12--13 & $35.56\pm1.62$ & $37.85\pm1.31$ & $33.30\pm1.44$ & $28.50\pm1.78$ & $20.81\pm3.01$\\
\kern 0.25em13--14 & $38.79\pm1.93$ & $36.68\pm1.51$ & $34.21\pm1.80$ & $30.97\pm2.50$ & $36.20\pm6.19$\\
 \hline
\end{tabular}
\label{tab7TeV:Fb}
\end{table}
%%%%%%%%%%%%%%%%%%%%%%%%%%%%%%%%%%%%%%%%%%%%%%%%%%%%%%%%%%
%\kern 0.75em
%\kern 0.25em

%%%%%%%%%%%%%%%%%%%%%%%%%%%%%%%%%%%%%%%%%%%%%%%%%%%%%%%%%%
\begin{table}[htp]
\caption{Ratios of production cross-sections at $13\tev$ between \psitwos mesons and \jpsi mesons in bins of \pt for prompt production and for those from $\bquark$-hadron decays integrated in the rapidity range $2.0<y<4.5$. The statistical and systematic uncertainties are added in quadrature.}
\centering
%\small
\begin{tabular}{l|cc}
\hline
\pt(\mygevc)& Prompt & From $\bquark$-hadron decays \\
\hline
\kern 0.5em2--3 & $0.135\pm0.010$ & $0.241\pm0.018$\\
\kern 0.5em3--4 & $0.158\pm0.007$ & $0.263\pm0.013$\\
\kern 0.5em4--5 & $0.182\pm0.012$ & $0.278\pm0.019$\\
\kern 0.5em5--6 & $0.201\pm0.011$ & $0.317\pm0.018$\\
\kern 0.5em6--7 & $0.225\pm0.023$ & $0.311\pm0.031$\\
\kern 0.5em7--8 & $0.259\pm0.016$ & $0.341\pm0.021$\\
\kern 0.5em8--9 & $0.271\pm0.017$ & $0.374\pm0.024$\\
\kern 0.5em9--10 & $0.284\pm0.012$ & $0.365\pm0.018$\\
10--11 & $0.310\pm0.019$ & $0.370\pm0.025$\\
11--12 & $0.305\pm0.017$ & $0.357\pm0.023$\\
12--13 & $0.315\pm0.029$ & $0.380\pm0.037$\\
13--14 & $0.294\pm0.026$ & $0.361\pm0.033$\\

 \hline
\end{tabular}
\label{tab:Ratio_jpsi_PT}
\end{table}
%%%%%%%%%%%%%%%%%%%%%%%%%%%%%%%%%%%%%%%%%%%%%%%%%%%%%%%%%%

%%%%%%%%%%%%%%%%%%%%%%%%%%%%%%%%%%%%%%%%%%%%%%%%%%%%%%%%%%
\begin{table}[htp]
\caption{Ratios of production cross-sections between \psitwos mesons and \jpsi mesons at $13\tev$ in bins of $y$ for prompt production and for those from $\bquark$-hadron decays integrated in the transverse momentum range \mbox{$2<\pt<14\gevc$}. The statistical and systematic uncertainties are added in quadrature.}
\centering
%\small
\begin{tabular}{c|cc}
\hline
$y$ & Prompt & From $\bquark$-hadron decays \\
\hline
2.0--2.5 & $0.175\pm0.011$ & $0.288\pm0.018$\\
2.5--3.0 & $0.168\pm0.010$ & $0.289\pm0.017$\\
3.0--3.5 & $0.164\pm0.009$ & $0.279\pm0.017$\\
3.5--4.0 & $0.159\pm0.010$ & $0.261\pm0.017$\\
4.0--4.5 & $0.150\pm0.012$ & $0.237\pm0.019$\\
\hline
\end{tabular}
\label{tab:Ratio_jpsi_Y}
\end{table}
%%%%%%%%%%%%%%%%%%%%%%%%%%%%%%%%%%%%%%%%%%%%%%%%%%%%%%%%%%

%%%%%%%%%%%%%%%%%%%%%%%%%%%%%%%%%%%%%%%%%%%%%%%%%%%%%%%%%%
\begin{table}[htp]
\caption{Ratios of production cross-sections between $13\tev$ and $7\tev$ in bins of \pt for prompt \psitwos and \psitwos-from-$\bquark$ mesons integrated in the rapidity range $2.0<y<4.5$. The statistical and systematic uncertainties are added in quadrature.}
\centering
%\small
\begin{tabular}{l|cc}
\hline
\pt(\mygevc)& Prompt \psitwos & \psitwos-from-$\bquark$ \\
\hline
3.5--4 & $1.63\pm0.09$ & $1.95\pm0.13$\\
\kern 0.75em4--5 & $1.52\pm0.12$ & $1.83\pm0.15$\\
\kern 0.75em5--6 & $1.61\pm0.11$ & $1.97\pm0.15$\\
\kern 0.75em6--7 & $1.67\pm0.19$ & $1.98\pm0.23$\\
\kern 0.75em7--8 & $1.72\pm0.12$ & $2.05\pm0.14$\\
\kern 0.75em8--9 & $1.97\pm0.13$ & $2.20\pm0.15$\\
\kern 0.75em9--10 & $1.79\pm0.08$ & $2.33\pm0.14$\\
\kern 0.25em10--11 & $2.00\pm0.13$ & $2.28\pm0.17$\\
\kern 0.25em11--12 & $1.87\pm0.12$ & $2.39\pm0.15$\\
\kern 0.25em12--13 & $1.83\pm0.20$ & $2.57\pm0.24$\\
\kern 0.25em13--14 & $2.18\pm0.25$ & $2.49\pm0.27$\\
 \hline
\end{tabular}
\label{tab:Ratio_7TeV_PT}
\end{table}

\begin{table}[htbp]
\caption{Ratios of cross-sections between measurements at $13\tev$ and $7\tev$ in different bins of $y$ for prompt \psitwos and \psitwos-from-$\bquark$ mesons integrated in the transverse momentum range \mbox{$3.5<\pt<14\gevc$}. The statistical and systematic uncertainties are added in quadrature.}
\centering
%\small
\begin{tabular}{c|cc}
\hline
$y$ & Prompt \psitwos & \psitwos-from-$b$ \\
\hline
2.0--2.5 & $1.81\pm0.14$ & $2.00\pm0.17$\\
2.5--3.0 & $1.54\pm0.11$ & $1.89\pm0.14$\\
3.0--3.5 & $1.54\pm0.10$ & $1.94\pm0.13$\\
3.5--4.0 & $1.54\pm0.10$ & $2.03\pm0.13$\\
4.0--4.5 & $1.69\pm0.10$ & $2.17\pm0.14$\\
 \hline
\end{tabular}
\label{tab:Ratio_7TeV_Y}
\end{table}

\pagebreak

\clearpage
\section{Scaling factors for alternative polarisation \mbox{scenarios}}
\label{sec:tables_pol}

\begin{table}[htp]
\caption{Multiplicative scaling factors needed to obtain the prompt \psitwos differential cross-sections from unpolarised cross-section measurements at $13\tev$ as reported in Table \ref{tab:results_prompt} under the assumption of fully transverse polarisation ($\alpha = +1$).}
\centering
%\small
\begin{tabular}{l|ccccc}
\hline
\pt(\mygevc)& $2.0<y<2.5$& $2.5<y<3.0$& $3.0<y<3.5$& $3.5<y<4.0$& $4.0<y<4.5$\\
\hline
\kern 0.5em2--3 & $1.084$ & $1.081$ & $0.997$ & $0.984$ & $0.993$\\
\kern 0.5em3--4 & $1.202$ & $1.047$ & $0.964$ & $0.950$ & $0.948$\\
\kern 0.5em4--5 & $1.164$ & $1.034$ & $0.961$ & $0.937$ & $0.921$\\
\kern 0.5em5--6 & $1.138$ & $1.016$ & $0.959$ & $0.940$ & $0.922$\\
\kern 0.5em6--7 & $1.100$ & $1.012$ & $0.970$ & $0.939$ & $0.923$\\
\kern 0.5em7--8 & $1.110$ & $0.996$ & $0.965$ & $0.937$ & $0.915$\\
\kern 0.5em8--9 & $1.120$ & $0.993$ & $0.962$ & $0.943$ & $0.937$\\
\kern 0.5em9--10  & $1.037$ & $0.985$ & $0.962$ & $0.940$ & $0.941$\\
10--11 & $1.070$ & $0.941$ & $0.941$ & $0.960$ & $0.910$\\
11--12 & $ 1.022$ & $0.952$ & $0.958$ & $0.926$ & $0.925$\\
12--13 & $ 1.096$ & $0.940$ & $0.932$ & $0.944$ & $0.897$\\
13--14 & $ 1.007$ & $0.964$ & $0.949$ & $0.932$ & $0.924$\\  \cline{6-6}
14--15 &   $ 1.060$ &   $0.944$ &   $0.939$ &   $0.924$ &  \multirow{2}{*}{$0.909$}   \\  \cline{5-5}
15--16 &   $ 0.932$ &   $0.940$ &   $0.910$ & \multirow{2}{*}{$0.931$} &   \\  \cline{6-6}
16--17 &    $ 1.051$ &    $0.932$ &    $0.930$ & & \multirow{4}{*}{$0.850$} \\ \cline{4-5}
17--18 &    $ 1.080$ &    $0.946$ & \multirow{3}{*}{$0.924$} & \multirow{3}{*}{$0.911$} &   \\  \cline{2-3}
18--19 & \multirow{2}{*}{$0.969$} & \multirow{2}{*}{$0.934$} &    &    &   \\
19--20 &    &    &    &    &   \\ \cline{2-4}
 \hline
\end{tabular}
\label{tab:corr_factors_13_+1}
\end{table}

\begin{table}[htp]
\caption{Multiplicative scaling factors needed to obtain the prompt \psitwos differential cross-sections from unpolarised cross-section measurements at $13\tev$ as reported in Table \ref{tab:results_prompt} under the assumption of fully longitudinal polarisation ($\alpha = -1$).}
\centering
%\small
\begin{tabular}{l|ccccc}
\hline
\pt(\mygevc)& $2.0<y<2.5$& $2.5<y<3.0$& $3.0<y<3.5$& $3.5<y<4.0$& $4.0<y<4.5$\\
\hline
\kern 0.5em2--3 & $0.846$ & $0.870$ & $0.987$ & $1.016$ & $1.008$\\
\kern 0.5em3--4 & $0.754$ & $0.909$ & $1.039$ & $1.079$ & $1.092$\\
\kern 0.5em4--5 & $0.777$ & $0.934$ & $1.053$ & $1.092$ & $1.123$\\
\kern 0.5em5--6 & $0.806$ & $0.950$ & $1.051$ & $1.091$ & $1.132$\\
\kern 0.5em6--7 & $0.829$ & $0.958$ & $1.040$ & $1.069$ & $1.129$\\
\kern 0.5em7--8 & $0.843$ & $0.977$ & $1.041$ & $1.079$ & $1.102$\\
\kern 0.5em8--9 & $0.865$ & $0.992$ & $1.054$ & $1.073$ & $1.120$\\
\kern 0.5em9--10  & $0.908$ & $1.004$ & $1.061$ & $1.066$ & $1.129$\\
10--11 & $ 0.882$ & $1.024$ & $1.068$ & $1.098$ & $1.096$\\
11--12 & $ 0.944$ & $1.035$ & $1.077$ & $1.069$ & $1.114$\\
12--13 & $ 0.941$ & $1.050$ & $1.079$ & $1.076$ & $1.126$\\
13--14 & $ 0.992$ & $1.069$ & $1.067$ & $1.100$ & $1.115$\\  \cline{6-6}
14--15 &   $ 0.966$ &   $1.076$ &   $1.085$ &   $1.114$ &  \multirow{2}{*}{$1.143$}   \\  \cline{5-5}
15--16 &   $ 0.964$ &   $1.082$ &   $1.142$ & \multirow{2}{*}{$1.102$} &   \\  \cline{6-6}
16--17 &    $ 1.011$ &    $1.099$ &    $1.102$ &     & \multirow{4}{*}{$1.013$} \\ \cline{4-5}
17--18 &    $ 0.992$ &    $1.074$ & \multirow{3}{*}{$1.114$} & \multirow{3}{*}{$1.140$} &   \\  \cline{2-3}
18--19 & \multirow{2}{*}{$1.037$} & \multirow{2}{*}{$1.094$} &    &    &   \\
19--20 &    &    &    &    &   \\ \cline{2-4}
 \hline
\end{tabular}
\label{tab:corr_factors_13_-1}
\end{table}

\begin{table}[htp]
\caption{Multiplicative scaling factors needed to obtain the prompt \psitwos differential cross-sections from unpolarised cross-section measurements at $13\tev$ as reported in Table \ref{tab:results_prompt} under the assumption of 20 \%  longitudinal polarisation ($\alpha = -0.2$), corresponding to a conservative limit of the \psitwos polarisation measured at $7\tev$~\cite{LHCb-PAPER-2013-067}.}
\centering
%\small
\begin{tabular}{l|ccccc}
\hline
\pt(\mygevc)& $2.0<y<2.5$& $2.5<y<3.0$& $3.0<y<3.5$& $3.5<y<4.0$& $4.0<y<4.5$\\
\hline
\kern 0.5em2--3 & $0.975$ & $0.976$ & $0.997$ & $1.001$ & $0.998$\\
\kern 0.5em3--4 & $0.955$ & $0.983$ & $1.005$ & $1.013$ & $1.012$\\
\kern 0.5em4--5 & $0.955$ & $0.990$ & $1.010$ & $1.012$ & $1.012$\\
\kern 0.5em5--6 & $0.963$ & $0.988$ & $1.007$ & $1.014$ & $1.018$\\
\kern 0.5em6--7 & $0.960$ & $0.990$ & $1.008$ & $1.004$ & $1.020$\\
\kern 0.5em7--8 & $0.976$ & $0.991$ & $1.006$ & $1.007$ & $1.004$\\
\kern 0.5em8--9 & $0.993$ & $0.998$ & $1.010$ & $1.008$ & $1.025$\\
\kern 0.5em9--10 & $0.973$ & $0.998$ & $1.013$ & $1.003$ & $1.033$\\
10--11 & $ 0.981$ & $0.984$ & $1.004$ & $1.030$ & $0.999$\\
11--12 & $ 0.986$ & $0.996$ & $1.019$ & $0.996$ & $1.016$\\
12--13 & $ 1.023$ & $0.995$ & $1.004$ & $1.010$ & $1.005$\\
13--14 & $ 1.005$ & $1.021$ & $1.008$ & $1.014$ & $1.016$\\  \cline{6-6}
14--15 & $ 1.024$ & $1.010$ & $1.011$ & $1.016$ &  \multirow{2}{*}{$1.019$}\\  \cline{5-5}
15--16 & $ 0.948$ & $1.011$ & $1.020$ & \multirow{2}{*}{$1.014$} &   \\  \cline{6-6}
16--17 & $ 1.049$ & $1.013$ & $1.013$ & & \multirow{4}{*}{$0.929$} \\ \cline{4-5}
17--18 & $ 1.086$ & $1.009$ & \multirow{3}{*}{$1.015$} & \multirow{3}{*}{$1.019$} &   \\  \cline{2-3}
18--19 & \multirow{2}{*}{$1.004$} & \multirow{2}{*}{$1.013$} &    &    &   \\  
19--20 &    &    &    &    &   \\ \cline{2-4}
\hline
\end{tabular}
\label{tab:corr_factors_13_-0.2}
\end{table}

%%%%%%%%%%%%%%%%%%%%%%%%%%%%%%%%%%%%%%%%%%%%%%%%%%%%%%%%%%
\begin{table}[htp]
\caption{Multiplicative scaling factors needed to obtain the prompt \psitwos differential cross-sections from unpolarised cross-section measurements at $7\tev$ as reported in Table \ref{tab7TeV:results_prompt} under the assumption of fully transverse  polarisation ($\alpha = +1$).}
\centering
%\small
\begin{tabular}{l|ccccc}
\hline
\pt(\mygevc)& $2.0<y<2.5$& $2.5<y<3.0$& $3.0<y<3.5$& $3.5<y<4.0$& $4.0<y<4.5$\\
\hline
3.5--4 & $1.087$ & $1.288$ & $1.164$ & $1.111$ & $1.079$\\
\kern 0.75em4--5 & $1.435$ & $1.260$ & $1.157$ & $1.103$ & $1.067$\\
\kern 0.75em5--6 & $1.397$ & $1.231$ & $1.156$ & $1.102$ & $1.066$\\
\kern 0.75em6--7 & $1.362$ & $1.220$ & $1.144$ & $1.102$ & $1.071$\\
\kern 0.75em7--8 & $1.305$ & $1.186$ & $1.142$ & $1.097$ & $1.072$\\
\kern 0.75em8--9 & $1.287$ & $1.184$ & $1.114$ & $1.088$ & $1.074$\\
\kern 0.75em9--10 & $1.250$ & $1.150$ & $1.124$ & $1.086$ & $1.071$\\
\kern 0.25em10--11 & $1.350$ & $1.218$ & $1.188$ & $1.162$ & $1.106$\\
\kern 0.25em11--12 & $1.347$ & $1.209$ & $1.091$ & $1.143$ & $1.085$\\
\kern 0.25em12--13 & $1.246$ & $1.222$ & $1.127$ & $1.139$ & $1.103$\\
\kern 0.25em13--14 & $1.253$ & $1.162$ & $1.181$ & $1.115$ & $1.057$\\
 \hline
\end{tabular}
\label{tab:corr_factors_7_+1}
\end{table}
%%%%%%%%%%%%%%%%%%%%%%%%%%%%%%%%%%%%%%%%%%%%%%%%%%%%%%%%%%%%%%%%%

%%%%%%%%%%%%%%%%%%%%%%%%%%%%%%%%%%%%%%%%%%%%%%%%%%%%%%%%%%
\begin{table}[htp]
\caption{Multiplicative scaling factors needed to obtain the prompt \psitwos differential cross-sections from unpolarised cross-section measurements at $7\tev$ as reported in Table \ref{tab7TeV:results_prompt} under the assumption of fully longitudinal  polarisation ($\alpha = -1$).}
\centering
%\small
\begin{tabular}{l|ccccc}
\hline
\pt(\mygevc)& $2.0<y<2.5$& $2.5<y<3.0$& $3.0<y<3.5$& $3.5<y<4.0$& $4.0<y<4.5$\\
\hline
3.5--4 & $0.858$ & $0.673$ & $0.784$ & $0.847$ & $0.893$\\
\kern 0.75em4--5 & $0.583$ & $0.697$ & $0.792$ & $0.854$ & $0.908$\\
\kern 0.75em5--6 & $0.596$ & $0.716$ & $0.788$ & $0.851$ & $0.906$\\
\kern 0.75em6--7 & $0.616$ & $0.726$ & $0.802$ & $0.851$ & $0.893$\\
\kern 0.75em7--8 & $0.627$ & $0.734$ & $0.780$ & $0.837$ & $0.874$\\
\kern 0.75em8--9 & $0.647$ & $0.734$ & $0.814$ & $0.850$ & $0.877$\\
\kern 0.75em9--10 & $0.676$ & $0.769$ & $0.808$ & $0.846$ & $0.873$\\
\kern 0.25em10--11 & $0.743$ & $0.821$ & $0.833$ & $0.868$ & $0.896$\\
\kern 0.25em11--12 & $0.745$ & $0.845$ & $0.940$ & $0.879$ & $0.882$\\
\kern 0.25em12--13 & $0.769$ & $0.820$ & $0.874$ & $0.861$ & $0.908$\\
\kern 0.25em13--14 & $0.802$ & $0.872$ & $0.834$ & $0.922$ & $0.947$\\
 \hline
\end{tabular}
\label{tab:corr_factors_7_-1}
\end{table}
%%%%%%%%%%%%%%%%%%%%%%%%%%%%%%%%%%%%%%%%%%%%%%%%%%%%%%%%%%%%%%%%%

%%%%%%%%%%%%%%%%%%%%%%%%%%%%%%%%%%%%%%%%%%%%%%%%%%%%%%%%%%
\begin{table}[htp]
\caption{Multiplicative scaling factors needed to obtain the prompt \psitwos differential cross-sections from unpolarised cross-section measurements at $7\tev$ as reported in Table \ref{tab7TeV:results_prompt} under the assumption of 20 \%  longitudinal polarisation ($\alpha = -0.2$), corresponding to a conservative limit of the \psitwos polarisation measured at $7\tev$~\cite{LHCb-PAPER-2013-067}}
\centering
%\small
\begin{tabular}{l|ccccc}
\hline
\pt(\mygevc)& $2.0<y<2.5$& $2.5<y<3.0$& $3.0<y<3.5$& $3.5<y<4.0$& $4.0<y<4.5$\\
\hline
3.5--4 & $0.980$ & $0.944$ & $0.967$ & $0.979$ & $0.986$\\
\kern 0.75em4--5 & $0.922$ & $0.949$ & $0.969$ & $0.980$ & $0.989$\\
\kern 0.75em5--6 & $0.923$ & $0.951$ & $0.967$ & $0.979$ & $0.988$\\
\kern 0.75em6--7 & $0.928$ & $0.953$ & $0.969$ & $0.978$ & $0.986$\\
\kern 0.75em7--8 & $0.917$ & $0.945$ & $0.957$ & $0.969$ & $0.978$\\
\kern 0.75em8--9 & $0.928$ & $0.944$ & $0.963$ & $0.972$ & $0.980$\\
\kern 0.75em9--10 & $0.930$ & $0.953$ & $0.963$ & $0.969$ & $0.977$\\
\kern 0.25em10--11 & $1.013$ & $1.014$ & $1.007$ & $1.017$ & $1.008$\\
\kern 0.25em11--12 & $1.016$ & $1.023$ & $1.020$ & $1.013$ & $0.988$\\
\kern 0.25em12--13 & $0.991$ & $1.013$ & $1.003$ & $1.001$ & $1.010$\\
\kern 0.25em13--14 & $1.011$ & $1.018$ & $1.003$ & $1.022$ & $1.005$\\
 \hline
\end{tabular}
\label{tab:corr_factors_7_-0.2}
\end{table}
%%%%%%%%%%%%%%%%%%%%%%%%%%%%%%%%%%%%%%%%%%%%%%%%%%%%%%%%%%%%%%%%%
\clearpage
\addcontentsline{toc}{section}{References}
\setboolean{inbibliography}{true}
\bibliographystyle{LHCb}
\bibliography{main,LHCb-PAPER,LHCb-CONF,LHCb-DP,LHCb-TDR,local}

\newpage

% Author List ---------------------------
\newpage
% LHCb collaboration author list
% Data extracted on August 6th, 2019 at 1:42pm for reference date 11-Dec-2018
\centerline
{\large\bf LHCb collaboration}
\begin
{flushleft}
\small
R.~Aaij$^{30}$,
C.~Abell{\'a}n~Beteta$^{47}$,
B.~Adeva$^{44}$,
M.~Adinolfi$^{51}$,
C.A.~Aidala$^{77}$,
Z.~Ajaltouni$^{8}$,
S.~Akar$^{62}$,
P.~Albicocco$^{21}$,
J.~Albrecht$^{13}$,
F.~Alessio$^{45}$
% ...

{\footnotesize \it

% ...
$ ^{8}$Universit{\'e} Clermont Auvergne, CNRS/IN2P3, LPC, Clermont-Ferrand, France\\
% ...
$ ^{13}$Fakult{\"a}t Physik, Technische Universit{\"a}t Dortmund, Dortmund, Germany\\
% ...
$ ^{21}$INFN Laboratori Nazionali di Frascati, Frascati, Italy\\
% ...
$ ^{30}$Nikhef National Institute for Subatomic Physics, Amsterdam, Netherlands\\
% ...
$ ^{44}$Instituto Galego de F{\'\i}sica de Altas Enerx{\'\i}as (IGFAE), Universidade de Santiago de Compostela, Santiago de Compostela, Spain\\
$ ^{45}$European Organization for Nuclear Research (CERN), Geneva, Switzerland\\
% ...
$ ^{47}$Physik-Institut, Universit{\"a}t Z{\"u}rich, Z{\"u}rich, Switzerland\\
% ...
$ ^{51}$H.H. Wills Physics Laboratory, University of Bristol, Bristol, United Kingdom\\
% ...
$ ^{62}$University of Cincinnati, Cincinnati, OH, United States\\
% ...
$ ^{64}$Syracuse University, Syracuse, NY, United States\\
% ...
$ ^{77}$University of Michigan, Ann Arbor, United States, associated to $^{64}$\\
% ...
}
\end{flushleft}

%The author list for journal publications is generated from the
%Membership Database shortly after 'approval to go to paper' has been
%given.  It will be sent to you by email shortly after a paper number
%has been assigned.  The author list should be included in the draft used for 
%first and second circulation, to allow new members of the collaboration to verify
%that they have been included correctly. Occasionally a misspelled
%name is corrected, or associated institutions become full members.
%Therefore an updated author list will be sent to you after the final
%EB review of the paper.  In case line numbering doesn't work well
%after including the authorlist, try moving the \verb!\bigskip! after
%the last author to a separate line.
%
%
%The authorship for Conference Reports should be ``The LHCb
%collaboration'', with a footnote giving the name(s) of the contact
%author(s), but without the full list of collaboration names.

\end{document}